\documentclass[10pt,aps,prd,superscriptaddress,nofootinbib,nobibnotes,longbibliography,floatfix,twocolumn]{revtex4-2}

\usepackage{comment}
\usepackage{booktabs,makecell}
\usepackage{graphicx}
\usepackage{amsmath,amssymb}
\usepackage{amsfonts}
\usepackage{xspace} 
\usepackage[usenames]{color}
\usepackage{dcolumn}
\usepackage{bm}
\usepackage{mathrsfs}
\usepackage[colorlinks=true,citecolor=blue]{hyperref}
\usepackage[all]{hypcap} 
\usepackage[utf8]{inputenc} 
\usepackage{slashed}
\usepackage{multirow}
\usepackage{float}
\usepackage{rotating}
\usepackage{longtable}
\definecolor{orange}{rgb}{1,0.5,0}
\usepackage{tabularx}
\usepackage{placeins}
\usepackage[export]{adjustbox}

\usepackage[normalem]{ulem}
\usepackage{aas_macros}

\usepackage[dvipsnames]{xcolor}

\usepackage{tabularray}
\usepackage{mathtools}
\usepackage{multirow}
\usepackage{orcidlink}
\hypersetup{
    colorlinks = true,
    linkcolor = {blue},
    citecolor = {blue},
    urlcolor = {blue},
    linkbordercolor = {white},
    }
    
\definecolor{initgreen}{HTML}{3CB44B}
\definecolor{inityellow}{HTML}{FFE119}

\newcommand{\dataavailability}[1]{%
  \section*{Data Availability}
  #1
}

\begin{document}

\title{Black hole spectroscopy of collapsing and merging neutron stars}

\newcommand{\affilaei}{Max Planck Institute for Gravitational Physics (Albert Einstein Institute), Am Mühlenberg 1, Potsdam 14476, Germany}
\newcommand{\affilpotsdam}{Institut für Physik und Astronomie, Universität Potsdam, Haus 28, Karl-Liebknecht-Str. 24-25, Potsdam, Germany}

\author{Oliver Steppohn\, \orcidlink{0009-0002-8172-3552}}
\email{oliver.steppohn@uni-potsdam.de}
\affiliation{\affilpotsdam}

\author{Sebastian H.\,V\"olkel\,\orcidlink{0000-0002-9432-7690}}
\email{sebastian.voelkel@aei.mpg.de}
\affiliation{\affilaei}

\author{Tim Dietrich\,\orcidlink{0000-0003-2374-307X}}
\email{tim.dietrich@uni-potsdam.de}
\affiliation{\affilpotsdam}
\affiliation{\affilaei}

\date{\today}

\begin{abstract}
Black hole spectroscopy is an important pillar when studying gravitational waves from black holes and
enables tests of general relativity. Most of the gravitational wave signals observed over the last decade
originate from binary black hole systems. Binary neutron star or black hole neutron-star systems are rarer
but of particular interest for the next-generation ground-based gravitational wave detectors. These events
offer the exciting possibility of studying matter effects on the ringdown of ``dirty black holes.'' In this work,
we ask the question; \textit{Does matter matter?} Using numerical-relativity, we simulate a wide range of
collapsing neutron stars producing matter environments, both in isolated scenarios and in binary mergers.
Qualitatively, the resulting ringdown signals can be classified into ``clean,'' ``modified,'' and ``distorted''
cases, depending on the amount of matter that is present. We apply standard strategies for extracting
quasinormal modes of clean signals, using both theory-agnostic and theory-specific assumptions. Even in
the presence of matter, possible modifications of quasinormal modes seem to be dominated by ringdown
modeling systematics. We find that incorporating multiple quasinormal modes allows one to drastically
reduce mismatches and errors in estimating the final black hole mass at early times. If not treated carefully,
deviations in the fundamental quasinormal mode might artificially be overestimated and falsely attributed
to the presence of matter or violations of general relativity.
\end{abstract}

\maketitle

\section{Introduction}\label{introduction}

One decade ago, gravitational-wave (GW) astronomy reached a major milestone with the observation of a binary black hole (BBH) merger, GW150914~\cite{Abbott2016GW150914,Abbott2016Results}. The signal observed by LIGO detectors originated from the collision of two stellar mass black holes (BHs), which spiraled around each other and migrated inwards as they lost energy from the emission of GWs. 
Since then, about 100 BBH signals have been observed~\cite{LIGOScientific:2018mvr,LIGOScientific:2020ibl,LIGOScientific:2020ibl,KAGRA:2021duu,KAGRA:2021vkt}, and the high number of alerts in the current observing run shows the potential for precision science based on GW observations~\cite{Akyuz:2025ype}. 
While most transients turned out to be BBH mergers, there were also systems that contained at least one neutron star (NS). The most famous was the binary neutron star (BNS) merger GW170817~\cite{LIGOScientific:2017vwq}, which was connected with a multimessenger detection of a short gamma-ray burst, the gamma-ray burst afterglow, and a kilonova~\cite{LIGOScientific:2017ync}. In addition to GW170817, there has been GW190425~\cite{LIGOScientific:2020aai}, likely a BNS merger event with unusually high NS masses, and numerous black hole-neutron star (BHNS) detections, e.g., \cite{LIGOScientific:2021qlt,LIGOScientific:2024elc}. 
Observing these GW transients enables us to learn more about the composition of NSs (see~\cite{Koehn:2024set,Chatziioannou:2024tjq} for recent reviews) and to perform tests on the validity of general relativity (GR) \cite{TestsGR2016,TestsGR2018,TestsGR2019}.

One can also use GWs to test the \textit{no-hair theorem}~\cite{Israel1967,Carter1971,Robinson1975} of BHs, which suggest that they are solely characterized by three parameters: mass, angular momentum, and electrical charge, via the Schwarzschild, Kerr, and Kerr-Newman solutions~\cite{Schwarzschild:1916uq,Kerr:1963ud,Newman:1965my} (in increasing generality). 
Departures from this theorem raise the possibility that BHs may have hair, which would challenge our perspectives on BHs and GR.

By concentrating on the \textit{ringdown}, it is possible to test the no-hair theorem's assumptions. A BH formed from the dynamical processes involving compact objects ``rings down'' from an initial state of nonequilibrium and emits GW radiation in form of damped sinusoids. 
Pioneering works first targeted nonrotating BHs~\cite{Regge1957,Zerilli1970,Zerilli1970b,Vishveshwara1970}, and then rotating ones~\cite{Teulkosky1973}. 
They discovered how a BH responds to an incoming GW pulse: 
it oscillates with characteristic damped frequencies. 
The damping is due to GWs in the vicinity of the horizon falling into the BH, while GWs far away leave the system. 
The ringdown is a consequence of both conditions. This stage is anticipated to occur as a result of the coalescence of compact objects or in other instances where BH formation occurs, such as the collapse of a perturbed NS. 
Seminal studies probing the nonlinear BH ringdown with numerical relativity can be found in Ref.~\cite{Buonanno:2006ui,Berti:2007fi,London:2014cma}.

The aforementioned characteristic oscillations of a BH are called \textit{quasinormal modes} (QNMs), and we refer to Refs.~\cite{Kokkotas:1999bd,Nollert:1999ji,Berti:2009kk,Konoplya:2011qq,Berti:2018vdi,Berti2016,Franchini:2023eda,Berti:2025hly} for reviews on this topic. 
QNM frequencies and damping times are labeled by harmonics $(\ell,m)$ and an overtone index $n$. 
For astrophysical BHs, they only depend on the mass and spin of the remnant BH, and there are infinitely many. 
Extracting more than one complex QNM, thus, allows for constraining mass and spin, while any additional QNM can be used for consistency checks. 
This is one key promise of \textit{black hole spectroscopy} \cite{1980ApJ...239..292D,Dreyer2003,Berti:2005ys}. 

In vacuum and in GR, the QNM spectrum is well known~\cite{Berti:2009kk}, while it requires strong simplifications to examine the QNM content in matter environments to make the equations decouple and make the analysis tractable. Maintaining the vacuum frequencies at a constant value from the start, allows for a simplified extraction of the complex-valued amplitudes $C_{\ell mn}$ from the signal. The amplitudes are fitted using standard least-squares fitting procedures~\cite{Giesler2019,Cheung2023,Baibhav2023}, while the variable projection method~\cite{Giesler2024Overtones,varpro} is employed for the nonlinear fitting problem. Fitting clean vacuum waveforms~\cite{Boyle2019,Scheel:2025jct} with these techniques opens the possibility of extracting dozens of QNMs, although their stability and robustness are still not universally defined and depend on the methods used~\cite{Gao2025}. Additional caveats pertain to the fit's starting time, which is the point at which perturbation theory is considered a valid description and the BH is described by a Kerr spacetime~\cite{Bhagwat2017}.

The presence of matter is expected in astrophysical environments and could potentially affect our ability to perform precision GW astronomy. A comprehensive examination of this issue has been conducted in Ref.~\cite{Barausse2014} for tractable cases. Because this subject is technically complex, there have been attempts to calculate QNM frequencies from modified/perturbed potentials that are intended to resemble deviations in the underlying perturbation equations~\cite{Nollert1996,Leung1997,Leung1999,Cheung:2021bol}.  
More recently, studies considering an analytical solution for a nonrotating BH that is surrounded by an accretion disk were able to extract QNM frequencies of supermassive BHs, e.g., \cite{Cardoso2021,Capuano:2024qhv,Pezzella2024}. 
A thin shell of matter was recently introduced as a modification to the Regge-Wheeler potential of a Schwarzschild BH in Ref.~\cite{Laeuger2025}. They demonstrate how a thin shell is the cause of secondary GW emission and can dramatically shift QNM frequencies for larger mass distributions. 
A method for phenomenologically inferring accretion disk properties from BNS mergers with features after the ringdown was presented in Ref.~\cite{Dhani:2025xno}, and remnant properties have been studied in Ref.~\cite{Dhani:2025axt}. Additionally, Ref.~\cite{Bandyopadhyay:2023ohl} used BNS merger simulations to attest whether QNM amplitudes can aid in determining pre-merger binary properties such as the mass ratio.

In this work\footnote{Our article is based on the Master thesis `Black hole spectroscopy of collapsing and merging neutron stars' by O.~Steppohn leading to overlap in the presented results and figures, and, partially, in existing text passages.}, we use a statistically significant set of numerical-relativity simulations to explore whether environmental effects impact precision black hole spectroscopy. 
This will assist in comprehending the extent to which the dynamical behavior of matter affects the BH ringdown beyond any theoretical approximations. Initially, we evolve several isolated, rotating NS initial data that are induced with small pressure perturbations, which ultimately leads to gravitational collapse. From these simulations, we conduct a systematic assessment of matter effects on the BH ringdown.
Systems can be divided into three groups based on our analysis: \texttt{Class I} systems exhibit a clean ringdown, \texttt{Class II} systems exhibit additional features that are excited during the initial ringdown signal, and \texttt{Class III} systems exhibit extremely dynamic matter environments that completely distort the signal and eliminate typical ringdown characteristics. 
In addition, we generate and evaluate BNS merger simulations to broaden our analysis to scenarios that are more astrophysically relevant. 
The ringdown's modeling systematics are likely to dominate the deviations of the fundamental mode frequency from the Kerr predictions, as demonstrated by theory-specific and theory-agnostic assumptions, such as those in~\cite{Nee:2023osy,Baibhav2023}. 
The inclusion of multiple QNMs is essential for the significant reduction of mismatches and errors in the early time extraction of the BH mass. This is comparable to the analysis of vacuum numerical-relativity simulations~\cite{Giesler2019,Cheung2023,Baibhav2023,Giesler2024Overtones,Gao2025}. 
If not carefully accounted for, deviations in the fundamental QNM may be exaggerated and mistakenly linked to the presence of matter or false violations of GR. 

Our work is structured as follows. In Sec.~\ref{numsetup}, we outline the numerical setups used for the gravitational collapse simulations and for the evolution of the BNS mergers. Section~\ref{classAnalysis} introduces the aforementioned ringdown classification scheme applied to the isolated NS cases and shows selected single and binary ringdown simulations. In Sec.~\ref{ringdown_analysis}, we outline our QNM fitting algorithm and follow up with results in Sec.~\ref{moreResults}, for both, isolated and BNS simulations. 
Our conclusions can be found in Sec.~\ref{conclusion}. 
Throughout the paper, we use geometric units and set $G=c=M_\odot=1$ (exceptions are indicated throughout).

\section{Numerical setup}
\label{numsetup}

For our simulations, we are using the \texttt{BAM} code \cite{Bruegmann:2006ulg,Thierfelder:2011yi}; cf.~for a detailed introduction to the topic of numerical-relativity~\cite{Pretorius:2005gq,baumgarte2010numerical,rezzolla2013relativistic,ShibateNRbook,Palenzuela2020}. 
Details about the simulations, specific to \texttt{BAM}, will be explained in Sec.~\ref{NUMSETBAM}. 
In the following, we shortly describe the construction of the differentially rotating neutron star (dRNS) models and the initial data for our BNS systems. Afterwards, we present key ingredients for the dynamical evolutions. 

\subsection{Initial data}
\subsubsection{Isolated stars with differential rotation}\label{dRNSinit}

In order to create the initial data for the dRNS systems, we start by considering the stationary and axisymmetric line element
\begin{align}
    ds^2=-&e^{\gamma+\rho}dt^2+e^{\gamma-\rho}r^2\sin^2{\theta}(d\phi-\omega dt)^2\\
    &+e^{2\mu}(dr^2+r^2d\theta^2),\nonumber
\end{align}
where $\gamma$, $\rho$, $\omega$, and $\mu$ are the metric potentials. For the matter, we assume a perfect fluid stress-energy tensor and a polytropic equation of state (EOS) relating pressure $p$ and energy density 
$\varepsilon$ via
\begin{align}
p = \kappa \varepsilon^{1+1/N} \,,
\end{align}
with the polytropic index $N=1$ and the polytropic constant $\kappa=100$. Although such an EOS is simplified, it still allows us to study a variety of different sources and reduces possible uncertainties due to the usage of more complicated nuclear physics EOSs. 

For the rotation law, we use the \textit{j-constant} law
\begin{align}\label{eq:rotlaw}
    j(\Omega)=A^2(\Omega_c-\Omega), 
\end{align}
where $j$ is the specific angular momentum, $A$ is the rotation parameter and $\Omega_c$ is the angular velocity at the center of the star. This rotation law is related to the metric potentials via
\begin{align}
    A^2(\Omega_c-\Omega)=\frac{(\Omega-\omega)r^2\sin^2{\theta}e^{2(\beta-\nu)}}{1-(\Omega-\omega)r^2\sin^2{\theta}e^{2(\beta-\nu)}}.
\end{align}
While we only use the $j$-constant rotation law for our analysis, we also want to point towards other choices that have been suggested in the literature, e.g., \cite{Galeazzi2011,Uryu2016,Uryu2017,Bauswein2017}. 

The equilibrium models are then computed using the iterative Komatsu-Eriguchi-Hachisu (KEH) method~\cite{Komatsu1989}. The publicly available \texttt{RNS} code \cite{Stergioulas1995,Stergioulas2003} implements this scheme and is able to compute equilibrium models of rotating NSs, however, only for stars with uniform rotation. 
We use a modified version of \texttt{RNS} which can handle differential rotation, using modifications to the KEH method based on \cite{Cook1992,Cook1994}.

For our ringdown analysis, we want to probe a large parameter space of models to identify possible trends. 
Following the approach outlined above, we have three free parameters that determine our equilibrium models (for a given EOS): the rescaled rotation parameter $\hat{A}$ (see the definition in \cite{Cook1992,Cook1994}), the axis ratio $r_p/r_e$ ($r_p$ is the polar and $r_e$ the equatorial radius), and the central energy density $\rho_c$. We cover the parameter space with $\hat{A}$ in the interval $[0.1,1.0]$ in steps of $0.1$, and the axis ratio in the interval $[0.35,0.95]$ in steps of $0.05$, resulting in $[\hat{A}]\times [r_p/r_e]=130$ models. 

Each pair $(\hat{A},r_p/r_e)$ has its own sequence in the $M-\varepsilon_c$ plane over different central energy densities. To get the most massive models at the extremal points, we run \texttt{RNS} for a given pair $(\hat{A},r_p/r_e)$ over a range of central energy densities to create the sequence for this particular model. We use 20 different densities in the range $[0.0005,0.005]$ and then apply quadratic interpolation to find the maximum. All of the 130 models, grouped by the rotation parameter $\hat{A}$, are further described in Appendix~\ref{app:RNSseq}. Corresponding values for the mass $M$, the equatorial radius $R_e$, and the central angular velocity $\Omega_c$ are listed in Table~\ref{tab:inittab}.  The data for all models, and instructions on how to visualize them, is made publicly available under~\cite{osteppohn2025repo}.

\subsubsection{BNS initial data}
For BNS evolutions, we construct initial data using the pseudospectral code \texttt{SGRID}~\cite{Tichy:2009yr,Tichy:2012rp,Dietrich:2015pxa,Tichy:2019ouu}, which is one of the standard approaches for BNS simulations performed with \texttt{BAM}. 
\texttt{SGRID} solves the Einstein constraint equations using the extended conformal thin sandwich approach~\cite{York:1998hy,Pfeiffer:2002iy}. 
For the construction of the initial data, we use the piecewise-polytropic fit of~\cite{Read:2008iy} of the SLy EOS~\cite{Douchin:2001sv} and add thermal effects through an additional thermal pressure following \cite{Bauswein:2010dn} with $\Gamma_{\rm th} = 1.75$. \par

\subsection{Evolution}\label{NUMSETBAM}
We use \texttt{BAM}~\cite{Bruegmann:2006ulg,Thierfelder:2011yi,Dietrich:2015iva,Bernuzzi:2016pie} for our dynamical evolutions, where we employ the Z4c formulation of the field equations of GR~\cite{Bernuzzi:2009ex,Hilditch:2012fp}. For the gauge, we are using the moving puncture gauge (1+log-slicing and $\Gamma$-driver shift conditions~\cite{Bona:1994dr,Alcubierre:2002kk,vanMeter:2006vi}). The matter variables are evolved using the Valencia formulation of general-relativistic hydrodynamics~\cite{Marti:1991wi,Banyuls:1997zz,Anton:2005gi}. 

BAM's grid infrastructure consists of a variable number of cell-centered nested grids. Each level contains one or more Cartesian boxes with a constant grid spacing and number of points. Due to the 2:1 refinement strategy, the resolution in each finer level is twice the resolution of the parent level. Inner levels can move dynamically according to the technique of `moving boxes’ and follow the motion of the NSs in our BNS cases. 

For the time evolution, we use the method of lines with a fourth-order Runge-Kutta scheme and a Courant factor of $0.25$. The spacetime sector uses a finite difference scheme with centered fourth-order stencils to compute derivatives. Hydrodynamics variables are instead modeled by a finite volumes formalism with high-resolution shock-capturing schemes to compute numerical fluxes between cells. As approximate Riemann solvers, we use the Harten, Lax, and van Leer (HLL) \cite{doi:10.1137/1025002} method for the BNSs and the local Lax Friedrich Riemann solver~\cite{NESSYAHU1990408, 2000JCoPh.160..241K} for the dRNS setups. Reconstruction is done with the WENOZ scheme~\cite{2008JCoPh.227.3191B}. In order to enforce the conservation of energy-momentum and baryon number, we apply the flux corrections of the conservative adaptive mesh refinement implemented in \cite{Dietrich:2015iva}. BHs in our simulations are identified by the presence of an apparent horizon (AH), which is defined as the outermost 2-surface whose outgoing future null geodesics have zero expansion everywhere. To ensure that the BHs, formed during the BNS merger, are sufficiently well resolved, we add additional refinement levels shortly before the BH formation. Details about the simulation setups can be found in Table~\ref{tab:numset}.

\begin{table}[ht]
    \caption{Resolution parameters for our simulations.
    }
    \label{tab:numset}
    \begin{tabularx}{0.9\linewidth}{>{\raggedright\arraybackslash}X r}
        \toprule
        \textbf{Parameter} & \textbf{Value} \\
        \midrule
        \multicolumn{2}{l}{\textbf{dRNS}} \\
        Number of refinement levels & 9 \\
        Number of grid points       & 128 \\
        Grid spacing                & 10.0 \\
        \addlinespace
        \multicolumn{2}{l}{\textbf{BNS}} \\
        Number of refinement levels & 7 \\
        Level after BH formation    & 9 \\
        Number of grid points       & 128 \\
        Moving boxes                & 2 \\
        Grid points (moving box)    & 128 \\
        Grid spacing                & 15.0 \\
        \bottomrule
    \end{tabularx}
\end{table}

\section{Waveform classification and qualitative analysis}\label{classAnalysis}

In the following, we discuss the waveform classification scheme and provide a qualitative analysis of the simulations. 
In Sec.~\ref{classification_a}, we first define the classification scheme, which, due to the large amount of simulations, is based on isolated NSs. 
We then investigate three particular BNS systems in Sec.~\ref{classification_b}.

\subsection{Isolated NS simulations}\label{classification_a}
All the 130 models use the numerical setup outlined in Sec.~\ref{NUMSETBAM}. The collapse of the isolated NSs is induced by changing the polytropic prefactor from $\kappa=100$ to $\kappa=99.8$.

\begin{figure}
    \centering    \includegraphics[width=0.5\textwidth]{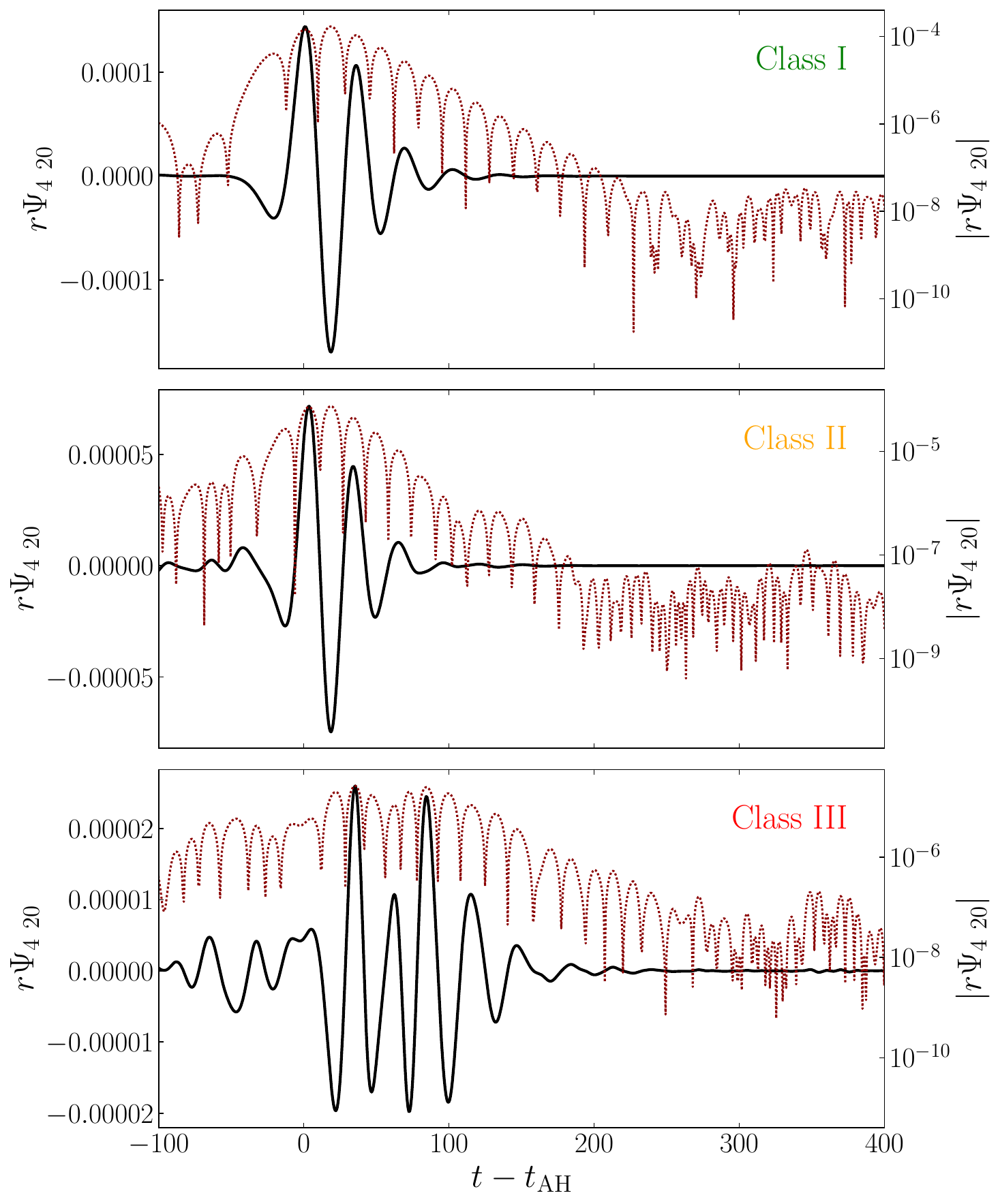}
    \caption{  
    Each panel represents an exemplary model of each class. Shown in each panel is the real part of the curvature scalar $r\Psi_{4}$ for the (2,0)-mode in linear scale (black solid line) and the modulus of $|r\Psi_{4}|$ in log-scale (red dashed line) plotted against the time with respect to AH formation $t_\mathrm{AH}$.}
    \label{fig:Classcomp}
\end{figure}

Our ringdown classification is defined as follows. Among the 130 ringdown signals, we observed three distinct signatures. 
In some cases, we observe a clean ringdown signal that is only cut off once it reaches the numerical noise floor. We define such signals to be of \texttt{Class I}. 
On the other hand, a signal is of \texttt{Class II}, if we initially find a clean ringdown, which at some point during the ringdown phase gets modified by matter exciting additional oscillations that make the initial ringdown features unrecognizable. 
Lastly, a signal is of \texttt{Class III}, if it does not exhibit any similarity to a clean ringdown at any time.
We show the Weyl scalar $r\Psi_4$ for the dominant $(\ell,m)=(2,0)$-mode incorporating the GW information for one representative waveform of each class in Fig.~\ref{fig:Classcomp}. 
Each model is equipped with an identifier labeling the corresponding rotation parameter and axis ratio. 
We provide a full list with more details in Appendix~\ref{app:RNSseq}.

\begin{figure}
    \centering
    \includegraphics[width=0.5\textwidth]{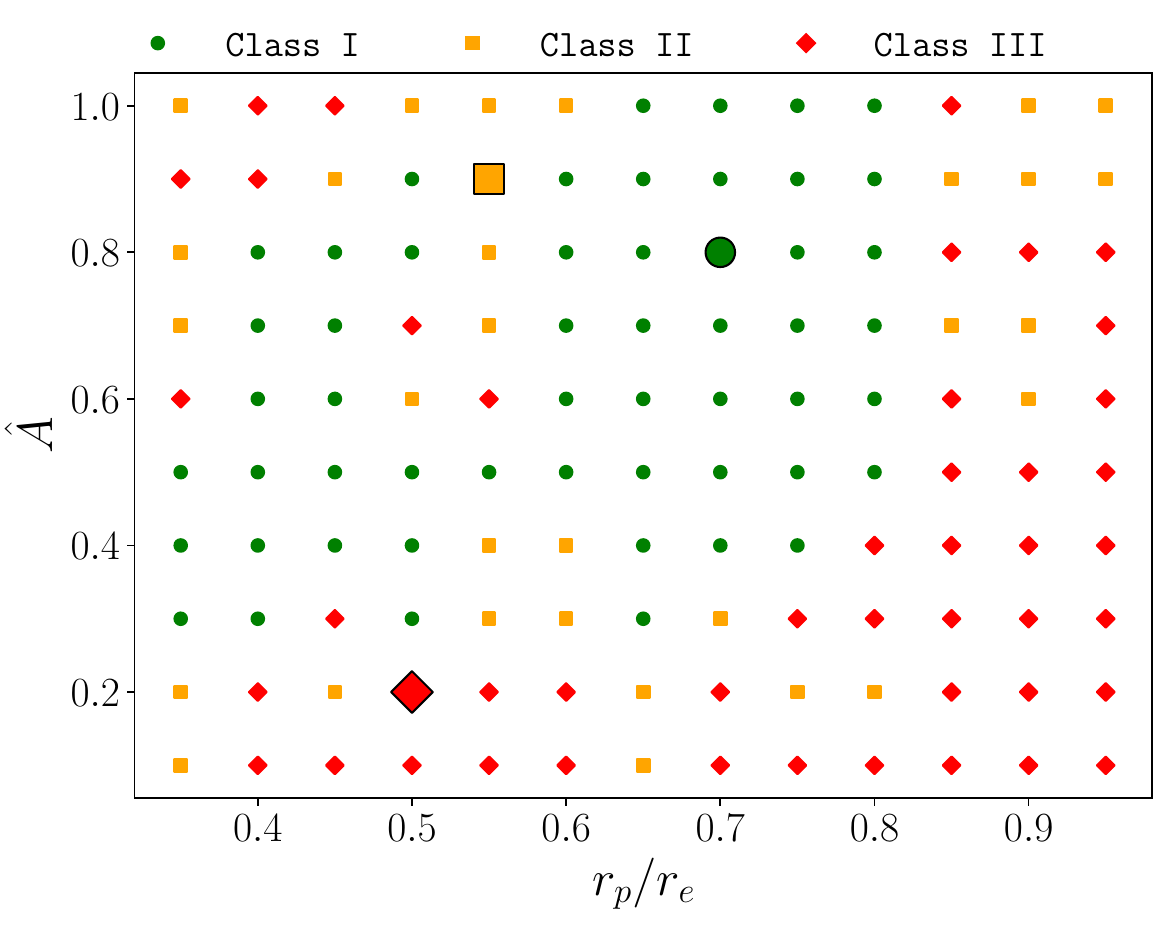}
    \caption{Classification of the 130 dRNS models shown in the $r_p/r_e$ versus $\hat{A}$ plane. Green circles denote \texttt{Class I}, orange boxes \texttt{Class II}, and red diamonds \texttt{Class III} models (as indicated in the legend). We highlight the markers of the models analyzed in detail below.}
    \label{fig:classification}
\end{figure}

We classified all 130 dRNS models through visual inspection (our classification for each model is given in Table~\ref{tab:inittab}). 
In particular, the distinction between \texttt{Class I} and \texttt{Class II} is quite subjective. 
Our distinguishing criterion between these two classes is that if there were no clear additional oscillations as for the \texttt{Class II} model in Fig.~\ref{fig:Classcomp}, i.e., something that looks like a clean \texttt{Class I} ringdown, then we would look at the amount of peaks present in the modulus of $|r\Psi_{4\ 20}|$. 
If the model had a suspiciously small number of peaks, then we classified it as a member of \texttt{Class II}, since matter cannot produce high-amplitude oscillations that erase previous ringdown features but rather is of the same amplitude and cuts off these later peaks. 
In Fig.~\ref{fig:classification}, we show the distribution of classes in the $\hat{A}-r_p/r_e$ plane. The most models are of \texttt{Class I} (53), followed by \texttt{Class III} (48), and the fewest amount of models are found for \texttt{Class II} (29). Looking back at Fig.~\ref{fig:classification}, we noticed a couple of trends. We find that the smaller the rotation parameter gets, the fewer \texttt{Class I} models we can find, indicating that a larger differential rotation disfavors the presence of a clean ringdown. Especially for models with $\hat{A}\leq0.2$, we find no \texttt{Class I} models, but all are either of \texttt{Class II} or \texttt{III}, since typically these are the systems with larger rotation and more complex collapse dynamics. Also, we found that models with $r_p/r_e\geq0.85$ are exclusively of \texttt{Class II} or \texttt{Class III}. For the \texttt{Class I} models, we identified two ``islands'' at intermediate and small axis ratios with $\hat{A}>0.2$. However, we will not attempt to draw particular conclusions about possible connections between $\hat{A}$ and $r_p/r_e$ and our classification, since generally the dynamics of each simulation are complex.

Furthermore, we find that, although generally robust and clearly dependent on the presence of matter outside of the formed horizon, the classification for each particular model depends on the exact numerical setup (e.g., resolution, reconstruction, Riemann solver).\footnote{To be more explicit, for the 130 isolated NS simulations in this work, we first performed respective simulations at a lower resolution. We found that for about half of the models a different collapse dynamics compared to the high-resolution runs was present. This in turn influenced the ringdown morphology and frequency extraction. The relatively large difference between the different resolutions comes from the fact that the employed setups probe the collapse dynamics near the critical point of stability and is introduced through a small perturbation. Larger perturbations would reduce the influence of the resolution, but would also introduce larger spurious constraint violations. Most importantly, the obtained classification of the ringdown signal is directly linked to the outflow of matter independent of the resolution, i.e., setups for which a large amount of matter is released before merger (independent of the resolution) will lead to \texttt{Class III} evolutions, while setups with a small amount of mass ejection will lead to \texttt{Class I} setups. Furthermore, we find that different reconstruction schemes and Riemann solvers seemed to have only a negligible influence on the results. The BNS cases on the other hand were robust, and the numerical setup only influenced the time of merger, but not the classification.}

Next, we focus on one model of each class in more detail and perform a thorough analysis to extract important features. 

\subsubsection{\texttt{Class I}}\label{C8}
As a representative example of \texttt{Class I}, we selected model C8 (following the naming convention of Appendix~\ref{app:RNSseq}) for a qualitative analysis. 
It has the intrinsic initial parameters $\hat{A}=0.8$, $r_p/r_e=0.7$, a gravitational mass of $1.867M_{\odot}$ and a dimensionless spin of $a=0.525$. 

\begin{figure}
    \centering
    \includegraphics[width=\columnwidth]{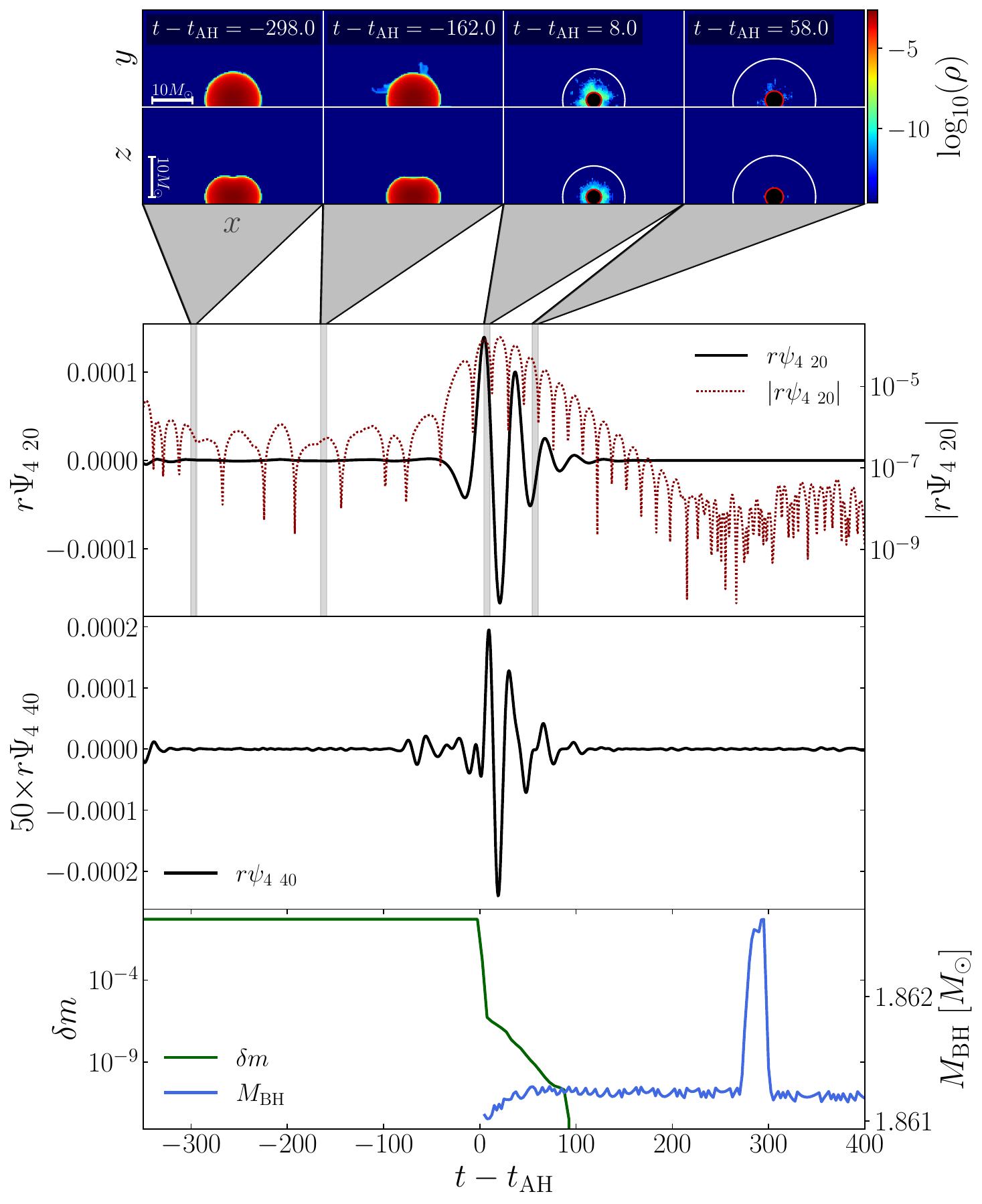}
    \caption{The upper row shows the base 10 logarithm of the maximum density on the fifth refinement level in the $xy$-plane at four different time slices for the \texttt{Class I} model C8, while the lower row depicts the $xz$-plane. In the last two panels of each row, the red solid line denotes the AH of the BH, and the white semicircle shows the position of the BH's lightring. We highlight the corresponding times in the waveform panel of the $(l,m)=(2,0)$-mode, directly below. This panel shows the $(2,0)$ waveform in linear scale (black solid line) and its modulus in log scale (red dotted line) plotted against time with respect to the point of AH formation. Below the $(2,0)$-mode, the $(l,m)=(4,0)$-mode is shown with increased amplitude by a factor of 50. On the last panel, below the $(4,0)$-mode, shown are respectively the baryonic mass on the fifth refinement level outside a spherical approximation of the AH (green) and the mass of the BH (blue).}
    \label{fig:C8}
\end{figure}

Above the top panel of Fig.~\ref{fig:C8}, we show the time evolution of energy density at different time slices. We clearly see the oblateness of the star in the $xz$-plane due to its rotation and axis ratio. At $t-t_\mathrm{AH}=-162.0$ ($t_\mathrm{AH}$ is the time when an AH is found first) the star sheds some mass in the equatorial plane. Once the gravitational field gets stronger, a BH with mass $M_\mathrm{BH}=1.861M_\odot$ and spin $a_\mathrm{BH}=0.519$ forms. Shortly afterwards, at $t-t_\mathrm{AH}=8.0$, a small asymmetric sphere of matter remains in close proximity to the final BH. We indicate the AH of the BH with a red solid line. At $t-t_\mathrm{AH}=58.0$, almost all matter is within the BH. In the last two time slices, we also show the location of the lightring (white line). We determine the lightring by solving the geodesic equation, shooting geodesics at different starting positions to find a stable orbit. Due to gauge effects, notably the $\Gamma$-driver shift, it grows over time, even though the BH mass stays roughly constant.

The ringdown signal is shown in the top and middle panels of Fig.~\ref{fig:C8}. 
First, we analyze the waveform of the (2,0)-mode in terms of the Weyl scalar $r\Psi_4$ extracted at $r=200$. 
We count 11 clear peaks in the modulus of $r\Psi_4$ for this mode, until the numerical noise floor is reached. 
Below the (2,0)-mode, we also show the real part of $r\Psi_4$ for the (4,0)-mode, which is fifty times smaller in amplitude than the (2,0)-mode. 
Therefore, the signal is not as smooth as for the (2,0)-mode, since the low-amplitude signal is closer to the noise floor. 

In the bottom panel, we visualize the baryonic mass outside a spherical approximation of the AH on the fifth refinement level and the BH mass over the same timescale. 
It is apparent that on a timescale of $\sim100$ after the BH forms, all the remaining matter after the collapse falls into the BH, and the baryonic mass outside is of order $\sim10^{-6}-10^{-8}$ within this time range. 
On the other hand, the mass of the BH increases slightly before reaching a constant level for the rest of the ringdown signal. 
\footnote{Also, notice the "bump" in the BH mass at roughly $300$ after AH formation. This is not a physical increase of the BH mass. This increase in the size of the BH is due to the $\Gamma$-driver shift, which we confirmed by performing some dedicated simulations with a different shift condition. Once the AH grows and covers the boundary region between refinement levels, we see a numerical issue in computing the exact location of the AH, which leads to an increased error in the computation of the BH mass. Once the horizon has grown enough to be fully outside the refinement boundary, the error drops again.}

\subsubsection{\texttt{Class II}}\label{B5}

For the study of \texttt{Class II} setups, we have chosen the model B5 as a representative example. 
Model B5 has a rotation parameter of $\hat{A}=0.9$ and an axis ratio of $r_p/r_e=0.55$. 
The gravitational mass and spin are $2.109M_{\odot}$ and $0.695$, respectively. 
It is more oblate than our previous \texttt{Class I} model, C8, and also has a higher initial spin. 

\begin{figure}
    \centering
    \includegraphics[width=\columnwidth]{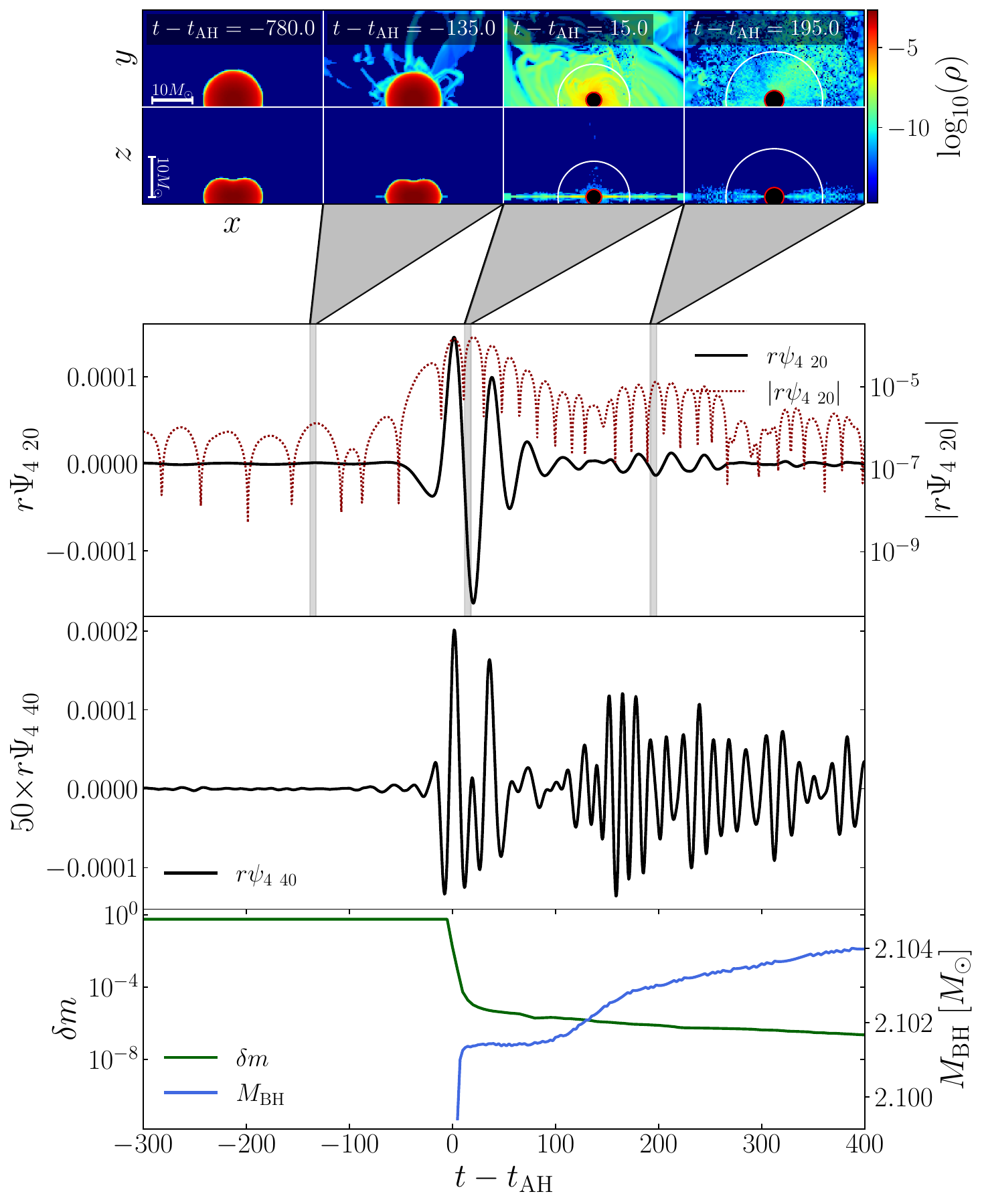}
    \caption{Same layout as Fig.~\ref{fig:C8} for the \texttt{Class II} model B5. The scaling factor of the (4,0)-mode and the extraction radius/refinement level are the same as for the C8 model. For details, see the text below.}
    \label{fig:B5}
\end{figure}

In Fig.~\ref{fig:B5}, we see that the star loses some of its material inside the equatorial plane at around $t-t_\mathrm{AH}=-135.0$. 
Once the star collapses, we retain a much heavier disk, which is confined inside the equatorial plane and can reach a baryonic mass of around $\sim10^{-5}-10^{-6}$. The BH reaches a mass of $M_\mathrm{BH}=2.104M_\odot$ and a spin of $a_\mathrm{BH}=0.684$, which is respectively larger than for the C8 model.
During the ringdown, the dense matter will start to fall into the BH, and at $t-t_\mathrm{AH}=195.0$ we only find densities of order $10^{-9}$ around the remnant, which corresponds to a total baryonic mass of $10^{-6}$. 
However, it is this increased amount of material that leads to additional high-amplitude oscillations in the waveform at around the same time.

This becomes apparent in the second half of Fig.~\ref{fig:B5}. 
Again, we show the Weyl scalar $r\Psi_4$ of the (2,0)-mode and the (4,0)-mode extracted at $r=200$ on the coarsest level. 
However, looking at the modulus of $r\Psi_4$ for the (2,0)-mode, we see that after around halfway through the ringdown, matter is responsible for exciting oscillations that overlap the initial ringdown in amplitude and break any signatures that could have been seen at later times. 
These oscillations are even more pronounced and influential in the (4,0)-mode, where the overall amplitude is lower by a factor of 50. 
The baryonic mass on the fifth refinement level outside the approximated AH shows a different profile than in the \texttt{Class I} example. 
Instead of all matter plunging into the BH over a comparatively small timescale, we retain a disk mass of around $10^{-6}$. 
Still, matter falls back to the BH, as can be seen from the slight negative gradient in the $\delta m$ time evolution. 

The exact behavior is best seen when looking at the BH mass at the bottom of Fig.~\ref{fig:B5}. 
Between $0$ and $100$ after horizon formation, we see that the BH mass is unchanged. 
This is also the time span in which we observe the initial unmodified ringdown. 
However, shortly after, the BH mass increases as matter falls back. 
This is when the additional excitations in the ringdown signal can be seen.
Looking at the evolution of the BH mass, we can clearly correlate the onset of the perturbations in the waveform to the onset of BH growth due to back-falling matter. Whether the additional observed features are a result of matter falling into the black hole or from the emission of secondary GWs (as in~\cite{Laeuger2025}), can not be clearly answered with the help of our simulations. In total, the BH mass increases by about 0.1\% from the onset of matter backfall at around $t-t_\mathrm{AH}=100$ to the end at around $t-t_\mathrm{AH}=300$.

\subsubsection{\texttt{Class III}}

As an example for the \texttt{Class III} setups, we consider model I4. 
With a rotation parameter of $\hat{A}=0.2$, it has the highest amount of differential rotation out of all models analyzed previously. 
It also has the smallest axis ratio with $r_p/r_e=0.5$, but maybe surprisingly, the smallest dimensionless spin with $a=0.401$. 
Its gravitational mass is fairly low with $1.79M_{\odot}$. 

\begin{figure}
    \centering
    \includegraphics[width=\columnwidth]{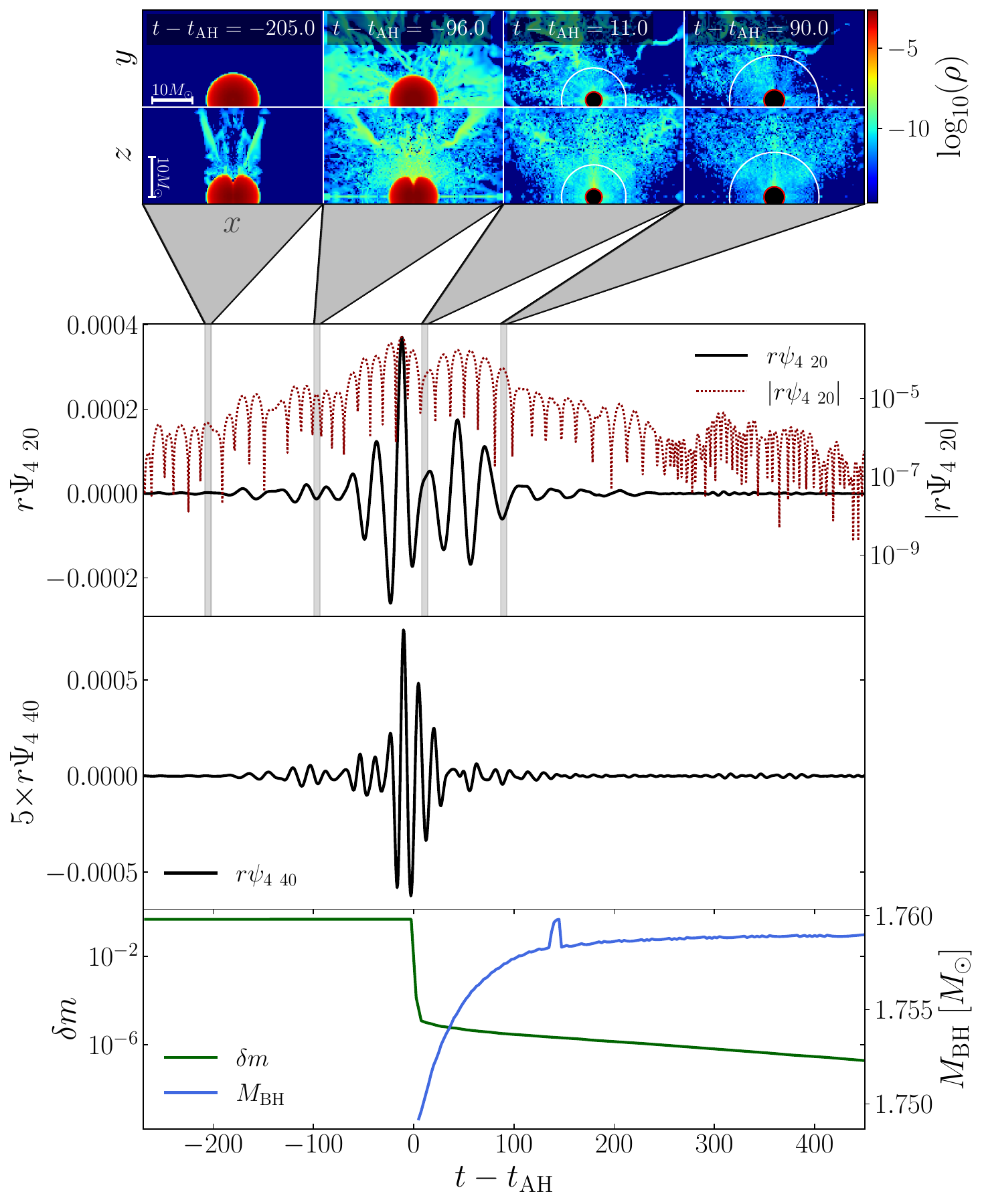}
    \caption{The conceptual content is the same as in Fig.~\ref{fig:C8} and Fig.~\ref{fig:B5}. However, for the (4,0)-mode of the model I4, we use a scaling factor of 5 for the amplitude instead of 50 as before.}
    \label{fig:I4}
\end{figure}

The density evolution is shown above the top panel of Fig.~\ref{fig:I4}. 
Before the actual collapse, we see some matter outflows in the polar regions at $t-t_\mathrm{AH}=205.0$ before the BH forms (we observed a consistent behavior of \texttt{Class II} models showing only equatorial outflows and \texttt{Class III} models showing both equatorial and polar outflows). 
At $t-t_\mathrm{AH}=-96.0$, we find even more massive outflows in the $xz$-plane, but also additional shedded mass in the $xy$-plane. 
Shortly after, the star collapses to a BH with mass $M_\mathrm{BH}=1.759M_\odot$ and spin $a_\mathrm{BH}=0.336$. 
What remains around the remnant are chunks of matter with densities of order $\sim10^{-9}$, which correspond to a disk mass of $\sim10^{-6}$. 
Between the last two snapshots at $t-t_\mathrm{AH}=11.0$ and $t-t_\mathrm{AH}=90.0$, we can clearly see the backflow of matter towards the BH. 
This material is constantly exciting the BH.

Looking at the (2,0)-mode of the GWs (second panel of Fig.~\ref{fig:I4}), we see GW emission not only during the ringdown phase but even before. In fact, we are unable to comment on the exact ringdown signatures, as they are all suppressed by the excitations of matter. For example, at $\sim80$ after horizon formation, we find a big excitation by matter, fully dominating the early ringdown peaks. Considering the third panel of Fig.~\ref{fig:I4}, we see that the (4,0)-mode has a much larger amplitude compared to the \texttt{Class I} and \texttt{Class II} models. Despite this difference, we find that the matter profile looks similar to the \texttt{Class II} model, with a gradual decrease as matter falls into the BH continuously. In total, the BH mass increases by about 0.5\% over a timescale of around $150$. The small "bump" is again due to the AH crossing a refinement boundary.

\subsection{BNS simulations}\label{classification_b}
In the following, to apply our classification to more realistic astrophysical scenarios, we investigate three representative BNS simulations and discuss how they fit in the previously defined classification scheme. 
Studies of the BH ringdown in BNS simulations have been presented in, e.g., ~\cite{Haas:2016cop,Bandyopadhyay:2023ohl}.
Studies of BHNS systems can be found in~\cite{steppohnthesis} and are in agreement with the findings of~\cite{Topolski:2024jva}.

\subsubsection{Equal mass $M_\mathrm{tot}=3.8M_{\odot}$}

\begin{figure}
    \centering
    \includegraphics[width=\columnwidth]{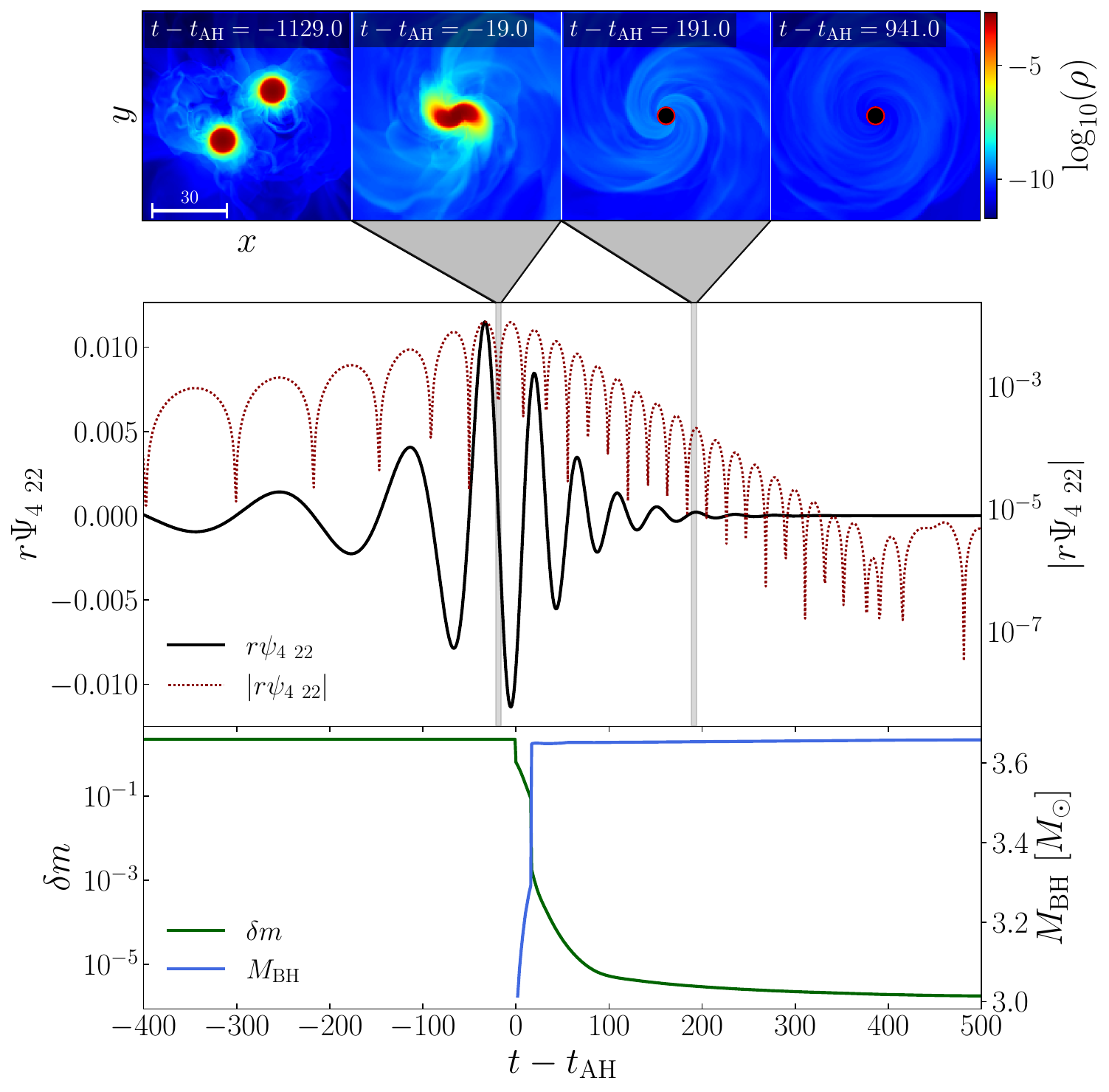}
    \caption{Shown is a conceptually similar qualitative representation to Fig.~\ref{fig:C8} for the heavy $M_\mathrm{tot}=3.8M_{\odot}$ BNS system. Here, however, we only focus on the density evolution in the $xy$-plane, show the ringdown of the (2,2)-mode and neglect the higher $\ell$-mode signal. Time is again reported with respect to the point of AH formation and the BH horizon is denoted with a red solid line.}
    \label{fig:M38SLyequ}
\end{figure}

We start by considering a heavy equal mass BNS system with a total mass of $M_{\text{total}}=3.8M_{\odot}$. 
In Fig.~\ref{fig:M38SLyequ}, we show a similar plot as in Sec.~\ref{classAnalysis}. Due to the change in symmetry (isolated to binary), we now report the dominant (2,2)-mode of $r\Psi_4$ (extracted at $r=400$), and the baryonic mass outside of the AH, as well as the BH mass. 
Due to the high mass of the system, no hypermassive NS is formed, but the system undergoes a prompt collapse to a BH with a mass of $M_\mathrm{BH}=2.958M_\odot$ after formation, which increases shortly after reaches a plateau of $M_\mathrm{BH}=3.659M_\odot$. Likewise, the spin increases from initially $a_\mathrm{BH}=0.628$ after merger to $a_\mathrm{BH}=0.726$. 
A snapshot of the BH during the ringdown is shown at $t-t_\mathrm{AH}=191.0$. 
In this case, only small amounts of baryonic mass of the order of $10^{-5}$ remain outside. 
However, we still see that matter falls into the BH for around half of the ringdown signal which results in the BH mass increasing by around 16\% in the early stages after formation. 
Because of this considerable change in mass, it is remarkable that the ringdown signal is very clean and long. 
The ringdown signal itself extends over a range of $\sim300$ after BH formation and shows 14-15 clear ringdown peaks in the modulus. Relating back to the previous classification scheme, this corresponds to a \texttt{Class I} ringdown. 

\subsubsection{Unequal mass $M_\mathrm{tot}=3.2M_{\odot}$}

Since in most realistic astrophysical scenarios both NSs will not have equal mass, we now investigate the ringdown of an unequal-mass BNS system \cite{Shibata2003,Lehner2016} with total mass $M_{\text{tot}}=3.2M_{\odot}$ and individual masses $M_1=1.8M_{\odot}$, and $M_2=1.4M_{\odot}$. 
We extract the waveform at $r=300$. 

\begin{figure}
    \centering
    \includegraphics[width=\columnwidth]{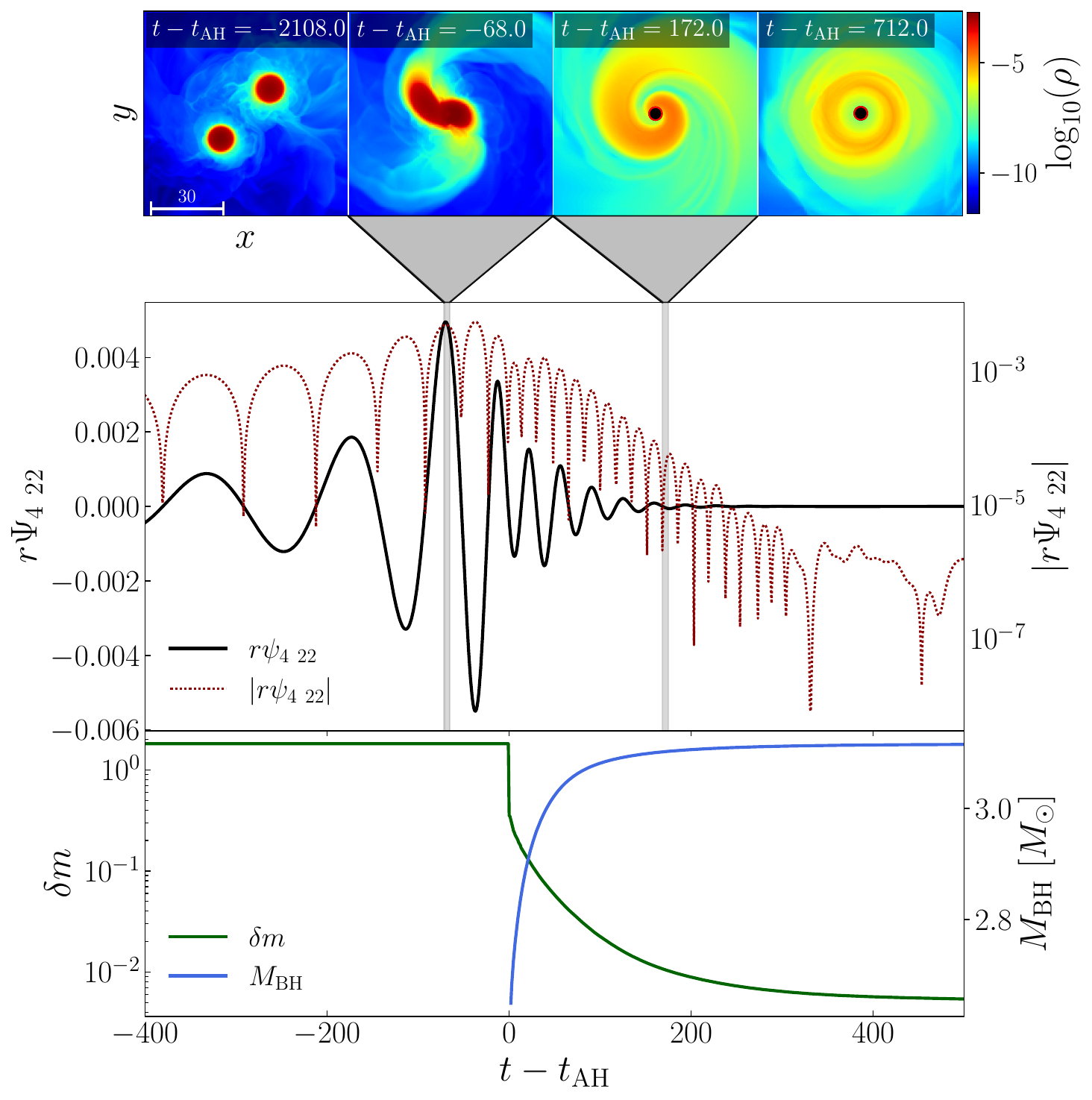}
    \caption{Same as Fig.~\ref{fig:M38SLyequ}, for the $M_{\text{total}}=3.2M_{\odot}$ unequal mass BNS system.}
    \label{fig:M32SLyunequ}
\end{figure}

The simulation is shown in Fig.~\ref{fig:M32SLyunequ}. 
Due to the different component masses, the heavier NS disrupts the lighter one earlier, as seen in the snapshot at $t-t_\mathrm{AH}=-68.0$. 
In our simulation (not visible here), the stars merge at a bit earlier than in the previous case, but BH formation is slightly delayed. After accreting some of the surrounding matter, the BH reaches a mass of $M_\mathrm{BH}=3.116M_\odot$ and a spin of $a_\mathrm{BH}=0.778$. It is interesting that the ringdown does not have a typical envelope, but rather some peaks that are slightly lifted up or down with respect to the others. 
We suspect that it is due to the heavy disk with baryonic mass of around $10^{-2}$ surrounding the remnant. 
The BH mass grows around 13\% during the ringdown phase, which might introduce additional slight perturbations of the BH, which lead to lifted peaks. Although the ringdown looks like \texttt{Class I}, some peaks look odd and a possible \texttt{Class II} ringdown could also be a valid description due to the large baryonic mass outside. The QNM extraction is, however, still possible and the results are reported in Sec.~\ref{UnequM32ring}. 

\subsubsection{Equal mass $M_\mathrm{tot}=2.7M_{\odot}$ }

As our last example, we consider an equal mass binary with a lower total mass of $M_{\text{tot}}=2.7M_{\odot}$. 
The waveform is extracted at $r=300$ on the second refinement level, as for example, the resolution on the zeroth refinement level would not have been sufficient to resolve the ringdown. 

\begin{figure}
    \centering
    \includegraphics[width=\columnwidth]{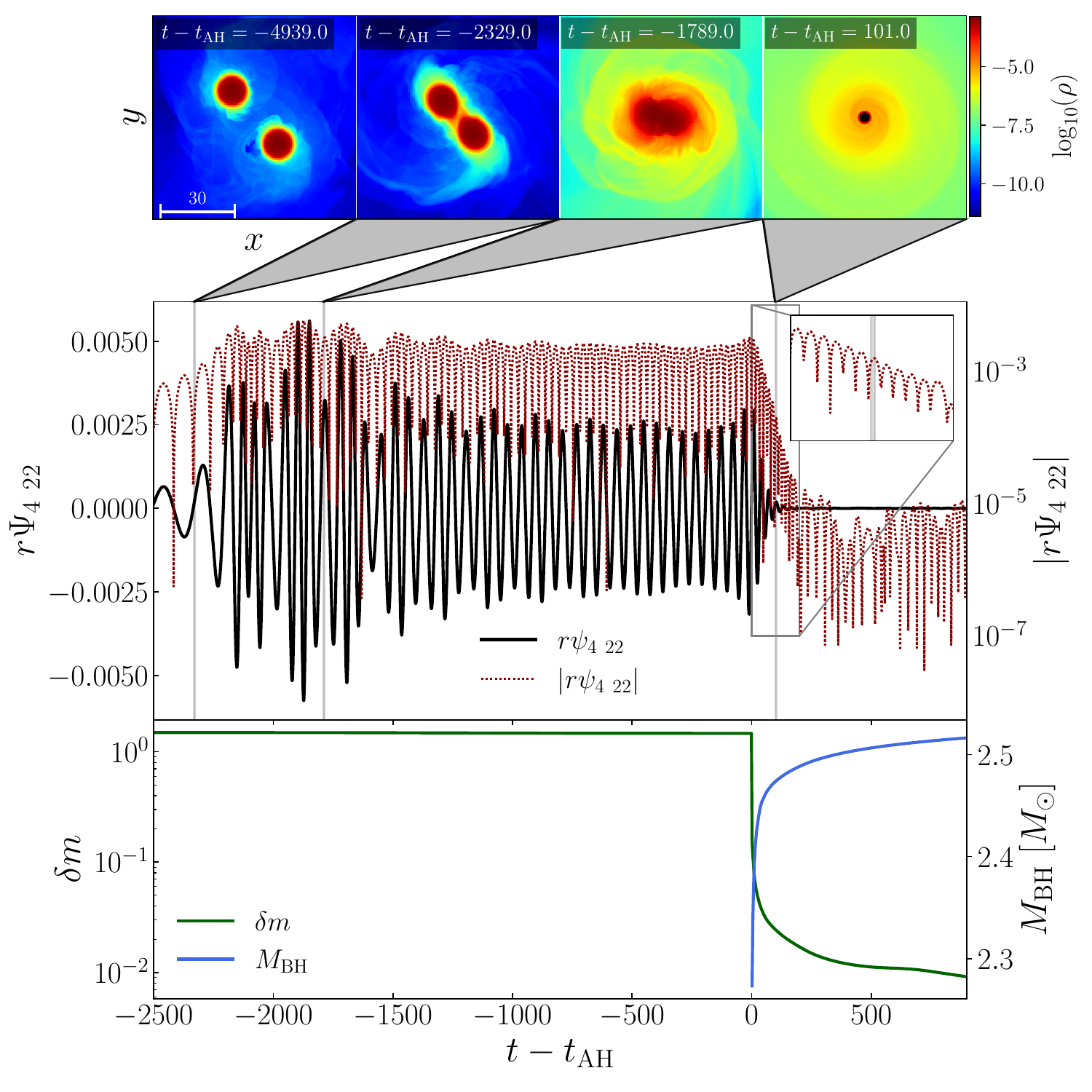}
    \caption{Same as Figs.~\ref{fig:M38SLyequ} and~\ref{fig:M32SLyunequ} for the light $M_\mathrm{tot}=2.7M_\odot$ binary, except, that we show the entire postmerger phase of the (2,2)-mode and the ringdown in an inset.}
    \label{fig:M27SLyequ}
\end{figure}

Looking at the density evolution in Fig.~\ref{fig:M27SLyequ}, we find that at around $t-t_\mathrm{AH}\approx-2300$, both NSs come into contact shortly before the merger. 
However, due to the low mass of the individual NSs, we do not observe a prompt collapse to a BH, but rather a phase where a hypermassive NS~\cite{Baumgarte1999,Hotokezaka2011,Bauswein2011} forms, which survives for $\sim2300$, until it collapses. 
The mass outside the AH is around $10^{-2}$, and the BH mass only increases by about 8\% in the range of $\sim100$ after its formation. After the BH settles into equilibrium, its mass reached $M_\mathrm{BH}=2.522M_\odot$ and its spin $a_\mathrm{BH}=0.661$.
In the waveform, we see that the modulus of $r\Psi_{4\ 22}$ in log-scale does not show a lot of ringdown peaks and the ringdown is rather short compared to the two previous cases. For that reason we rather classify it as \texttt{Class II} since matter seems to suppress the length of the signal significantly.

\section{Ringdown analysis}
\label{ringdown_analysis}

In the following, we briefly outline our setup for performing ringdown spectroscopy. 
For additional details regarding modeling pitfalls, in particular for theory-agnostic fits, we refer to Ref.~\cite{Cheung2023}.

\subsection{Ringdown models}\label{ringdownmodels}

We model the ringdown signal as a superposition of damped sinusoids, which for given QNM label $(\ell,m)$ is, in general, described as
\begin{equation}\label{equ:dampsin}
    h_{\ell m}=\sum_{n=0}^{\infty}C_{\ell mn}e^{-i\omega_{\ell mn}\bar{t}}\,.
\end{equation}
Here, $C_{\ell mn}$ are complex-valued excitation coefficients for each mode, $\omega_{\ell mn}$ are the corresponding complex-valued frequencies for which we use the convention $\omega_{\ell mn} = \mathrm{Re}(\omega_{\ell mn}) - i \mathrm{Im}(\omega_{\ell mn})$ with 
$\mathrm{Im}(\omega_{\ell mn}) > 0 $, and $\bar{t}$ is either the time with respect to the peak of the signal, $t-t_\mathrm{peak}$, or to the time of AH formation, $t-t_\mathrm{AH}$ (we explicitly state which of the two is used for the ringdown analysis). 
For a given set $(\ell,m)$, infinitely many overtone modes $n$ exist. 
In our analysis, we will vary the number of included QNMs. 
However, since our simulations include matter and are, thus, not as precise as pure vacuum simulations, we only use the most dominant ones. 

Within GR and the Teukolsky equation, all QNMs are known from perturbation theory and only depend on the mass $M$ and dimensionless spin $a$ of the final BH~\cite{Teukolsky:1974yv,Chandrasekhar:1975zza,Leaver1985}. 
We call this the "theory-specific" ringdown model because $\omega_{\ell m n} = \omega_{\ell m n}(M,a)$ is determined by GR. 
The corresponding Kerr QNM frequencies for a given spin are taken from Refs.~\cite{Berti:2005ys,Berti:2009kk}. 
We use cubic interpolation to inject the frequencies for a set $(M,a)$ to the fitting procedure described below. 

The alternative approach for modeling the same ringdown signal Eq.~\eqref{equ:dampsin} is called ``theory-agnostic''. 
Within this method, each $\omega_{\ell m n}$ will be described by an independent real and imaginary part. 
In the theory-agnostic case, the total number of free parameters is $4\times N$ (where N is the number of included QNMs). In the theory-specific case, it is only $2\times N +2$, which has profound impact on the stability of the fitting procedure when investigating a larger number of QNMs. 

The excitation coefficients are system-specific and not entirely intrinsic properties of the final BH. 
For example, a compact binary merger has a dominant $(2,2)$ mode, and, on the contrary, a collapsing NS has a dominant $(2,0)$ mode, as reported in Sec.~\ref{classAnalysis}. 
Other modes might weakly contribute, but are often subdominant and neglected in the following discussion.

In the theory-specific 
case, we define the relative error of the BH mass and spin as 
\begin{align}
\delta M &= \frac{M_\mathrm{BH}-M_\mathrm{fit}}{M_\mathrm{BH}}\,,\label{equ:deltaMdefinition}\\
\delta a &= \frac{a_\mathrm{BH}-a_\mathrm{fit}}{a_\mathrm{BH}}\,,\label{equ:deltaadefinition}
\end{align}
where $M_\mathrm{BH}$ and $a_\mathrm{BH}$ are determined in the numerical-relativity simulation from the AH finder~\cite{Thornburg:2003sf}, where the BH spin is simply calculated via $a_\mathrm{BH}=J_\mathrm{BH}/M_\mathrm{BH}^2$ with the BHs angular momentum $J_\mathrm{BH}$. 

The relative errors of the real and imaginary parts are given with respect to the Kerr predictions for mass and spin from the simulation. 
In this study, we only focus on the difference of the real and imaginary part of the fundamental mode to the Kerr predictions
\begin{align}
\delta \omega_R &= \frac{\mathrm{Re}({\omega}_\mathrm{BH}^{2m0})-\mathrm{Re}({\omega}_\mathrm{fit}^{2m0})}{\mathrm{Re}({\omega}_\mathrm{BH}^{2m0})}\,,\label{equ:deltaomegaRdefinition}\\
\delta \omega_I &= \frac{\mathrm{Im}({\omega}_\mathrm{BH}^{2m0})-\mathrm{Im}({\omega}_\mathrm{fit}^{2m0})}{\mathrm{Im}({\omega}_\mathrm{BH}^{2m0})}\,,\label{equ:deltaomegaIdefinition}
\end{align}
where $m=0$ for isolated NSs, and $m=2$ for BNSs.

\subsection{Selecting the ringdown signal}

We determine the ringdown signal from the simulations by locating the time where the waveform reaches its maximum, which is called $t_\mathrm{peak}$ from hereon (an exception are the BNS simulations in Sec.~\ref{UnequM32ring} and~\ref{M27ringdownanalysis}, where no prompt collapse occurs; for these we refer to the time of AH formation $t_\mathrm{AH}$). 
During the actual ringdown analysis, we vary the start of the fitting time $t_\mathrm{0}$ to check the robustness of the extracted parameters. 
Higher overtones, i.e., modes with large $n$, decay faster than modes with lower $n$. 
Therefore, overtones are harder to extract and the fundamental mode is the most robust and stable QNM (see Appendix~\ref{app:Modestab} for some details). 

The very late times are dominated by power-law tails~\cite{Price:1971fb,Leaver:1986gd,Gundlach:1993tp,Barack:1998bw} within linear perturbation theory. 
In our analysis, we cannot identify power-law tails, because the numerical noise cuts off this late-time part of the ringdown. 
Different end times $t_\mathrm{end}$ have also been investigated and we concluded that the extracted parameters are not sensitive to it. In the ringdown analysis, we explicitly state the fitting window used.

\subsection{Fitting routine and the mismatch}

For performing the fits, we use a simple least-squares minimization algorithm incorporated inside the \texttt{curve\_fit} function in the \texttt{scipy} package~\cite{2020SciPy-NMeth} in \texttt{Python}. 
As input, the minimization algorithm receives the damped sinusoidal ringdown model of Eq.~\eqref{equ:dampsin}. 
We reduce the risk of ending in a local minimum during the fitting procedure by repeating the analysis 100 times, for each starting time $t_0$, by considering randomized initial guesses. 
We use bounds for the parameters such that all amplitudes $A_{\ell m n}$ are positive, and that constrain the phase to be within $\phi_{\ell m n} \in [0, 2\pi]$.

For theory-agnostic fits, we use $\mathrm{Re}({\omega})\in[0,2]$ for the real part of the QNM frequencies. To help \texttt{curve\_fit} with finding the respective modes by their damping time, we employ separate bounds for the imaginary part of the QNM frequency of each mode. Since simulations result in different remnant masses and spins, we need to be careful with the bounds. In Table~\ref{tab:imagbounds}, we group the bounds on the imaginary part of the frequency for the different simulations motivated by the Kerr predictions. These bounds are sufficient to use up to three modes in our ringdown analysis, and it gives us a better estimate of the QNM deviations via Eqs.~\eqref{equ:deltaomegaRdefinition} and~\eqref{equ:deltaomegaIdefinition} and incorporates the entire range of mass and spin of our remnant BH.

For the theory-specific fits, we use the above boundaries for the amplitude and phase and add boundaries for the mass and the spin, i.e., $M\in[M_\mathrm{BH}-2\,M_{\odot},M_\mathrm{BH}+2\,M_{\odot}]$ and $a\in[0.01,0.99]$. 
The bounds on the mass ensure stability of the fit at all times, since at later times the fitting often became unstable for wider bounds.

\begin{table}[ht]
    \caption{Bounds on the imaginary part of the respective QNM frequency for the different simulations in this work.
    }
    \label{tab:imagbounds}
    \begin{tabularx}{0.9\linewidth}{Xccc}
        \toprule \textbf{Simulation}&$\mathrm{Im}\left(\omega_{2m0}\right)$&$\mathrm{Im}\left(\omega_{2m1}\right)$&$\mathrm{Im}\left(\omega_{2m2}\right)$\\
         \midrule
         \multicolumn{2}{l}{\textbf{dRNS}}\\
         \texttt{Class I/II}&[0, 0.07]&[0.071, 0.15]&[0.151, 0.35]\\
         \addlinespace
         \multicolumn{2}{l}{\textbf{BNS}}\\
         $M_\mathrm{tot}=3.8M_{\odot}$&[0, 0.04]&[0.041, 0.80]&[0.810, 0.35]\\
         $M_\mathrm{tot}=3.2M_{\odot}$&[0, 0.04]&[0.041, 0.12]&[0.121, 0.35]\\
         $M_\mathrm{tot}=2.7M_{\odot}$&[0, 0.05]&[0.051, 0.15]&[0.151, 0.35]\\
         \bottomrule
    \end{tabularx}
\end{table}

The \texttt{curve\_fit} routine minimizes the sum of squares of the residuals between the fit and the ringdown data
\begin{align}\label{chi2fit}
    \mathrm{SSR}=\sum_{i=1}^{n}\left(y_i-h(t_i,A_{lmn},\phi_{lmn},\omega_{lmn})\right)^2,
\end{align}
where $y_i$ is the $i$th data and $t_i$ the $i$th time point, and $h$ is our ringdown model in Eq.~\eqref{equ:dampsin} together with the fitted parameters. \texttt{curve\_fit} can take the error on the data as an argument. For the systems considered, we constrained the error in the data to about $10^{-7}$, which was usually the onset of the numerical noise floor. 
Although this error is not related to a measurement, it provides a qualitative estimate of how numerical errors might propagate to the fitted parameters. On the other hand, the error on the fitted parameters can be computed via
\begin{align}
    e(\theta)=\sqrt{\text{diag}(\mathrm{M}_{\text{cov}}(\theta))},
\end{align}
where $\mathrm{M}_{\text{cov}}$ is the covariance matrix obtained from \texttt{curve\_fit}, which incorporates correlations between individual fitting parameters $\theta$.

One important property in GW modeling is the mismatch $\mathcal{M}$, which is defined as
\begin{align}
\mathcal{M} = 1 - \frac{\left<h_1, h_2 \right> }{\sqrt{\left<h_1, h_1 \right> \left<h_2, h_2 \right> }}\,,
\end{align}
with 
\begin{align}
\left<h_1, h_2 \right> 
= \int_{t_0}^{t_\mathrm{end}} h_1(t) h_2(t) \text{d}t\,.
\end{align}
It is related, but not the same as Eq.~\eqref{chi2fit}. 
Note that the constant error used in the \texttt{curve\_fit} routine does not change the mismatch, while a time- or frequency-dependent error would need to be included. 
The smallest mismatch is used to determine the best-fit results among the 100 random guesses at a particular starting time $t_0$. 

\section{Applications}\label{moreResults}
In the following, we analyze the ringdown signal of the \texttt{Class I} model C8 
in Sec.~\ref{ringdown_c8}. 
We then provide a more condensed summary of the ringdown analysis for the other \texttt{Class I} models in Sec.~\ref{ringdown_all}. The early part of the \texttt{Class II} model B5 is investigated in Sec.~\ref{B5ringdownanalysis}. 
The ringdown analysis of our three selected BNS simulations are shown in Secs.~\ref{M38ringdownanalysis}-\ref{M27ringdownanalysis}.

\subsection{QNMs of simulation C8}\label{ringdown_c8}

For the \texttt{Class I} model C8, we count 11 clear peaks in the modulus of $r\Psi_{4\ 20}$ (see Fig.~\ref{fig:C8}) until reaching the numerical noise floor, spanning over a time range of roughly $\sim 150$. 
Since we have no indication how much the matter at the beginning of the signal affects the QNM spectrum, we first apply a theory-agnostic fit to the (2,0)-signal, and then a theory-specific one afterward. 

For the remainder of this section, we study three different cases of Eq.~\eqref{equ:dampsin}: a one-mode ringdown model corresponding to the fundamental mode $(2,0,0)$, a two-mode model corresponding to the fundamental mode and the first overtone $(2,0,1)$, and lastly a three-mode model corresponding to the fundamental mode, and first and second overtone $(2,0,2)$. To investigate the effect of the starting time on the ringdown analysis we vary $t_0$.

\begin{figure}
    \centering
    \includegraphics[width=\linewidth]{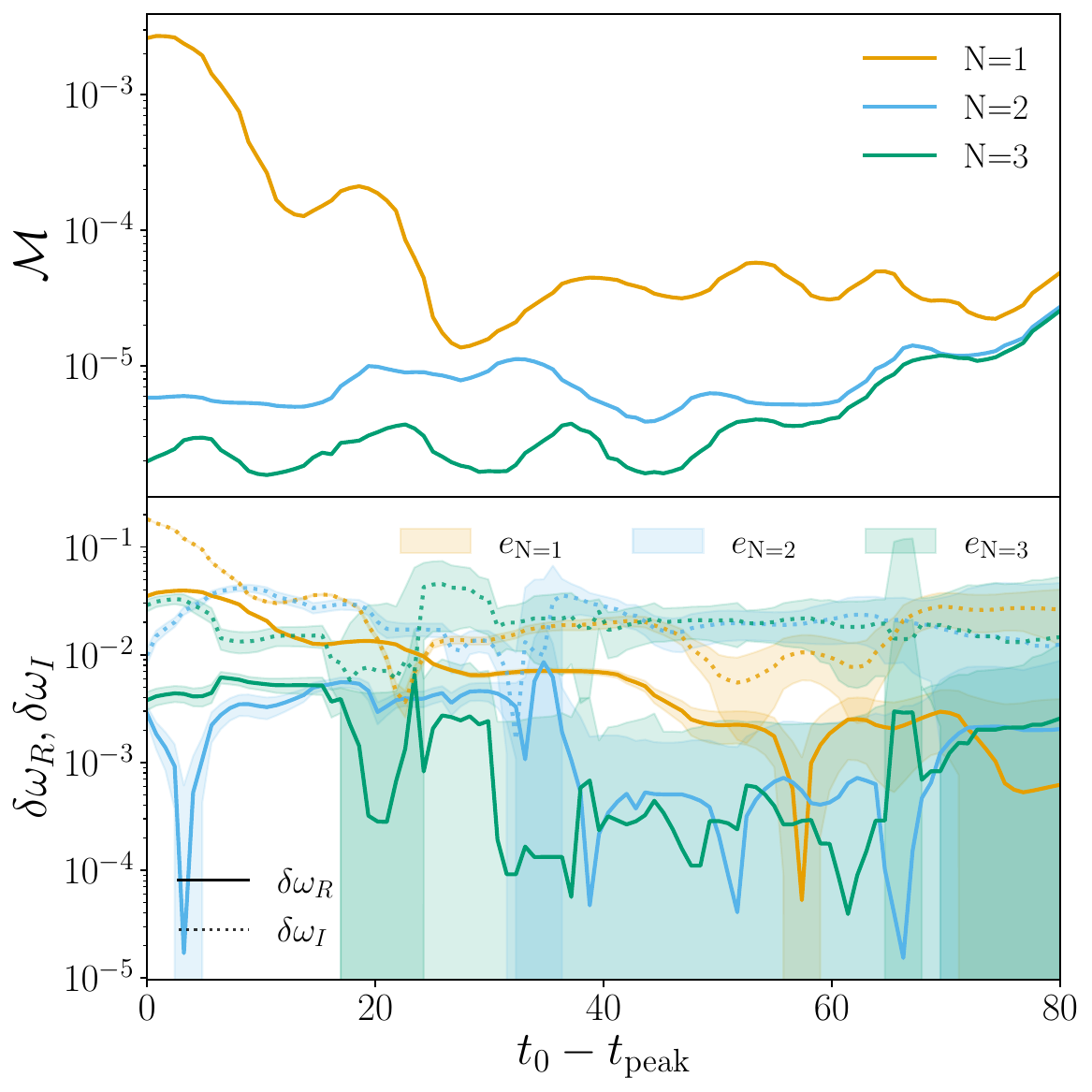}
    \caption{Agnostic fits of the \texttt{Class I} ringdown from model C8 (see Sec.~\ref{C8}), with 1, 2, and 3 modes respectively (colors indicated as in the legend of the top panel). Shown on top is the mismatch for each ringdown model as we vary the starting time of the fit. On the bottom panel, we depict the deviation of the real/imaginary part of the fitted ringdown frequencies of the fundamental mode from the vacuum values, defined in Eq.~\eqref{equ:deltaomegaRdefinition} and~\eqref{equ:deltaomegaIdefinition}. Errors on the fit are shown as bands around the fitted value.}
    \label{fig:C8agnostic}
\end{figure}

The results of the theory-agnostic fits are shown in Fig.~\ref{fig:C8agnostic}. 
The mismatch, as expected, decreases when including more modes. 
The stability of the amplitudes and phases can be found in Appendix~\ref{app:Modestab}. 
We can observe a stable amplitude and phase of the fundamental mode across all simulations of this work. 
However, stability of overtones for \texttt{Class I/II} models is not observed, with partial stability in the BNS cases. In the bottom panel of Fig.~\ref{fig:C8agnostic}, we report the deviation of the real/imaginary part of the fundamental QNM from the vacuum results.

The model with $N=1$ shows large deviations for $\delta\omega_R$ at the beginning of the ringdown signal of about $\sim3-5\%$. 
At later starting times, we see that $\delta\omega_R$ approaches the vacuum values. 
The significant deviation of $\delta\omega_R$ at the beginning (for $N=1$) might lead to the conclusion that the small amount of matter present at the beginning of the ringdown has a large influence on the QNM frequency of the fundamental mode. 
However, the difference compared to the vacuum frequency of the fundamental mode at the beginning vanishes once more QNMs are included (for $N=2$ and $N=3$). 
Thus, the deviation of the fundamental mode frequency for the $N=1$ model is not due to matter, but rather due to the lack of modes in our fitting model.
For $N=2$ and $N=3$, $\delta\omega_R$ is around $0.5\%$ at early and even well below $0.1\%$ at late times. A similar behavior is seen for $\delta\omega_I$, although the deviation to the predicted value is larger than $\delta\omega_R$ with $\gtrsim1\%$. The fitted values are also complemented with error bands, which widen at later starting times due to fewer data points and smaller signal amplitudes. The uncertainty in the fundamental mode parameters also increases when including more modes. 

\begin{figure}
    \centering
    \includegraphics[width=\linewidth]{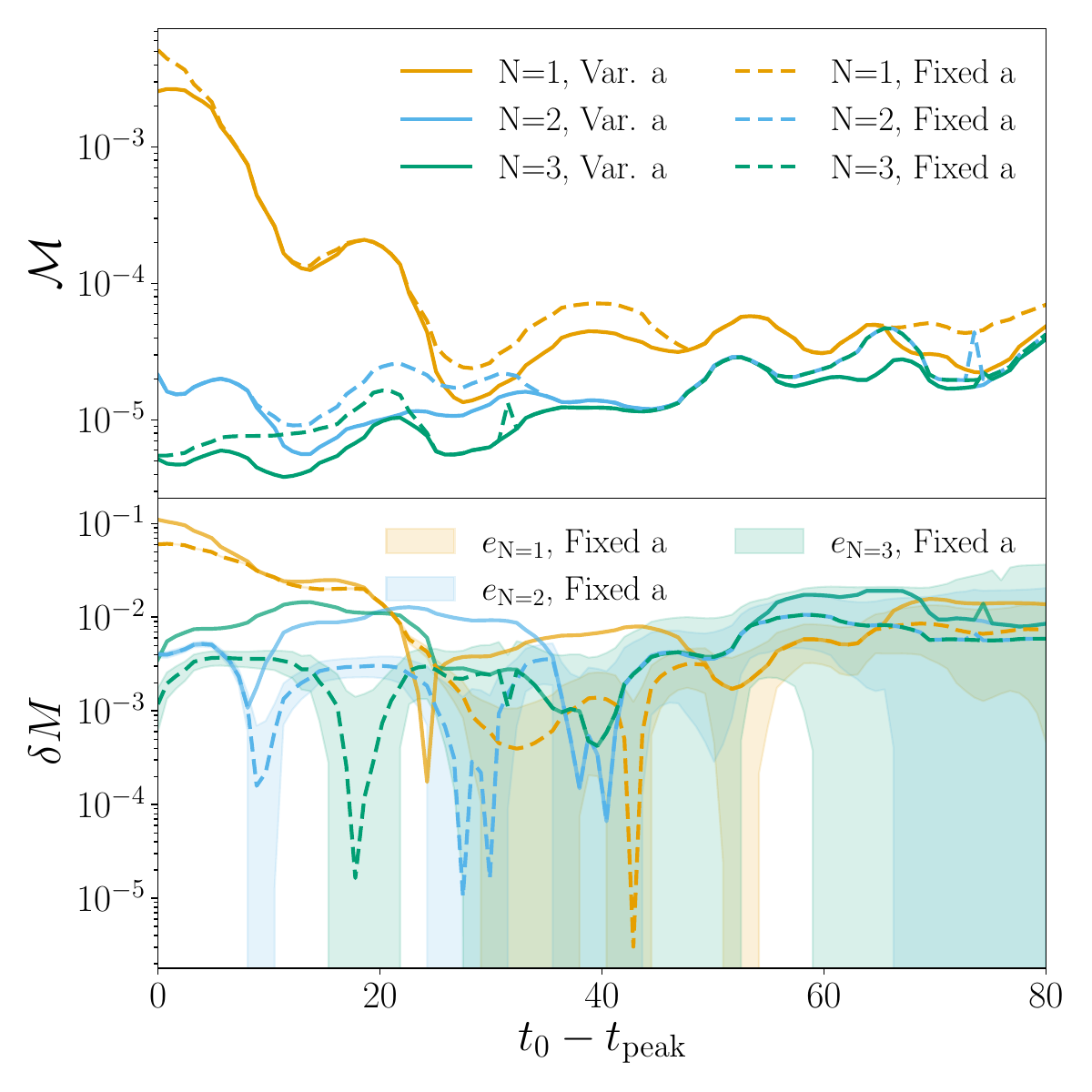}
    \caption{The difference of the C8 theory-specific fits for our three ringdown models (consult Sec.~\ref{ringdown_analysis}) once with a freely varying spin and the other time for a fixed spin. \textit{Upper panel:} Mismatches for all cases, as indicated in the legend. \textit{Lower panel:} Difference between the BH mass from the simulation and the fitted mass, defined in Eq.~\eqref{equ:deltaMdefinition}. Errors are only drawn as bands around the fits with fixed spin.}
    \label{fig:mode_comp}
\end{figure}

Next we perform the theory-specific analysis and include up to three QNMs. 
Since spin measurements are, in general, more difficult, we perform the analysis in two ways. 
First, we vary the spin in the full physical range, second, we constrain the spin in a range $a_{\text{Fit}}=a_{\text{BH}}\pm0.05$. 
The latter is motivated if one would be able to provide constraints on the final spin from an independent inspiral-merger analysis. The results are reported in Fig.~\ref{fig:mode_comp}.

For fixed spin, the mismatches are larger at early and intermediate starting times, since we force the least-squares algorithm to stay within tight bounds while it longs to explore large values for the spin to decrease the mismatch. Therefore, the mismatches increase at certain times. However, for most of the starting time parameter space, the mismatches for the two cases are nearly identical over all models. It is also apparent that if we fix the spin to vary in these tight bounds, the mass deviation from the true mass of the simulation is much smaller at early starting times for the higher-mode models, and for the one-mode model at intermediate starting times. If in the future we were able to infer the spin of the remnant BH by some measurements, then this would give us a better estimate of the mass, as we would also intuitively expect.

\subsection{QNMs of combined \texttt{Class I} simulations}\label{ringdown_all}

In the following, we provide a condensed analysis of all 53 \texttt{Class I} models. We limit the length of the ringdown signal to $110$ after the peak, so that we fit the frequencies in the same time window for all cases, and get a more coherent picture. 
The fits are performed for each model at 60 different starting times in the range from $0$ to $60$, to leave enough signal for the later starting times. To explore the possibility that nonvacuum effects impact the fundamental mode, we report the corresponding baryonic mass outside the BH as a qualitative measure.  

\begin{figure}
    \centering
    \includegraphics[width=\linewidth]{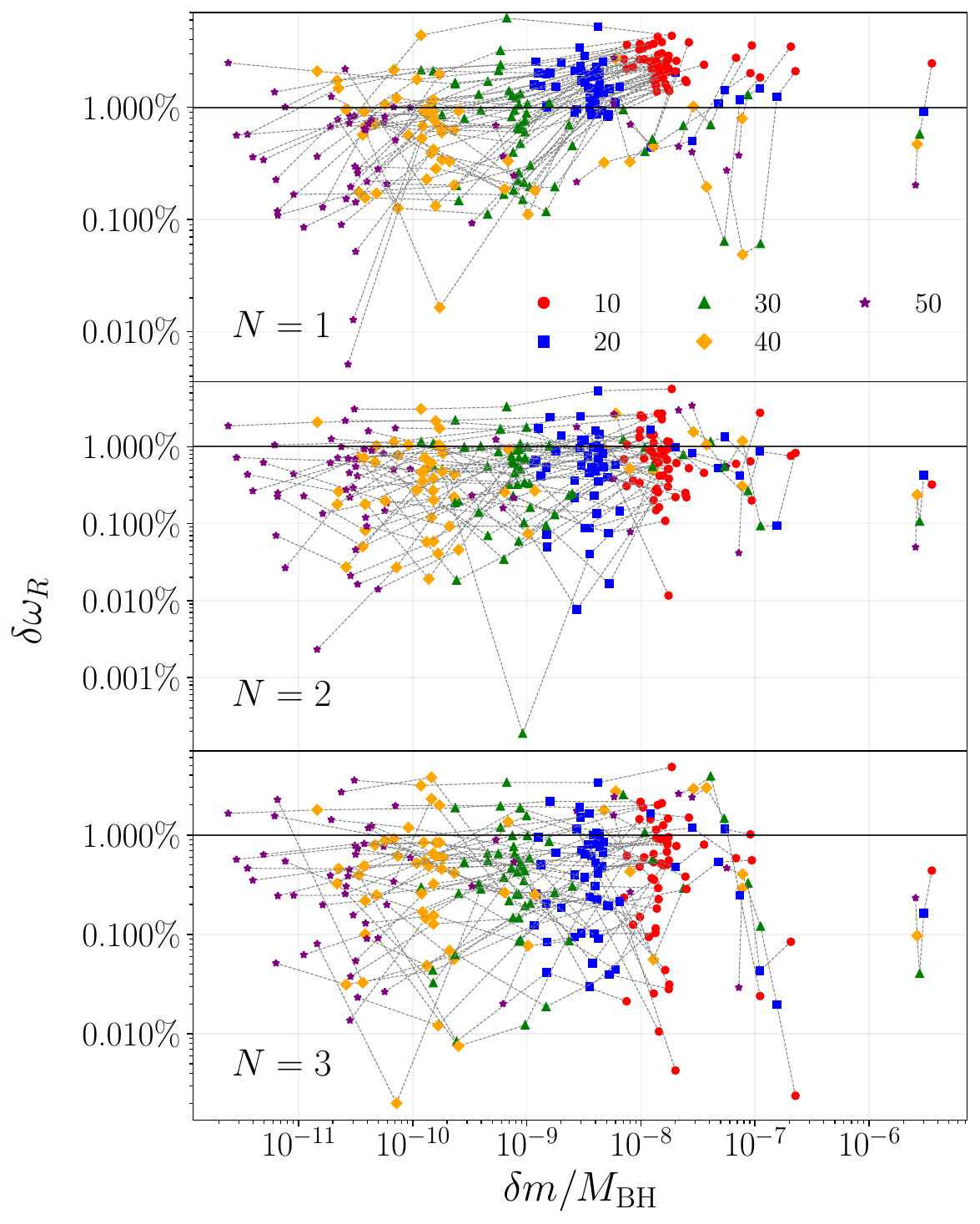}
    \caption{
    Percentage deviations of $\delta\omega_R$ in log scale against the baryonic mass outside the BH $\delta m$ scaled by the BH mass $M_\mathrm{BH}$. 
    The legend indicates markers for the QNM fit at different starting times after the peak. 
    Each marker is connected by a gray dotted line for each model to trace the evolution of the frequency deviation. 
    A thin black line is drawn to indicate a threshold of 1\%. 
    \textit{Top:} Ringdown model with just the fundamental mode $(N=1)$. 
    \textit{Middle:} Ringdown model with the fundamental mode and first overtone $(N=2)$. 
    \textit{Bottom:} Ringdown model with fundamental mode and the first two overtones $(N=3)$.}
    \label{fig:scalingclassI}
\end{figure}

In Fig.~\ref{fig:scalingclassI}, we show the theory-agnostic fits of all the \texttt{Class I} models for different $N$. 
For $N=1$, we find that, apart from some outliers, the deviation of the fundamental mode frequency at $t_0-t_{\text{peak}}=10$ is between 2\% and 4\% at baryonic masses between $10^{-7}$ and $10^{-8}$ scaled by the BH mass. 
This deviation decreases significantly once we consider later starting times. 
At $t_0-t_{\text{peak}}=50$, the frequency deviations for most models compared to the vacuum ones are around 0.1\% and 1\%. 
Including additional QNMs ($N>1$) shows the same behavior as in the previous section.  
The deviations at $t_0-t_{\text{peak}}=10$ now shift below 1\%. 
This indicates that the starting deviation is not due to the moderately low baryonic mass content at the ringdown beginning, but, again, due to the lack of modes. 
It further improves for ringdown models with $N=3$ modes. 

\begin{figure}
    \centering
    \includegraphics[width=\linewidth]{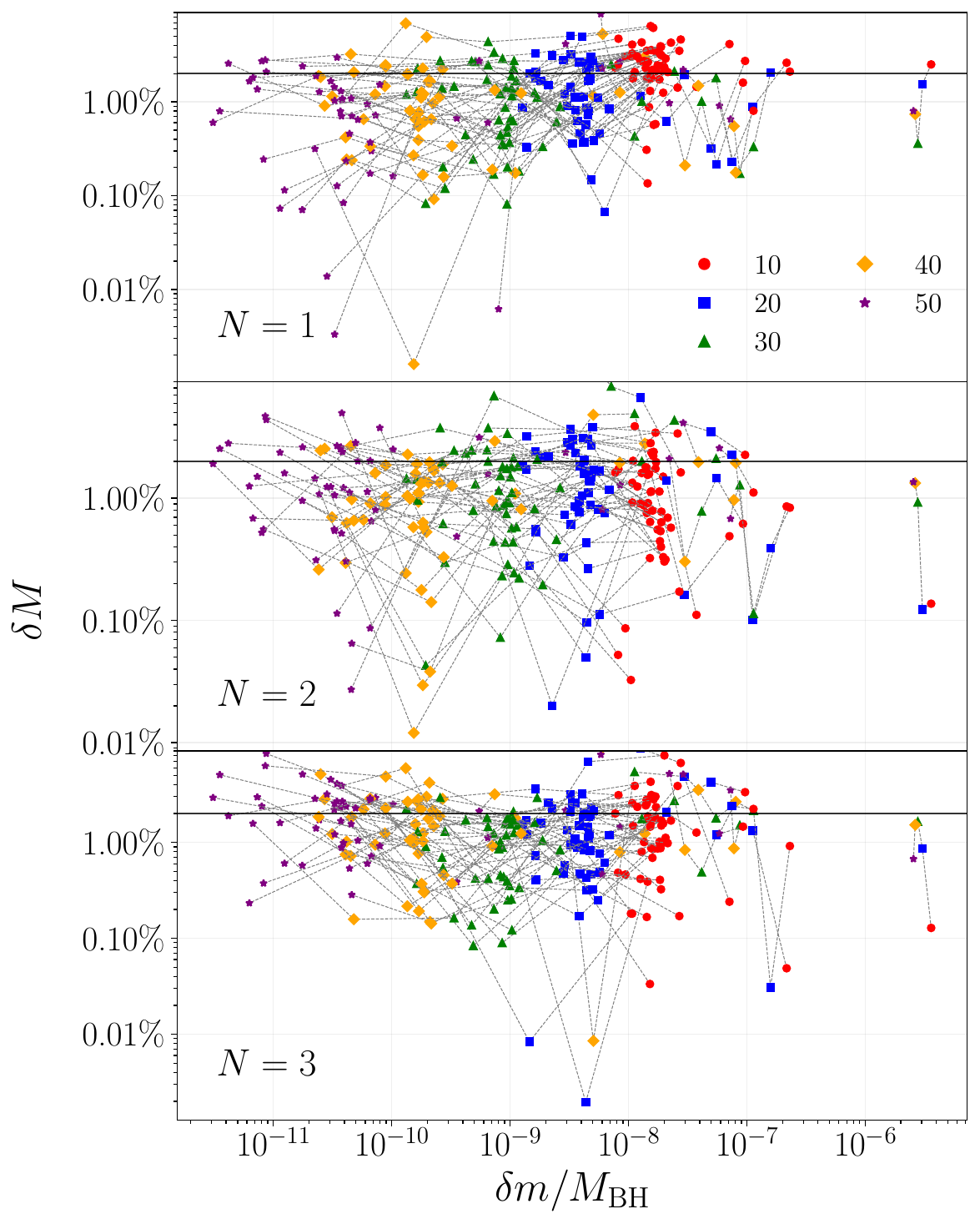}
    \caption{A similar depiction of \texttt{Class I} models as in Fig.~\ref{fig:scalingclassI} for the theory-specific case. The threshold line is now set at 2\%.}
    \label{fig:scalingclassI_ts}
\end{figure}

All the theory-specific \texttt{Class I} fits can be found in Fig.~\ref{fig:scalingclassI_ts}. 
We first note, that the results are more noisy than in the theory-agnostic case, but a similar trend can be seen. 
For $N=1$, the inferred mass difference is between 2\% and 4\%. 
This difference decreases upon inclusion of more modes and drops below 2\% for most of the models at almost all starting times. 
An exception is the latest starting time at $t_0-t_\mathrm{peak}=50$, where some models seem to jump above the shown 2\% threshold line. 
This is most likely due to 
different signal lengths
between the models. 
A fixed length of $110$ may already include numerical noise
for some models and make the signal less informative. 

\subsection{Early B5 QNM spectrum}\label{B5ringdownanalysis}
\begin{figure}
    \centering
    \includegraphics[width=\columnwidth]{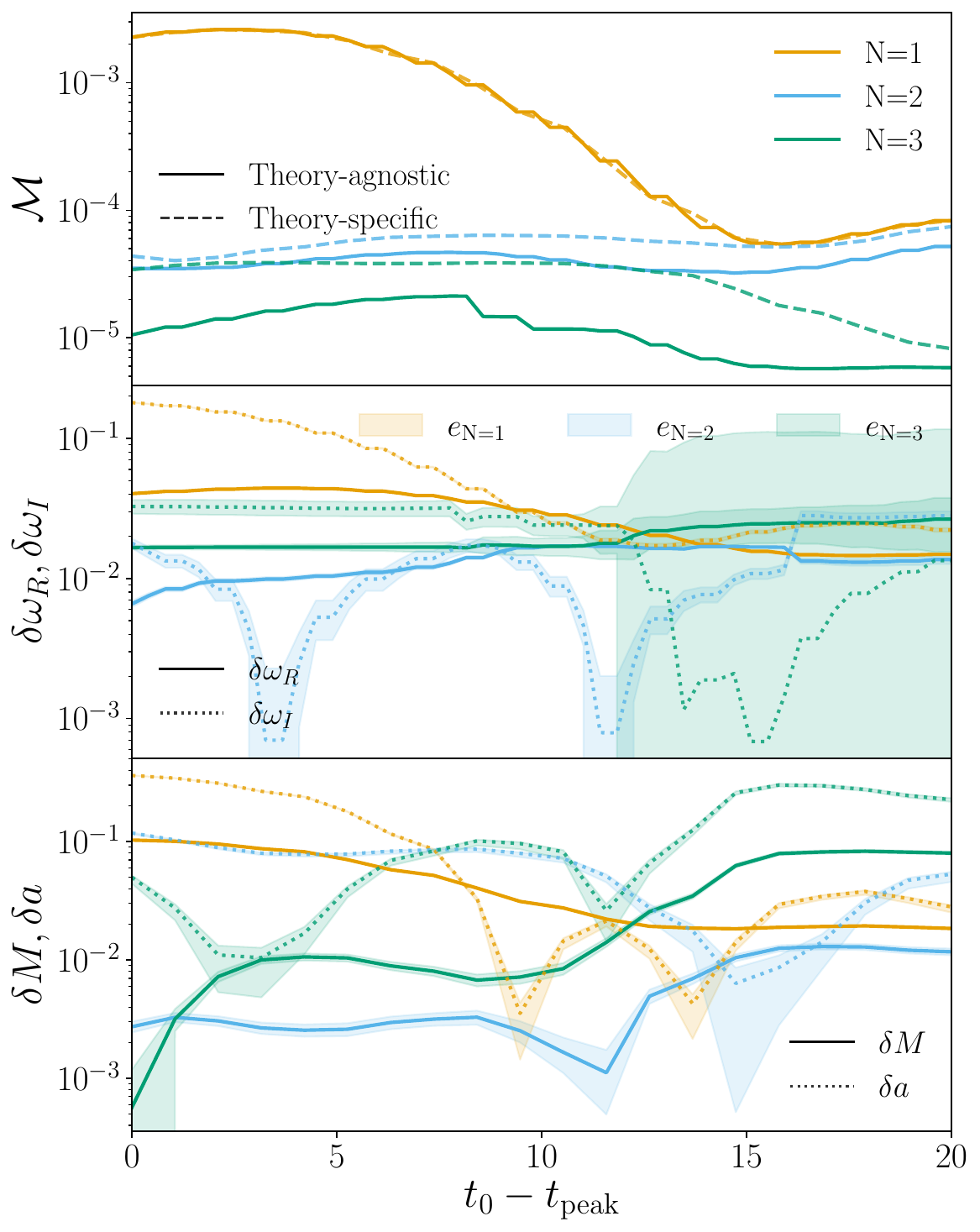}
    \caption{A combination of the the theory-agnostic fit in Fig.~\ref{fig:C8agnostic} and the theory-specific fit in Fig.~\ref{fig:mode_comp} for the \texttt{Class II} model B5. Here, we only focus on theory-specific fits where the spin can vary over the entire range. \textit{Upper panel:} Mismatches for the theory-agnostic fits (solid) and the theory-specific fits (dashed) of all three discussed ringdown models. \textit{Middle panel:} The deviation of the real part (solid) and imaginary part (dotted) of the fundamental mode frequency compared to vacuum in the theory-agnostic case. \textit{Bottom panel:} Mass (solid) and spin (dotted) difference compared to the "true" mass/spin for the theory-specific case.}
    \label{fig:B5fit}
\end{figure}

Although the ringdown signal of the \texttt{Class II} model B5 in Fig.~\ref{fig:B5} gets interrupted after some time, we are still able to analyze the QNM spectrum of the early ``clean'' part of the signal. 
In particular, we focus on a window of length $80$ after the peak of the signal and vary the starting time between $0$ and $20$. 
In Fig.~\ref{fig:B5fit}, we show the deviation of the fundamental mode in the theory-agnostic case as defined by Eqs.~\eqref{equ:deltaomegaRdefinition} and~\eqref{equ:deltaomegaIdefinition}, and similarly the deviation of the inferred mass and spin in the theory-specific case as defined by Eqs.~\eqref{equ:deltaMdefinition} and~\eqref{equ:deltaadefinition}.
The mismatch for the $N=1$ ringdown model in the theory-agnostic and -specific case are similar, but for the $N=2$ and $N=3$ models we observe that the theory-agnostic mismatch is always lower than the respective theory-agnostic one. 
This is expected because the agnostic parameter space is bigger and contains the theory-specific one as a subset. 

With the increased baryonic mass of around $10^{-5}$ around the remnant, we also observe larger differences for $\delta\omega_R$ in the theory-agnostic fit between 1\% and 2\% at all starting times for the higher-mode models ($N>1$), while $\delta\omega_I$ is slightly larger than $\delta\omega_R$ depending on the ringdown model and starting time. A similar behavior can be observed for the theory-specific fit, where the mass difference is even below 1\% for $N=2$ across all and $N=3$ during the first half of the starting times. The $N=1$ model only approaches the 1\% threshold at late times. Moreover, the spin is more uncertain, varying between 1\% and 10\%. Another interesting feature is that in the theory-agnostic fits, the $N=2$ model shows values for $\delta\omega_R$ that are always smaller than for the $N=3$ model, which stays roughly constant around 2\% for the majority of the starting time range. This is similar in the theory-specific case.

\subsection{Equal mass BNS with $M_{\text{total}}=3.8M_{\odot}$}\label{M38ringdownanalysis}

Similar to the previous ringdown analysis, we now investigate the three selected BNS ringdowns, starting with the heavy $M_\mathrm{tot}=3.8M_{\odot}$ binary. 
We chose a window of length $280$ from the peak of the signal and vary the starting time between $0$ and $150$. 

\begin{figure}
    \centering
    \includegraphics[width=\linewidth]{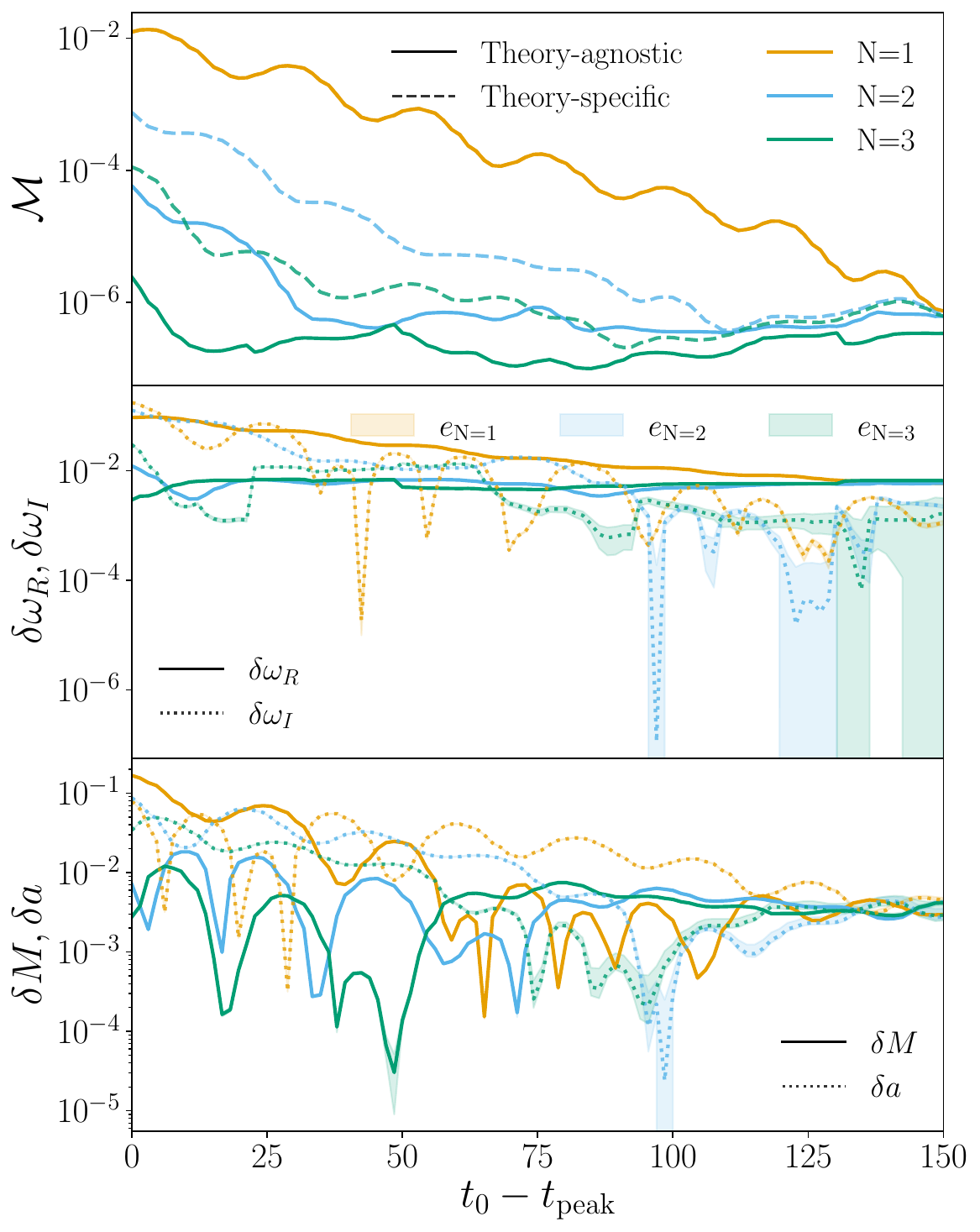}
    \caption{Same as Fig.~\ref{fig:B5fit} for the heavy $M_\mathrm{tot}=3.8M_{\odot}$ equal-mass system. Note that the mass errors in the bottom panel are so small that they are barely visible.}
    \label{fig:agnoSLyM38}
\end{figure}

The combined theory-agnostic and -specific fits are shown in Fig.~\ref{fig:agnoSLyM38}. As expected, the mismatches decrease when we increase the number of modes in the ringdown model. For $N=2$ and $N=3$, the mismatch approaches that of $N=1$ near the end of the initial time range, indicating that the modeled overtones reach their fitting bounds and do not improve the signal anymore (more details can be found in Appendix~\ref{app:Modestab}; we find a stable fundamental mode, while only observing a semistable first overtone for the first part of the analysis). We observe that for $N=1$, the mismatch between the vacuum and the fitted values for $\mathrm{Re}(\omega)$ decreases significantly once we move to later starting times, approaching around 1\% when fitting 150 time steps after the peak of the signal. Including additional modes leads to a 1\% difference even at early times. The deviation of $\mathrm{Im}(\omega)$ from the predictions closely resembles that of the real part, and is surprisingly accurate at late times. Mass and spin extractions show us a similar picture, with small deviations compared to the true values. Especially at early times, the fitting procedure seems to have trouble converging to the true spin, which improves at later times (this can be best seen in Fig.~\ref{fig:cycloid} of Appendix~\ref{app:Modestab}, where we show the starting time evolution in the $M-a$ plane). 

Given the relatively high mass of this BNS system, the ringdown frequency is about $\sim4.7$\,kHz, which is, even for the 3G detectors, e.g., \cite{Reitze:2019iox,Abac:2025saz}, outside the most sensitive region, which generally will make the detection of the ringdown only achievable for close events.  

\subsection{Unequal mass BNS with $M_{\text{total}}=3.2M_{\odot}$}\label{UnequM32ring}

\begin{figure}
    \centering
    \includegraphics[width=\linewidth]{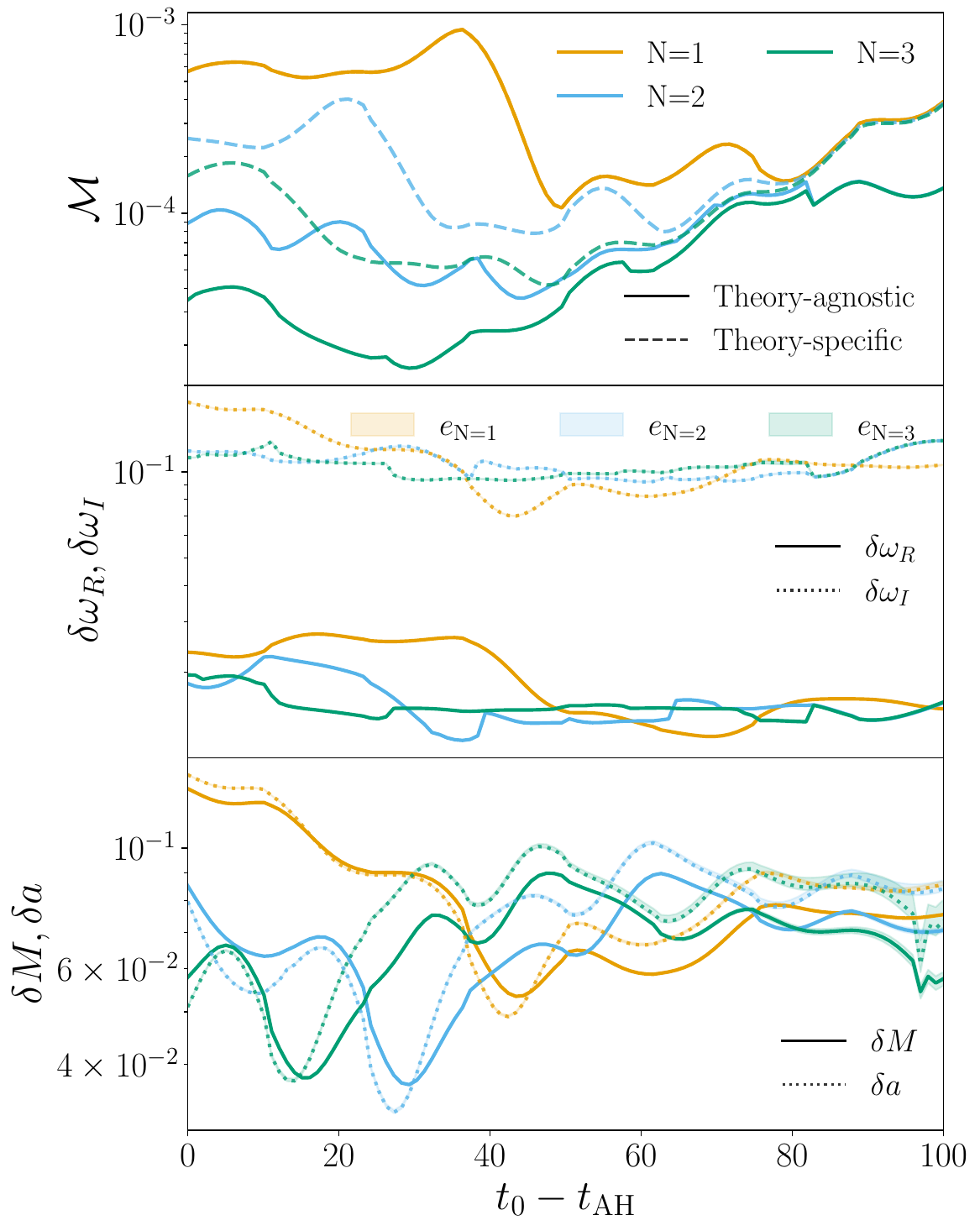}
    \caption{Same as Fig.~\ref{fig:B5fit} and the previous systems, for the unequal-mass $M_\mathrm{tot}=3.2M_{\odot}$ BNS.}
    \label{fig:agnoSLyM32u}
\end{figure}

In Fig.~\ref{fig:agnoSLyM32u}, we show results for the unequal-mass system with total mass $M_\mathrm{tot}=3.2M_{\odot}$. 
We chose a window of length $180$ and vary the starting time from $0$ to $100$. 
Note that the starting time is with respect to the time of BH formation, since collapse does not occur right away. 

One particularly interesting feature from the mismatches is that around $t-t_\mathrm{AH}=50$, the mismatches for all models seem to increase for the rest of the fitting window. This is likely due to the noise present during the last couple of oscillations in the ringdown signal of this system. Also, the theory-agnostic results reveal that across all ringdown models, $\mathrm{Re}(\omega)$ is recovered within a precision of 1-2\%, while $\mathrm{Im}(\omega)$ is about 10\% off the Kerr predictions for the most part. Interestingly, the theory-specific fits extract a spin deviation that closely follows that for the mass. In that case, besides the large offset for the $N=1$ model at early times, differences of the mass and spin values from the simulation oscillate a lot within the starting time range and stay within 4-10\%. Whether the large deviation is due to the baryonic matter outside or signal imperfections that impact the fitting procedure, cannot be said at this point. The frequency of the fundamental mode for the ringdown is about $\sim5.9$\,kHz. 

\subsection{Equal mass BNS with $M_{\text{total}}=2.7M_{\odot}$}\label{M27ringdownanalysis}
\begin{figure}
    \centering
    \includegraphics[width=\linewidth]{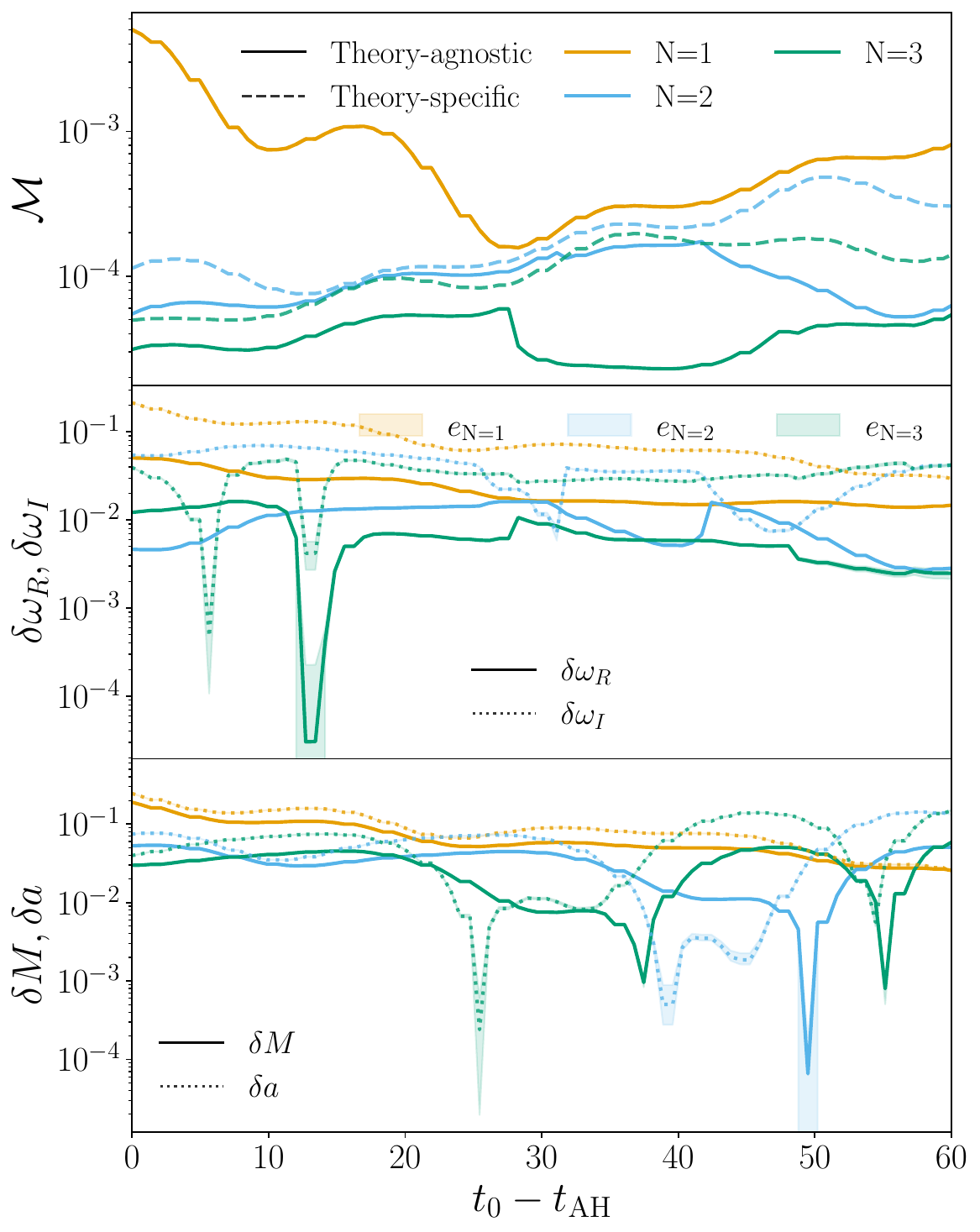}
    \caption{Same as Fig.~\ref{fig:B5fit} for the light $M_\mathrm{tot}=2.7M_{\odot}$ equal-mass BNS system.
    }
    \label{fig:agnoSLyM27}
\end{figure}

Lastly, the ringdown analysis of the lightest of the three BNS simulations, i.e., the $M_\mathrm{tot}=2.7M_{\odot}$ binary, is shown in Fig.~\ref{fig:agnoSLyM27}. Due to the shortness of the signal, we cannot extend the length of the signal to longer than $150$ after BH formation. To consider enough data points at later starting times, we set the maximum starting time to $60$ after AH formation. Starting with the theory-agnostic fits, we see that for the $N=1$ ringdown model, deviations to the prediction of $\mathrm{Re}(\omega)$ are around 5\% for most of the signal. This, of course, decreases once we consider more modes. For the $N=2$ and $N=3$ models, this reduces to about 1\% difference from the vacuum values, either due to the baryonic matter left outside or modeling specifications. The imaginary part for all ringdown models is above the 1\% threshold for most starting times. In general the fits are not as "clean" as for the $M_\mathrm{tot}=3.8M_{\odot}$ system, since most likely the baryonic matter outside affects the ringdown morphology which influences the parameter extraction. This is also visible, when looking at the theory-specific results in the bottom panel of Fig.~\ref{fig:agnoSLyM27}. At early times for the multimode models and late times for the one-mode model, we see deviations of 3-4\% for both mass and spin. With $\sim6.5$\,kHz the fundamental ringdown frequency is in the high-frequency regime making it unlikely to be detected. 

\section{Conclusions}\label{conclusion}

In traditional BH spectroscopy, one utilizes Teukolsky’s equation~\cite{Teulkosky1973} to investigate gravitational perturbations of a Kerr spacetime in vacuum~\cite{Kerr:1963ud}. 
The ringdown of BBH mergers is described by this well-established assumption, but BNS and BHNS mergers violate it. 
Assessing the impact of dynamical matter environments on the ringdown is, in general, nontrivial. 
Current GW detectors are not able to measure QNMs from collapsing NSs in such systems, mainly because their QNM frequencies are outside the sensitivity band. 
However, next-generation GW detectors like the Einstein Telescope~\cite{Punturo:2010zz,ET:2019dnz,Abac:2025saz} and Cosmic Explorer~\cite{Reitze:2019iox,Evans:2021gyd} might enable precision measurements of the ringdown of such systems, although the ringdown will most likely be seen in the less sensitive high-frequency regime. 

In this work, we addressed the question of whether matter matters for ringdown spectroscopy.
Using the numerical-relativity code \texttt{BAM}~\cite{Bruegmann:2006ulg,Thierfelder:2011yi}, we generated simulations of isolated, rotating NSs that were brought to collapse by introducing small perturbations, as well as BNS mergers. 
After examining the ringdown signals' morphology, we developed a qualitative classification system that divides them into three categories: ``clean'' (\texttt{Class I}), ``modified'' (\texttt{Class II}), and ``distorted'' (\texttt{Class III}) signals. 
We discuss the parameter space and individual cases in Sec.~\ref{classAnalysis}.

Signals of \texttt{Class I} are resembling a typical BH ringdown, but could in principle be prone to nonvacuum effects, which might be interpreted as false violations from GR. 
We therefore focused on using standard methods in BH spectroscopy to analyze such signals from collapsing isolated NSs and for BNS mergers in Sec.~\ref{moreResults}. 
Whether matter matters is highly case dependent. 
Even though they are unstable and physically questionable, adding more QNMs to our ringdown models significantly affects the accuracy of parameter estimation for important properties such as the remnant mass and its fundamental mode. 
Matter clearly matters for systems of \texttt{Class II} and \texttt{Class III} as they show clear deviations from a typical ringdown signal. 
\texttt{Class II} ringdowns admit qualitative similarities to cases presented in Ref.~\cite{Laeuger2025}, which studied oscillations of a matter shell surrounding a BH. 
Nevertheless, it is improbable that any confusion with a BH ringdown will occur, and if they are observed, they would not be employed for conventional BH spectroscopy. 
Phenomenological ringdown models may be used to extract disk properties for these systems~\cite{Dhani:2025xno}.

In our study, we find that 
using only the dominant QNM for the ringdown is not sufficient to obtain the correct BH mass below a percent level threshold. 
It appears that the presence of matter are not the causes of these deviations, at least not for waveforms of \texttt{Class I}.
By adding one or two overtones, mismatches are greatly reduced, and the reconstructed BH mass is often even accurate to within a few percent. 
This underlines that adding overtones does not only provide better fits, but moreover, improves the extraction of physical parameters.
Furthermore, it provides more consistent values for the amplitude and phase of the fundamental mode, albeit not as consistent as in the vacuum case or linear simulations.

Spins are, in general, more difficult to extract. 
Consequently, we examined two theory-specific cases. 
The first involved permitting the spin to vary freely within the physically permissible range, while the second involved restricting it in the vicinity of the correct value. 
The latter is motivated if one would obtain constraints on the final spin from an independent inspiral-merger analysis. 
Using early to intermediate starting times, the mass estimate can be enhanced by up to one order of magnitude by constraining the spin. However, it generally exacerbates mismatches. 
The effect of constraining the spin at late times is less relevant. 

There are some aspects of our work that could be extended in the future. 
Our analysis uses a least-squares minimization routine approximating the errors of the data points (that are always (implicitly) chosen in such an analysis), with those of the numerical-relativity noise floor. 
Bayesian analysis has been applied to real data in the context of determining overtones, which shows how the unclear starting time and impact of noise influence the analysis, see Refs.~\cite{Cotesta:2022pci,Isi:2023nif,Carullo:2023gtf,Finch:2022ynt} as an example discussing the difficulties. 
GW data analysis pipelines could be utilized to incorporate our numerical-relativity simulations as an injection, e.g., similar to Ref.~\cite{Dhani:2023ijt}, which investigates the detectability of black hole formation through BNS mergers with next generation detectors. 
Such an analysis would also yield measurement uncertainties regarding the ringdown parameters, which would be contingent upon the detector specification. 
Moreover, this would make it possible to measure the ratio of statistical to systematic errors, a crucial metric for determining the suitability of the underlying model used to analyze a signal. 
Using linear-signal analysis (LSA)~\cite{Finn:1992xs,Krolak:1993zy,Flanagan:1997kp,Cutler:2007mi}, in particular the bias formula~\cite{Flanagan:1997kp,Cutler:2007mi}, the bias ratio could also be explored by using the numerical-relativity simulations as an injection, and including more QNMs in a ringdown model, until systematic effects are smaller than statistical ones.
This has recently been explored in Ref.~\cite{Capuano:2025kkl} in the vacuum case, while the range of validity of LSA predictions for ringdown systematics has been explored in Ref.~\cite{Volkel:2025jdx} by explicitly comparing it to Bayesian analysis using Mark-chain-Monte-Carlo sampling.

\acknowledgments

O.\,S. thanks Anna Neuweiler for help regarding the creation of BNS initial data with \texttt{SGRID} and their evolution with \texttt{BAM}, as well as Ivan Markin for useful discussions related to issues with \texttt{BAM}. Additionally, O.\,S. thanks Henrique Gieg and Anna Puecher for help with computations on signal-to-noise ratio estimates.
S.\,H.\,V. thanks Arnab Dhani for useful discussions related to the detectability of our systems with next-generation GW detectors. 
S.\,H.\,V. acknowledges funding from the Deutsche Forschungsgemeinschaft (DFG): Project No. 386119226. 
T.~D.\ acknowledges funding from the European Union (ERC, SMArt, 101076369). 
Views and opinions expressed are those of the authors only and do not necessarily reflect those of the European Union or the European Research Council. Neither the European Union nor the granting authority can be held responsible for them.
Computations for this paper have been performed on the DFG-funded research cluster Jarvis at the University of Potsdam (INST 336/173-1; Proj. No.: 502227537).

\dataavailability{The data that support the findings of this article are
openly available~\cite{zenodorepo}.}

\bibliographystyle{apsrev4-2}
\bibliography{references}

\begin{thebibliography}{138}%
\makeatletter
\providecommand \@ifxundefined [1]{%
 \@ifx{#1\undefined}
}%
\providecommand \@ifnum [1]{%
 \ifnum #1\expandafter \@firstoftwo
 \else \expandafter \@secondoftwo
 \fi
}%
\providecommand \@ifx [1]{%
 \ifx #1\expandafter \@firstoftwo
 \else \expandafter \@secondoftwo
 \fi
}%
\providecommand \natexlab [1]{#1}%
\providecommand \enquote  [1]{``#1''}%
\providecommand \bibnamefont  [1]{#1}%
\providecommand \bibfnamefont [1]{#1}%
\providecommand \citenamefont [1]{#1}%
\providecommand \href@noop [0]{\@secondoftwo}%
\providecommand \href [0]{\begingroup \@sanitize@url \@href}%
\providecommand \@href[1]{\@@startlink{#1}\@@href}%
\providecommand \@@href[1]{\endgroup#1\@@endlink}%
\providecommand \@sanitize@url [0]{\catcode `\\12\catcode `\$12\catcode `\&12\catcode `\#12\catcode `\^12\catcode `\_12\catcode `\%12\relax}%
\providecommand \@@startlink[1]{}%
\providecommand \@@endlink[0]{}%
\providecommand \url  [0]{\begingroup\@sanitize@url \@url }%
\providecommand \@url [1]{\endgroup\@href {#1}{\urlprefix }}%
\providecommand \urlprefix  [0]{URL }%
\providecommand \Eprint [0]{\href }%
\providecommand \doibase [0]{https://doi.org/}%
\providecommand \selectlanguage [0]{\@gobble}%
\providecommand \bibinfo  [0]{\@secondoftwo}%
\providecommand \bibfield  [0]{\@secondoftwo}%
\providecommand \translation [1]{[#1]}%
\providecommand \BibitemOpen [0]{}%
\providecommand \bibitemStop [0]{}%
\providecommand \bibitemNoStop [0]{.\EOS\space}%
\providecommand \EOS [0]{\spacefactor3000\relax}%
\providecommand \BibitemShut  [1]{\csname bibitem#1\endcsname}%
\let\auto@bib@innerbib\@empty
\bibitem [{\citenamefont {Abbott}\ \emph {et~al.}(2016{\natexlab{a}})\citenamefont {Abbott} \emph {et~al.}}]{Abbott2016GW150914}%
  \BibitemOpen
  \bibfield  {author} {\bibinfo {author} {\bibfnamefont {B.~P.}\ \bibnamefont {Abbott}} \emph {et~al.} (\bibinfo {collaboration} {LIGO Scientific and Virgo Collaborations}),\ }\href {https://doi.org/10.1103/PhysRevLett.116.061102} {\bibfield  {journal} {\bibinfo  {journal} {Phys. Rev. Lett.}\ }\textbf {\bibinfo {volume} {116}},\ \bibinfo {pages} {061102} (\bibinfo {year} {2016}{\natexlab{a}})}\BibitemShut {NoStop}%
\bibitem [{\citenamefont {Abbott}\ \emph {et~al.}(2016{\natexlab{b}})\citenamefont {Abbott} \emph {et~al.}}]{Abbott2016Results}%
  \BibitemOpen
  \bibfield  {author} {\bibinfo {author} {\bibfnamefont {B.~P.}\ \bibnamefont {Abbott}} \emph {et~al.} (\bibinfo {collaboration} {LIGO Scientific and Virgo Collaborations}),\ }\href {https://doi.org/10.1103/PhysRevD.93.122003} {\bibfield  {journal} {\bibinfo  {journal} {Phys. Rev. D}\ }\textbf {\bibinfo {volume} {93}},\ \bibinfo {pages} {122003} (\bibinfo {year} {2016}{\natexlab{b}})}\BibitemShut {NoStop}%
\bibitem [{\citenamefont {Abbott}\ \emph {et~al.}(2019{\natexlab{a}})\citenamefont {Abbott} \emph {et~al.}}]{LIGOScientific:2018mvr}%
  \BibitemOpen
  \bibfield  {author} {\bibinfo {author} {\bibfnamefont {B.~P.}\ \bibnamefont {Abbott}} \emph {et~al.} (\bibinfo {collaboration} {LIGO Scientific and Virgo Collaborations}),\ }\href {https://doi.org/10.1103/PhysRevX.9.031040} {\bibfield  {journal} {\bibinfo  {journal} {Phys. Rev. X}\ }\textbf {\bibinfo {volume} {9}},\ \bibinfo {pages} {031040} (\bibinfo {year} {2019}{\natexlab{a}})}\BibitemShut {NoStop}%
\bibitem [{\citenamefont {Abbott}\ \emph {et~al.}(2021{\natexlab{a}})\citenamefont {Abbott} \emph {et~al.}}]{LIGOScientific:2020ibl}%
  \BibitemOpen
  \bibfield  {author} {\bibinfo {author} {\bibfnamefont {R.}~\bibnamefont {Abbott}} \emph {et~al.} (\bibinfo {collaboration} {LIGO Scientific and Virgo Collaborations}),\ }\href {https://doi.org/10.1103/PhysRevX.11.021053} {\bibfield  {journal} {\bibinfo  {journal} {Phys. Rev. X}\ }\textbf {\bibinfo {volume} {11}},\ \bibinfo {pages} {021053} (\bibinfo {year} {2021}{\natexlab{a}})}\BibitemShut {NoStop}%
\bibitem [{\citenamefont {Abbott}\ \emph {et~al.}(2023{\natexlab{a}})\citenamefont {Abbott} \emph {et~al.}}]{KAGRA:2021duu}%
  \BibitemOpen
  \bibfield  {author} {\bibinfo {author} {\bibfnamefont {R.}~\bibnamefont {Abbott}} \emph {et~al.} (\bibinfo {collaboration} {KAGRA, Virgo and LIGO Scientific Collaborations}),\ }\href {https://doi.org/10.1103/PhysRevX.13.011048} {\bibfield  {journal} {\bibinfo  {journal} {Phys. Rev. X}\ }\textbf {\bibinfo {volume} {13}},\ \bibinfo {pages} {011048} (\bibinfo {year} {2023}{\natexlab{a}})}\BibitemShut {NoStop}%
\bibitem [{\citenamefont {Abbott}\ \emph {et~al.}(2023{\natexlab{b}})\citenamefont {Abbott} \emph {et~al.}}]{KAGRA:2021vkt}%
  \BibitemOpen
  \bibfield  {author} {\bibinfo {author} {\bibfnamefont {R.}~\bibnamefont {Abbott}} \emph {et~al.} (\bibinfo {collaboration} {KAGRA, Virgo and LIGO Scientific Collaborations}),\ }\href {https://doi.org/10.1103/PhysRevX.13.041039} {\bibfield  {journal} {\bibinfo  {journal} {Phys. Rev. X}\ }\textbf {\bibinfo {volume} {13}},\ \bibinfo {pages} {041039} (\bibinfo {year} {2023}{\natexlab{b}})}\BibitemShut {NoStop}%
\bibitem [{\citenamefont {{Aky{\"u}z}}\ \emph {et~al.}()\citenamefont {{Aky{\"u}z}}, \citenamefont {{Correia}}, \citenamefont {{Garofalo}}, \citenamefont {{Kacanja}}, \citenamefont {{Jadhav Y}}, \citenamefont {{Roy}}, \citenamefont {{Soni}}, \citenamefont {{Tan}}, \citenamefont {{Capano}},\ and\ \citenamefont {{Nitz}}}]{Akyuz:2025ype}%
  \BibitemOpen
  \bibfield  {author} {\bibinfo {author} {\bibfnamefont {A.}~\bibnamefont {{Aky{\"u}z}}}, \bibinfo {author} {\bibfnamefont {A.}~\bibnamefont {{Correia}}}, \bibinfo {author} {\bibfnamefont {J.}~\bibnamefont {{Garofalo}}}, \bibinfo {author} {\bibfnamefont {K.}~\bibnamefont {{Kacanja}}}, \bibinfo {author} {\bibfnamefont {V.}~\bibnamefont {{Jadhav Y}}}, \bibinfo {author} {\bibfnamefont {L.}~\bibnamefont {{Roy}}}, \bibinfo {author} {\bibfnamefont {K.}~\bibnamefont {{Soni}}}, \bibinfo {author} {\bibfnamefont {H.}~\bibnamefont {{Tan}}}, \bibinfo {author} {\bibfnamefont {C.~D.}\ \bibnamefont {{Capano}}},\ and\ \bibinfo {author} {\bibfnamefont {A.~H.}\ \bibnamefont {{Nitz}}},\ }\href@noop {} {\ }\Eprint {https://arxiv.org/abs/2507.08778} {arXiv:2507.08778} \BibitemShut {NoStop}%
\bibitem [{\citenamefont {Abbott}\ \emph {et~al.}(2017{\natexlab{a}})\citenamefont {Abbott} \emph {et~al.}}]{LIGOScientific:2017vwq}%
  \BibitemOpen
  \bibfield  {author} {\bibinfo {author} {\bibfnamefont {B.~P.}\ \bibnamefont {Abbott}} \emph {et~al.} (\bibinfo {collaboration} {LIGO Scientific and Virgo Collaborations}),\ }\href {https://doi.org/10.1103/PhysRevLett.119.161101} {\bibfield  {journal} {\bibinfo  {journal} {Phys. Rev. Lett.}\ }\textbf {\bibinfo {volume} {119}},\ \bibinfo {pages} {161101} (\bibinfo {year} {2017}{\natexlab{a}})}\BibitemShut {NoStop}%
\bibitem [{\citenamefont {Abbott}\ \emph {et~al.}(2017{\natexlab{b}})\citenamefont {Abbott} \emph {et~al.}}]{LIGOScientific:2017ync}%
  \BibitemOpen
  \bibfield  {author} {\bibinfo {author} {\bibfnamefont {B.~P.}\ \bibnamefont {Abbott}} \emph {et~al.} (\bibinfo {collaboration} {LIGO Scientific, Virgo, Fermi GBM, INTEGRAL, IceCube, AstroSat Cadmium Zinc Telluride Imager Team, IPN, Insight-Hxmt, ANTARES, Swift, AGILE Team, 1M2H Team, Dark Energy Camera GW-EM, DES, DLT40, GRAWITA, Fermi-LAT, ATCA, ASKAP, Las Cumbres Observatory Group, OzGrav, DWF (Deeper Wider Faster Program), AST3, CAASTRO, VINROUGE, MASTER, J-GEM, GROWTH, JAGWAR, CaltechNRAO, TTU-NRAO, NuSTAR, Pan-STARRS, MAXI Team, TZAC Consortium, KU, Nordic Optical Telescope, ePESSTO, GROND, Texas Tech University, SALT Group, TOROS, BOOTES, MWA, CALET, IKI-GW Follow-up, H.E.S.S., LOFAR, LWA, HAWC, Pierre Auger, ALMA, Euro VLBI Team, Pi of Sky, Chandra Team at McGill University, DFN, ATLAS Telescopes, High Time Resolution Universe Survey, RIMAS, RATIR, SKA South Africa/MeerKAT Collaborations}),\ }\href {https://doi.org/10.3847/2041-8213/aa91c9} {\bibfield  {journal} {\bibinfo  {journal} {Astrophys. J.
  Lett.}\ }\textbf {\bibinfo {volume} {848}},\ \bibinfo {pages} {L12} (\bibinfo {year} {2017}{\natexlab{b}})}\BibitemShut {NoStop}%
\bibitem [{\citenamefont {Abbott}\ \emph {et~al.}(2020)\citenamefont {Abbott} \emph {et~al.}}]{LIGOScientific:2020aai}%
  \BibitemOpen
  \bibfield  {author} {\bibinfo {author} {\bibfnamefont {B.~P.}\ \bibnamefont {Abbott}} \emph {et~al.} (\bibinfo {collaboration} {LIGO Scientific and Virgo Collaborations}),\ }\href {https://doi.org/10.3847/2041-8213/ab75f5} {\bibfield  {journal} {\bibinfo  {journal} {Astrophys. J. Lett.}\ }\textbf {\bibinfo {volume} {892}},\ \bibinfo {pages} {L3} (\bibinfo {year} {2020})}\BibitemShut {NoStop}%
\bibitem [{\citenamefont {Abbott}\ \emph {et~al.}(2021{\natexlab{b}})\citenamefont {Abbott} \emph {et~al.}}]{LIGOScientific:2021qlt}%
  \BibitemOpen
  \bibfield  {author} {\bibinfo {author} {\bibfnamefont {R.}~\bibnamefont {Abbott}} \emph {et~al.} (\bibinfo {collaboration} {LIGO Scientific, KAGRA and Virgo Collaborations}),\ }\href {https://doi.org/10.3847/2041-8213/ac082e} {\bibfield  {journal} {\bibinfo  {journal} {Astrophys. J. Lett.}\ }\textbf {\bibinfo {volume} {915}},\ \bibinfo {pages} {L5} (\bibinfo {year} {2021}{\natexlab{b}})}\BibitemShut {NoStop}%
\bibitem [{\citenamefont {Abac}\ \emph {et~al.}(2024)\citenamefont {Abac} \emph {et~al.}}]{LIGOScientific:2024elc}%
  \BibitemOpen
  \bibfield  {author} {\bibinfo {author} {\bibfnamefont {A.~G.}\ \bibnamefont {Abac}} \emph {et~al.} (\bibinfo {collaboration} {LIGO Scientific, KAGRA, and Virgo Collaborations}),\ }\href {https://doi.org/10.3847/2041-8213/ad5beb} {\bibfield  {journal} {\bibinfo  {journal} {Astrophys. J. Lett.}\ }\textbf {\bibinfo {volume} {970}},\ \bibinfo {pages} {L34} (\bibinfo {year} {2024})}\BibitemShut {NoStop}%
\bibitem [{\citenamefont {Koehn}\ \emph {et~al.}(2025)\citenamefont {Koehn} \emph {et~al.}}]{Koehn:2024set}%
  \BibitemOpen
  \bibfield  {author} {\bibinfo {author} {\bibfnamefont {H.}~\bibnamefont {Koehn}} \emph {et~al.},\ }\href {https://doi.org/10.1103/PhysRevX.15.021014} {\bibfield  {journal} {\bibinfo  {journal} {Phys. Rev. X}\ }\textbf {\bibinfo {volume} {15}},\ \bibinfo {pages} {021014} (\bibinfo {year} {2025})}\BibitemShut {NoStop}%
\bibitem [{\citenamefont {{Chatziioannou}}\ \emph {et~al.}(2025)\citenamefont {{Chatziioannou}}, \citenamefont {{Cromartie}}, \citenamefont {{Gandolfi}}, \citenamefont {{Tews}}, \citenamefont {{Radice}}, \citenamefont {{Steiner}},\ and\ \citenamefont {{Watts}}}]{Chatziioannou:2024tjq}%
  \BibitemOpen
  \bibfield  {author} {\bibinfo {author} {\bibfnamefont {K.}~\bibnamefont {{Chatziioannou}}}, \bibinfo {author} {\bibfnamefont {H.~T.}\ \bibnamefont {{Cromartie}}}, \bibinfo {author} {\bibfnamefont {S.}~\bibnamefont {{Gandolfi}}}, \bibinfo {author} {\bibfnamefont {I.}~\bibnamefont {{Tews}}}, \bibinfo {author} {\bibfnamefont {D.}~\bibnamefont {{Radice}}}, \bibinfo {author} {\bibfnamefont {A.~W.}\ \bibnamefont {{Steiner}}},\ and\ \bibinfo {author} {\bibfnamefont {A.~L.}\ \bibnamefont {{Watts}}},\ }\href {https://doi.org/10.1103/ymsq-cfcw} {\bibfield  {journal} {\bibinfo  {journal} {Reviews of Modern Physics}\ }\textbf {\bibinfo {volume} {97}},\ \bibinfo {eid} {045007} (\bibinfo {year} {2025})}\BibitemShut {NoStop}%
\bibitem [{\citenamefont {Abbott}\ \emph {et~al.}(2016{\natexlab{c}})\citenamefont {Abbott} \emph {et~al.}}]{TestsGR2016}%
  \BibitemOpen
  \bibfield  {author} {\bibinfo {author} {\bibfnamefont {B.~P.}\ \bibnamefont {Abbott}} \emph {et~al.} (\bibinfo {collaboration} {LIGO Scientific and Virgo Collaborations}),\ }\href {https://doi.org/10.1103/PhysRevLett.116.221101} {\bibfield  {journal} {\bibinfo  {journal} {Phys. Rev. Lett.}\ }\textbf {\bibinfo {volume} {116}},\ \bibinfo {pages} {221101} (\bibinfo {year} {2016}{\natexlab{c}})},\ \bibinfo {note} {[Erratum: Phys.Rev.Lett. 121, 129902 (2018)]}\BibitemShut {NoStop}%
\bibitem [{\citenamefont {Abbott}\ \emph {et~al.}(2019{\natexlab{b}})\citenamefont {Abbott} \emph {et~al.}}]{TestsGR2018}%
  \BibitemOpen
  \bibfield  {author} {\bibinfo {author} {\bibfnamefont {B.~P.}\ \bibnamefont {Abbott}} \emph {et~al.} (\bibinfo {collaboration} {LIGO Scientific and Virgo Collaborations}),\ }\href {https://doi.org/10.1103/PhysRevLett.123.011102} {\bibfield  {journal} {\bibinfo  {journal} {Phys. Rev. Lett.}\ }\textbf {\bibinfo {volume} {123}},\ \bibinfo {pages} {011102} (\bibinfo {year} {2019}{\natexlab{b}})}\BibitemShut {NoStop}%
\bibitem [{\citenamefont {Abbott}\ \emph {et~al.}(2019{\natexlab{c}})\citenamefont {Abbott} \emph {et~al.}}]{TestsGR2019}%
  \BibitemOpen
  \bibfield  {author} {\bibinfo {author} {\bibfnamefont {B.~P.}\ \bibnamefont {Abbott}} \emph {et~al.} (\bibinfo {collaboration} {LIGO Scientific and Virgo Collaborations}),\ }\href {https://doi.org/10.1103/PhysRevD.100.104036} {\bibfield  {journal} {\bibinfo  {journal} {Phys. Rev. D}\ }\textbf {\bibinfo {volume} {100}},\ \bibinfo {pages} {104036} (\bibinfo {year} {2019}{\natexlab{c}})}\BibitemShut {NoStop}%
\bibitem [{\citenamefont {Israel}(1967)}]{Israel1967}%
  \BibitemOpen
  \bibfield  {author} {\bibinfo {author} {\bibfnamefont {W.}~\bibnamefont {Israel}},\ }\href {https://doi.org/10.1103/PhysRev.164.1776} {\bibfield  {journal} {\bibinfo  {journal} {Phys. Rev.}\ }\textbf {\bibinfo {volume} {164}},\ \bibinfo {pages} {1776} (\bibinfo {year} {1967})}\BibitemShut {NoStop}%
\bibitem [{\citenamefont {{Carter}}(1971)}]{Carter1971}%
  \BibitemOpen
  \bibfield  {author} {\bibinfo {author} {\bibfnamefont {B.}~\bibnamefont {{Carter}}},\ }\href {https://doi.org/10.1103/PhysRevLett.26.331} {\bibfield  {journal} {\bibinfo  {journal} {\prl}\ }\textbf {\bibinfo {volume} {26}},\ \bibinfo {pages} {331} (\bibinfo {year} {1971})}\BibitemShut {NoStop}%
\bibitem [{\citenamefont {Robinson}(1975)}]{Robinson1975}%
  \BibitemOpen
  \bibfield  {author} {\bibinfo {author} {\bibfnamefont {D.~C.}\ \bibnamefont {Robinson}},\ }\href {https://doi.org/10.1103/PhysRevLett.34.905} {\bibfield  {journal} {\bibinfo  {journal} {Phys. Rev. Lett.}\ }\textbf {\bibinfo {volume} {34}},\ \bibinfo {pages} {905} (\bibinfo {year} {1975})}\BibitemShut {NoStop}%
\bibitem [{\citenamefont {Schwarzschild}(1916)}]{Schwarzschild:1916uq}%
  \BibitemOpen
  \bibfield  {author} {\bibinfo {author} {\bibfnamefont {K.}~\bibnamefont {Schwarzschild}},\ }\href@noop {} {\bibfield  {journal} {\bibinfo  {journal} {Sitzungsber. Preuss. Akad. Wiss. Berlin (Math. Phys. )}\ }\textbf {\bibinfo {volume} {1916}},\ \bibinfo {pages} {189} (\bibinfo {year} {1916})}\BibitemShut {NoStop}%
\bibitem [{\citenamefont {Kerr}(1963)}]{Kerr:1963ud}%
  \BibitemOpen
  \bibfield  {author} {\bibinfo {author} {\bibfnamefont {R.~P.}\ \bibnamefont {Kerr}},\ }\href {https://doi.org/10.1103/PhysRevLett.11.237} {\bibfield  {journal} {\bibinfo  {journal} {Phys. Rev. Lett.}\ }\textbf {\bibinfo {volume} {11}},\ \bibinfo {pages} {237} (\bibinfo {year} {1963})}\BibitemShut {NoStop}%
\bibitem [{\citenamefont {Newman}\ \emph {et~al.}(1965)\citenamefont {Newman}, \citenamefont {Couch}, \citenamefont {Chinnapared}, \citenamefont {Exton}, \citenamefont {Prakash},\ and\ \citenamefont {Torrence}}]{Newman:1965my}%
  \BibitemOpen
  \bibfield  {author} {\bibinfo {author} {\bibfnamefont {E.~T.}\ \bibnamefont {Newman}}, \bibinfo {author} {\bibfnamefont {E.}~\bibnamefont {Couch}}, \bibinfo {author} {\bibfnamefont {K.}~\bibnamefont {Chinnapared}}, \bibinfo {author} {\bibfnamefont {A.}~\bibnamefont {Exton}}, \bibinfo {author} {\bibfnamefont {A.}~\bibnamefont {Prakash}},\ and\ \bibinfo {author} {\bibfnamefont {R.}~\bibnamefont {Torrence}},\ }\href {https://doi.org/10.1063/1.1704351} {\bibfield  {journal} {\bibinfo  {journal} {J. Math. Phys. (N.Y.)}\ }\textbf {\bibinfo {volume} {6}},\ \bibinfo {pages} {918} (\bibinfo {year} {1965})}\BibitemShut {NoStop}%
\bibitem [{\citenamefont {Regge}\ and\ \citenamefont {Wheeler}(1957)}]{Regge1957}%
  \BibitemOpen
  \bibfield  {author} {\bibinfo {author} {\bibfnamefont {T.}~\bibnamefont {Regge}}\ and\ \bibinfo {author} {\bibfnamefont {J.~A.}\ \bibnamefont {Wheeler}},\ }\href {https://doi.org/10.1103/PhysRev.108.1063} {\bibfield  {journal} {\bibinfo  {journal} {Phys. Rev.}\ }\textbf {\bibinfo {volume} {108}},\ \bibinfo {pages} {1063} (\bibinfo {year} {1957})}\BibitemShut {NoStop}%
\bibitem [{\citenamefont {Zerilli}(1970{\natexlab{a}})}]{Zerilli1970}%
  \BibitemOpen
  \bibfield  {author} {\bibinfo {author} {\bibfnamefont {F.~J.}\ \bibnamefont {Zerilli}},\ }\href {https://doi.org/10.1103/PhysRevLett.24.737} {\bibfield  {journal} {\bibinfo  {journal} {Phys. Rev. Lett.}\ }\textbf {\bibinfo {volume} {24}},\ \bibinfo {pages} {737} (\bibinfo {year} {1970}{\natexlab{a}})}\BibitemShut {NoStop}%
\bibitem [{\citenamefont {Zerilli}(1970{\natexlab{b}})}]{Zerilli1970b}%
  \BibitemOpen
  \bibfield  {author} {\bibinfo {author} {\bibfnamefont {F.~J.}\ \bibnamefont {Zerilli}},\ }\href {https://doi.org/10.1103/PhysRevD.2.2141} {\bibfield  {journal} {\bibinfo  {journal} {Phys. Rev. D}\ }\textbf {\bibinfo {volume} {2}},\ \bibinfo {pages} {2141} (\bibinfo {year} {1970}{\natexlab{b}})}\BibitemShut {NoStop}%
\bibitem [{\citenamefont {Vishveshwara}(1970)}]{Vishveshwara1970}%
  \BibitemOpen
  \bibfield  {author} {\bibinfo {author} {\bibfnamefont {C.~V.}\ \bibnamefont {Vishveshwara}},\ }\href {https://doi.org/10.1038/227936a0} {\bibfield  {journal} {\bibinfo  {journal} {Nature (London)}\ }\textbf {\bibinfo {volume} {227}},\ \bibinfo {pages} {936} (\bibinfo {year} {1970})}\BibitemShut {NoStop}%
\bibitem [{\citenamefont {{Teukolsky}}(1973)}]{Teulkosky1973}%
  \BibitemOpen
  \bibfield  {author} {\bibinfo {author} {\bibfnamefont {S.~A.}\ \bibnamefont {{Teukolsky}}},\ }\href {https://doi.org/10.1086/152444} {\bibfield  {journal} {\bibinfo  {journal} {\apj}\ }\textbf {\bibinfo {volume} {185}},\ \bibinfo {pages} {635} (\bibinfo {year} {1973})}\BibitemShut {NoStop}%
\bibitem [{\citenamefont {Buonanno}\ \emph {et~al.}(2007)\citenamefont {Buonanno}, \citenamefont {Cook},\ and\ \citenamefont {Pretorius}}]{Buonanno:2006ui}%
  \BibitemOpen
  \bibfield  {author} {\bibinfo {author} {\bibfnamefont {A.}~\bibnamefont {Buonanno}}, \bibinfo {author} {\bibfnamefont {G.~B.}\ \bibnamefont {Cook}},\ and\ \bibinfo {author} {\bibfnamefont {F.}~\bibnamefont {Pretorius}},\ }\href {https://doi.org/10.1103/PhysRevD.75.124018} {\bibfield  {journal} {\bibinfo  {journal} {Phys. Rev. D}\ }\textbf {\bibinfo {volume} {75}},\ \bibinfo {pages} {124018} (\bibinfo {year} {2007})}\BibitemShut {NoStop}%
\bibitem [{\citenamefont {Berti}\ \emph {et~al.}(2007)\citenamefont {Berti}, \citenamefont {Cardoso}, \citenamefont {Gonzalez}, \citenamefont {Sperhake}, \citenamefont {Hannam}, \citenamefont {Husa},\ and\ \citenamefont {Br{\"u}gmann}}]{Berti:2007fi}%
  \BibitemOpen
  \bibfield  {author} {\bibinfo {author} {\bibfnamefont {E.}~\bibnamefont {Berti}}, \bibinfo {author} {\bibfnamefont {V.}~\bibnamefont {Cardoso}}, \bibinfo {author} {\bibfnamefont {J.~A.}\ \bibnamefont {Gonzalez}}, \bibinfo {author} {\bibfnamefont {U.}~\bibnamefont {Sperhake}}, \bibinfo {author} {\bibfnamefont {M.}~\bibnamefont {Hannam}}, \bibinfo {author} {\bibfnamefont {S.}~\bibnamefont {Husa}},\ and\ \bibinfo {author} {\bibfnamefont {B.}~\bibnamefont {Br{\"u}gmann}},\ }\href {https://doi.org/10.1103/PhysRevD.76.064034} {\bibfield  {journal} {\bibinfo  {journal} {Phys. Rev. D}\ }\textbf {\bibinfo {volume} {76}},\ \bibinfo {pages} {064034} (\bibinfo {year} {2007})}\BibitemShut {NoStop}%
\bibitem [{\citenamefont {London}\ \emph {et~al.}(2014)\citenamefont {London}, \citenamefont {Shoemaker},\ and\ \citenamefont {Healy}}]{London:2014cma}%
  \BibitemOpen
  \bibfield  {author} {\bibinfo {author} {\bibfnamefont {L.}~\bibnamefont {London}}, \bibinfo {author} {\bibfnamefont {D.}~\bibnamefont {Shoemaker}},\ and\ \bibinfo {author} {\bibfnamefont {J.}~\bibnamefont {Healy}},\ }\href {https://doi.org/10.1103/PhysRevD.90.124032} {\bibfield  {journal} {\bibinfo  {journal} {Phys. Rev. D}\ }\textbf {\bibinfo {volume} {90}},\ \bibinfo {pages} {124032} (\bibinfo {year} {2014})},\ \bibinfo {note} {[Erratum: Phys.Rev.D 94, 069902 (2016)]}\BibitemShut {NoStop}%
\bibitem [{\citenamefont {Kokkotas}\ and\ \citenamefont {Schmidt}(1999)}]{Kokkotas:1999bd}%
  \BibitemOpen
  \bibfield  {author} {\bibinfo {author} {\bibfnamefont {K.~D.}\ \bibnamefont {Kokkotas}}\ and\ \bibinfo {author} {\bibfnamefont {B.~G.}\ \bibnamefont {Schmidt}},\ }\href {https://doi.org/10.12942/lrr-1999-2} {\bibfield  {journal} {\bibinfo  {journal} {Living Rev. Relativity}\ }\textbf {\bibinfo {volume} {2}},\ \bibinfo {pages} {2} (\bibinfo {year} {1999})}\BibitemShut {NoStop}%
\bibitem [{\citenamefont {Nollert}(1999)}]{Nollert:1999ji}%
  \BibitemOpen
  \bibfield  {author} {\bibinfo {author} {\bibfnamefont {H.-P.}\ \bibnamefont {Nollert}},\ }\href {https://doi.org/10.1088/0264-9381/16/12/201} {\bibfield  {journal} {\bibinfo  {journal} {Classical Quantum Gravity}\ }\textbf {\bibinfo {volume} {16}},\ \bibinfo {pages} {R159} (\bibinfo {year} {1999})}\BibitemShut {NoStop}%
\bibitem [{\citenamefont {Berti}\ \emph {et~al.}(2009)\citenamefont {Berti}, \citenamefont {Cardoso},\ and\ \citenamefont {Starinets}}]{Berti:2009kk}%
  \BibitemOpen
  \bibfield  {author} {\bibinfo {author} {\bibfnamefont {E.}~\bibnamefont {Berti}}, \bibinfo {author} {\bibfnamefont {V.}~\bibnamefont {Cardoso}},\ and\ \bibinfo {author} {\bibfnamefont {A.~O.}\ \bibnamefont {Starinets}},\ }\href {https://doi.org/10.1088/0264-9381/26/16/163001} {\bibfield  {journal} {\bibinfo  {journal} {Class. Quant. Grav.}\ }\textbf {\bibinfo {volume} {26}},\ \bibinfo {pages} {163001} (\bibinfo {year} {2009})},\ \Eprint {https://arxiv.org/abs/0905.2975} {arXiv:0905.2975 [gr-qc]} \BibitemShut {NoStop}%
\bibitem [{\citenamefont {Konoplya}\ and\ \citenamefont {Zhidenko}(2011)}]{Konoplya:2011qq}%
  \BibitemOpen
  \bibfield  {author} {\bibinfo {author} {\bibfnamefont {R.~A.}\ \bibnamefont {Konoplya}}\ and\ \bibinfo {author} {\bibfnamefont {A.}~\bibnamefont {Zhidenko}},\ }\href {https://doi.org/10.1103/RevModPhys.83.793} {\bibfield  {journal} {\bibinfo  {journal} {Rev. Mod. Phys.}\ }\textbf {\bibinfo {volume} {83}},\ \bibinfo {pages} {793} (\bibinfo {year} {2011})}\BibitemShut {NoStop}%
\bibitem [{\citenamefont {Berti}\ \emph {et~al.}(2018)\citenamefont {Berti}, \citenamefont {Yagi}, \citenamefont {Yang},\ and\ \citenamefont {Yunes}}]{Berti:2018vdi}%
  \BibitemOpen
  \bibfield  {author} {\bibinfo {author} {\bibfnamefont {E.}~\bibnamefont {Berti}}, \bibinfo {author} {\bibfnamefont {K.}~\bibnamefont {Yagi}}, \bibinfo {author} {\bibfnamefont {H.}~\bibnamefont {Yang}},\ and\ \bibinfo {author} {\bibfnamefont {N.}~\bibnamefont {Yunes}},\ }\href {https://doi.org/10.1007/s10714-018-2372-6} {\bibfield  {journal} {\bibinfo  {journal} {Gen. Relativ. Gravit.}\ }\textbf {\bibinfo {volume} {50}},\ \bibinfo {pages} {49} (\bibinfo {year} {2018})}\BibitemShut {NoStop}%
\bibitem [{\citenamefont {Berti}\ \emph {et~al.}(2016)\citenamefont {Berti}, \citenamefont {Sesana}, \citenamefont {Barausse}, \citenamefont {Cardoso},\ and\ \citenamefont {Belczynski}}]{Berti2016}%
  \BibitemOpen
  \bibfield  {author} {\bibinfo {author} {\bibfnamefont {E.}~\bibnamefont {Berti}}, \bibinfo {author} {\bibfnamefont {A.}~\bibnamefont {Sesana}}, \bibinfo {author} {\bibfnamefont {E.}~\bibnamefont {Barausse}}, \bibinfo {author} {\bibfnamefont {V.}~\bibnamefont {Cardoso}},\ and\ \bibinfo {author} {\bibfnamefont {K.}~\bibnamefont {Belczynski}},\ }\href {https://doi.org/10.1103/PhysRevLett.117.101102} {\bibfield  {journal} {\bibinfo  {journal} {Phys. Rev. Lett.}\ }\textbf {\bibinfo {volume} {117}},\ \bibinfo {pages} {101102} (\bibinfo {year} {2016})}\BibitemShut {NoStop}%
\bibitem [{\citenamefont {Franchini}\ and\ \citenamefont {V\"olkel}(2024)}]{Franchini:2023eda}%
  \BibitemOpen
  \bibfield  {author} {\bibinfo {author} {\bibfnamefont {N.}~\bibnamefont {Franchini}}\ and\ \bibinfo {author} {\bibfnamefont {S.~H.}\ \bibnamefont {V\"olkel}},\ }in\ \href {https://doi.org/https://doi.org/10.1007/978-981-97-2871-8} {\emph {\bibinfo {booktitle} {Recent Progress on Gravity Tests}}}\ (\bibinfo  {publisher} {Springer},\ \bibinfo {address} {Singapore},\ \bibinfo {year} {2024})\BibitemShut {NoStop}%
\bibitem [{\citenamefont {Berti}\ \emph {et~al.}()\citenamefont {Berti} \emph {et~al.}}]{Berti:2025hly}%
  \BibitemOpen
  \bibfield  {author} {\bibinfo {author} {\bibfnamefont {E.}~\bibnamefont {Berti}} \emph {et~al.},\ }\href@noop {} {\ }\Eprint {https://arxiv.org/abs/2505.23895} {arXiv:2505.23895} \BibitemShut {NoStop}%
\bibitem [{\citenamefont {{Detweiler}}(1980)}]{1980ApJ...239..292D}%
  \BibitemOpen
  \bibfield  {author} {\bibinfo {author} {\bibfnamefont {S.}~\bibnamefont {{Detweiler}}},\ }\href {https://doi.org/10.1086/158109} {\bibfield  {journal} {\bibinfo  {journal} {\apj}\ }\textbf {\bibinfo {volume} {239}},\ \bibinfo {pages} {292} (\bibinfo {year} {1980})}\BibitemShut {NoStop}%
\bibitem [{\citenamefont {Dreyer}\ \emph {et~al.}(2004)\citenamefont {Dreyer}, \citenamefont {Kelly}, \citenamefont {Krishnan}, \citenamefont {Finn}, \citenamefont {Garrison},\ and\ \citenamefont {Lopez-Aleman}}]{Dreyer2003}%
  \BibitemOpen
  \bibfield  {author} {\bibinfo {author} {\bibfnamefont {O.}~\bibnamefont {Dreyer}}, \bibinfo {author} {\bibfnamefont {B.~J.}\ \bibnamefont {Kelly}}, \bibinfo {author} {\bibfnamefont {B.}~\bibnamefont {Krishnan}}, \bibinfo {author} {\bibfnamefont {L.~S.}\ \bibnamefont {Finn}}, \bibinfo {author} {\bibfnamefont {D.}~\bibnamefont {Garrison}},\ and\ \bibinfo {author} {\bibfnamefont {R.}~\bibnamefont {Lopez-Aleman}},\ }\href {https://doi.org/10.1088/0264-9381/21/4/003} {\bibfield  {journal} {\bibinfo  {journal} {Classical Quantum Gravity}\ }\textbf {\bibinfo {volume} {21}},\ \bibinfo {pages} {787} (\bibinfo {year} {2004})}\BibitemShut {NoStop}%
\bibitem [{\citenamefont {Berti}\ \emph {et~al.}(2006)\citenamefont {Berti}, \citenamefont {Cardoso},\ and\ \citenamefont {Will}}]{Berti:2005ys}%
  \BibitemOpen
  \bibfield  {author} {\bibinfo {author} {\bibfnamefont {E.}~\bibnamefont {Berti}}, \bibinfo {author} {\bibfnamefont {V.}~\bibnamefont {Cardoso}},\ and\ \bibinfo {author} {\bibfnamefont {C.~M.}\ \bibnamefont {Will}},\ }\href {https://doi.org/10.1103/PhysRevD.73.064030} {\bibfield  {journal} {\bibinfo  {journal} {Phys. Rev. D}\ }\textbf {\bibinfo {volume} {73}},\ \bibinfo {pages} {064030} (\bibinfo {year} {2006})}\BibitemShut {NoStop}%
\bibitem [{\citenamefont {Giesler}\ \emph {et~al.}(2019)\citenamefont {Giesler}, \citenamefont {Isi}, \citenamefont {Scheel},\ and\ \citenamefont {Teukolsky}}]{Giesler2019}%
  \BibitemOpen
  \bibfield  {author} {\bibinfo {author} {\bibfnamefont {M.}~\bibnamefont {Giesler}}, \bibinfo {author} {\bibfnamefont {M.}~\bibnamefont {Isi}}, \bibinfo {author} {\bibfnamefont {M.~A.}\ \bibnamefont {Scheel}},\ and\ \bibinfo {author} {\bibfnamefont {S.~A.}\ \bibnamefont {Teukolsky}},\ }\href {https://doi.org/10.1103/PhysRevX.9.041060} {\bibfield  {journal} {\bibinfo  {journal} {Phys. Rev. X}\ }\textbf {\bibinfo {volume} {9}},\ \bibinfo {pages} {041060} (\bibinfo {year} {2019})}\BibitemShut {NoStop}%
\bibitem [{\citenamefont {Cheung}\ \emph {et~al.}(2024)\citenamefont {Cheung}, \citenamefont {Berti}, \citenamefont {Baibhav},\ and\ \citenamefont {Cotesta}}]{Cheung2023}%
  \BibitemOpen
  \bibfield  {author} {\bibinfo {author} {\bibfnamefont {M.~H.-Y.}\ \bibnamefont {Cheung}}, \bibinfo {author} {\bibfnamefont {E.}~\bibnamefont {Berti}}, \bibinfo {author} {\bibfnamefont {V.}~\bibnamefont {Baibhav}},\ and\ \bibinfo {author} {\bibfnamefont {R.}~\bibnamefont {Cotesta}},\ }\href {https://doi.org/10.1103/PhysRevD.109.044069} {\bibfield  {journal} {\bibinfo  {journal} {Phys. Rev. D}\ }\textbf {\bibinfo {volume} {109}},\ \bibinfo {pages} {044069} (\bibinfo {year} {2024})},\ \bibinfo {note} {[Erratum: Phys.Rev.D 110, 049902 (2024)]}\BibitemShut {NoStop}%
\bibitem [{\citenamefont {Baibhav}\ \emph {et~al.}(2023)\citenamefont {Baibhav}, \citenamefont {Cheung}, \citenamefont {Berti}, \citenamefont {Cardoso}, \citenamefont {Carullo}, \citenamefont {Cotesta}, \citenamefont {Del~Pozzo},\ and\ \citenamefont {Duque}}]{Baibhav2023}%
  \BibitemOpen
  \bibfield  {author} {\bibinfo {author} {\bibfnamefont {V.}~\bibnamefont {Baibhav}}, \bibinfo {author} {\bibfnamefont {M.~H.-Y.}\ \bibnamefont {Cheung}}, \bibinfo {author} {\bibfnamefont {E.}~\bibnamefont {Berti}}, \bibinfo {author} {\bibfnamefont {V.}~\bibnamefont {Cardoso}}, \bibinfo {author} {\bibfnamefont {G.}~\bibnamefont {Carullo}}, \bibinfo {author} {\bibfnamefont {R.}~\bibnamefont {Cotesta}}, \bibinfo {author} {\bibfnamefont {W.}~\bibnamefont {Del~Pozzo}},\ and\ \bibinfo {author} {\bibfnamefont {F.}~\bibnamefont {Duque}},\ }\href {https://doi.org/10.1103/PhysRevD.108.104020} {\bibfield  {journal} {\bibinfo  {journal} {Phys. Rev. D}\ }\textbf {\bibinfo {volume} {108}},\ \bibinfo {pages} {104020} (\bibinfo {year} {2023})}\BibitemShut {NoStop}%
\bibitem [{\citenamefont {Giesler}\ \emph {et~al.}(2025)\citenamefont {Giesler} \emph {et~al.}}]{Giesler2024Overtones}%
  \BibitemOpen
  \bibfield  {author} {\bibinfo {author} {\bibfnamefont {M.}~\bibnamefont {Giesler}} \emph {et~al.},\ }\href {https://doi.org/10.1103/PhysRevD.111.084041} {\bibfield  {journal} {\bibinfo  {journal} {Phys. Rev. D}\ }\textbf {\bibinfo {volume} {111}},\ \bibinfo {pages} {084041} (\bibinfo {year} {2025})}\BibitemShut {NoStop}%
\bibitem [{\citenamefont {{SXS Collaboration}}(2024)}]{varpro}%
  \BibitemOpen
  \bibfield  {author} {\bibinfo {author} {\bibnamefont {{SXS Collaboration}}},\ }\href@noop {} {\bibinfo {title} {varpro}},\ \bibinfo {howpublished} {\url{https://github.com/sxs-collaboration/varpro}} (\bibinfo {year} {2024}),\ \bibinfo {note} {accessed: 2025-05-19}\BibitemShut {NoStop}%
\bibitem [{\citenamefont {Boyle}\ \emph {et~al.}(2019)\citenamefont {Boyle} \emph {et~al.}}]{Boyle2019}%
  \BibitemOpen
  \bibfield  {author} {\bibinfo {author} {\bibfnamefont {M.}~\bibnamefont {Boyle}} \emph {et~al.},\ }\href {https://doi.org/10.1088/1361-6382/ab34e2} {\bibfield  {journal} {\bibinfo  {journal} {Classical Quantum Gravity}\ }\textbf {\bibinfo {volume} {36}},\ \bibinfo {pages} {195006} (\bibinfo {year} {2019})}\BibitemShut {NoStop}%
\bibitem [{\citenamefont {Scheel}\ \emph {et~al.}(2025)\citenamefont {Scheel} \emph {et~al.}}]{Scheel:2025jct}%
  \BibitemOpen
  \bibfield  {author} {\bibinfo {author} {\bibfnamefont {M.~A.}\ \bibnamefont {Scheel}} \emph {et~al.},\ }\href {https://doi.org/10.1088/1361-6382/adfd34} {\bibfield  {journal} {\bibinfo  {journal} {Classical Quantum Gravity}\ }\textbf {\bibinfo {volume} {42}},\ \bibinfo {pages} {195017} (\bibinfo {year} {2025})}\BibitemShut {NoStop}%
\bibitem [{\citenamefont {Gao}\ \emph {et~al.}(2025)\citenamefont {Gao} \emph {et~al.}}]{Gao2025}%
  \BibitemOpen
  \bibfield  {author} {\bibinfo {author} {\bibfnamefont {L.}~\bibnamefont {Gao}} \emph {et~al.},\ }\href {https://doi.org/10.1103/3jj6-jc8q} {\bibfield  {journal} {\bibinfo  {journal} {Phys. Rev. D}\ }\textbf {\bibinfo {volume} {112}},\ \bibinfo {pages} {024025} (\bibinfo {year} {2025})}\BibitemShut {NoStop}%
\bibitem [{\citenamefont {Bhagwat}\ \emph {et~al.}(2018)\citenamefont {Bhagwat}, \citenamefont {Okounkova}, \citenamefont {Ballmer}, \citenamefont {Brown}, \citenamefont {Giesler}, \citenamefont {Scheel},\ and\ \citenamefont {Teukolsky}}]{Bhagwat2017}%
  \BibitemOpen
  \bibfield  {author} {\bibinfo {author} {\bibfnamefont {S.}~\bibnamefont {Bhagwat}}, \bibinfo {author} {\bibfnamefont {M.}~\bibnamefont {Okounkova}}, \bibinfo {author} {\bibfnamefont {S.~W.}\ \bibnamefont {Ballmer}}, \bibinfo {author} {\bibfnamefont {D.~A.}\ \bibnamefont {Brown}}, \bibinfo {author} {\bibfnamefont {M.}~\bibnamefont {Giesler}}, \bibinfo {author} {\bibfnamefont {M.~A.}\ \bibnamefont {Scheel}},\ and\ \bibinfo {author} {\bibfnamefont {S.~A.}\ \bibnamefont {Teukolsky}},\ }\href {https://doi.org/10.1103/PhysRevD.97.104065} {\bibfield  {journal} {\bibinfo  {journal} {Phys. Rev. D}\ }\textbf {\bibinfo {volume} {97}},\ \bibinfo {pages} {104065} (\bibinfo {year} {2018})}\BibitemShut {NoStop}%
\bibitem [{\citenamefont {Barausse}\ \emph {et~al.}(2014)\citenamefont {Barausse}, \citenamefont {Cardoso},\ and\ \citenamefont {Pani}}]{Barausse2014}%
  \BibitemOpen
  \bibfield  {author} {\bibinfo {author} {\bibfnamefont {E.}~\bibnamefont {Barausse}}, \bibinfo {author} {\bibfnamefont {V.}~\bibnamefont {Cardoso}},\ and\ \bibinfo {author} {\bibfnamefont {P.}~\bibnamefont {Pani}},\ }\href {https://doi.org/10.1103/PhysRevD.89.104059} {\bibfield  {journal} {\bibinfo  {journal} {Phys. Rev. D}\ }\textbf {\bibinfo {volume} {89}},\ \bibinfo {pages} {104059} (\bibinfo {year} {2014})}\BibitemShut {NoStop}%
\bibitem [{\citenamefont {Nollert}(1996)}]{Nollert1996}%
  \BibitemOpen
  \bibfield  {author} {\bibinfo {author} {\bibfnamefont {H.-P.}\ \bibnamefont {Nollert}},\ }\href {https://doi.org/10.1103/PhysRevD.53.4397} {\bibfield  {journal} {\bibinfo  {journal} {Phys. Rev. D}\ }\textbf {\bibinfo {volume} {53}},\ \bibinfo {pages} {4397} (\bibinfo {year} {1996})}\BibitemShut {NoStop}%
\bibitem [{\citenamefont {Leung}\ \emph {et~al.}(1997)\citenamefont {Leung}, \citenamefont {Liu}, \citenamefont {Suen}, \citenamefont {Tam},\ and\ \citenamefont {Young}}]{Leung1997}%
  \BibitemOpen
  \bibfield  {author} {\bibinfo {author} {\bibfnamefont {P.~T.}\ \bibnamefont {Leung}}, \bibinfo {author} {\bibfnamefont {Y.~T.}\ \bibnamefont {Liu}}, \bibinfo {author} {\bibfnamefont {W.~M.}\ \bibnamefont {Suen}}, \bibinfo {author} {\bibfnamefont {C.~Y.}\ \bibnamefont {Tam}},\ and\ \bibinfo {author} {\bibfnamefont {K.}~\bibnamefont {Young}},\ }\href {https://doi.org/10.1103/PhysRevLett.78.2894} {\bibfield  {journal} {\bibinfo  {journal} {Phys. Rev. Lett.}\ }\textbf {\bibinfo {volume} {78}},\ \bibinfo {pages} {2894} (\bibinfo {year} {1997})}\BibitemShut {NoStop}%
\bibitem [{\citenamefont {Leung}\ \emph {et~al.}(1999)\citenamefont {Leung}, \citenamefont {Liu}, \citenamefont {Suen}, \citenamefont {Tam},\ and\ \citenamefont {Young}}]{Leung1999}%
  \BibitemOpen
  \bibfield  {author} {\bibinfo {author} {\bibfnamefont {P.~T.}\ \bibnamefont {Leung}}, \bibinfo {author} {\bibfnamefont {Y.~T.}\ \bibnamefont {Liu}}, \bibinfo {author} {\bibfnamefont {W.~M.}\ \bibnamefont {Suen}}, \bibinfo {author} {\bibfnamefont {C.~Y.}\ \bibnamefont {Tam}},\ and\ \bibinfo {author} {\bibfnamefont {K.}~\bibnamefont {Young}},\ }\href {https://doi.org/10.1103/PhysRevD.59.044034} {\bibfield  {journal} {\bibinfo  {journal} {Phys. Rev. D}\ }\textbf {\bibinfo {volume} {59}},\ \bibinfo {pages} {044034} (\bibinfo {year} {1999})}\BibitemShut {NoStop}%
\bibitem [{\citenamefont {Cheung}\ \emph {et~al.}(2022)\citenamefont {Cheung}, \citenamefont {Destounis}, \citenamefont {Macedo}, \citenamefont {Berti},\ and\ \citenamefont {Cardoso}}]{Cheung:2021bol}%
  \BibitemOpen
  \bibfield  {author} {\bibinfo {author} {\bibfnamefont {M.~H.-Y.}\ \bibnamefont {Cheung}}, \bibinfo {author} {\bibfnamefont {K.}~\bibnamefont {Destounis}}, \bibinfo {author} {\bibfnamefont {R.~P.}\ \bibnamefont {Macedo}}, \bibinfo {author} {\bibfnamefont {E.}~\bibnamefont {Berti}},\ and\ \bibinfo {author} {\bibfnamefont {V.}~\bibnamefont {Cardoso}},\ }\href {https://doi.org/10.1103/PhysRevLett.128.111103} {\bibfield  {journal} {\bibinfo  {journal} {Phys. Rev. Lett.}\ }\textbf {\bibinfo {volume} {128}},\ \bibinfo {pages} {111103} (\bibinfo {year} {2022})}\BibitemShut {NoStop}%
\bibitem [{\citenamefont {Cardoso}\ \emph {et~al.}(2022)\citenamefont {Cardoso}, \citenamefont {Destounis}, \citenamefont {Duque}, \citenamefont {Macedo},\ and\ \citenamefont {Maselli}}]{Cardoso2021}%
  \BibitemOpen
  \bibfield  {author} {\bibinfo {author} {\bibfnamefont {V.}~\bibnamefont {Cardoso}}, \bibinfo {author} {\bibfnamefont {K.}~\bibnamefont {Destounis}}, \bibinfo {author} {\bibfnamefont {F.}~\bibnamefont {Duque}}, \bibinfo {author} {\bibfnamefont {R.~P.}\ \bibnamefont {Macedo}},\ and\ \bibinfo {author} {\bibfnamefont {A.}~\bibnamefont {Maselli}},\ }\href {https://doi.org/10.1103/PhysRevD.105.L061501} {\bibfield  {journal} {\bibinfo  {journal} {Phys. Rev. D}\ }\textbf {\bibinfo {volume} {105}},\ \bibinfo {pages} {L061501} (\bibinfo {year} {2022})}\BibitemShut {NoStop}%
\bibitem [{\citenamefont {Capuano}\ \emph {et~al.}(2024)\citenamefont {Capuano}, \citenamefont {Santoni},\ and\ \citenamefont {Barausse}}]{Capuano:2024qhv}%
  \BibitemOpen
  \bibfield  {author} {\bibinfo {author} {\bibfnamefont {L.}~\bibnamefont {Capuano}}, \bibinfo {author} {\bibfnamefont {L.}~\bibnamefont {Santoni}},\ and\ \bibinfo {author} {\bibfnamefont {E.}~\bibnamefont {Barausse}},\ }\href {https://doi.org/10.1103/PhysRevD.110.084081} {\bibfield  {journal} {\bibinfo  {journal} {Phys. Rev. D}\ }\textbf {\bibinfo {volume} {110}},\ \bibinfo {pages} {084081} (\bibinfo {year} {2024})}\BibitemShut {NoStop}%
\bibitem [{\citenamefont {Pezzella}\ \emph {et~al.}(2025)\citenamefont {Pezzella}, \citenamefont {Destounis}, \citenamefont {Maselli},\ and\ \citenamefont {Cardoso}}]{Pezzella2024}%
  \BibitemOpen
  \bibfield  {author} {\bibinfo {author} {\bibfnamefont {L.}~\bibnamefont {Pezzella}}, \bibinfo {author} {\bibfnamefont {K.}~\bibnamefont {Destounis}}, \bibinfo {author} {\bibfnamefont {A.}~\bibnamefont {Maselli}},\ and\ \bibinfo {author} {\bibfnamefont {V.}~\bibnamefont {Cardoso}},\ }\href {https://doi.org/10.1103/PhysRevD.111.064026} {\bibfield  {journal} {\bibinfo  {journal} {Phys. Rev. D}\ }\textbf {\bibinfo {volume} {111}},\ \bibinfo {pages} {064026} (\bibinfo {year} {2025})}\BibitemShut {NoStop}%
\bibitem [{\citenamefont {Laeuger}\ \emph {et~al.}(2025)\citenamefont {Laeuger}, \citenamefont {Weller}, \citenamefont {Li},\ and\ \citenamefont {Chen}}]{Laeuger2025}%
  \BibitemOpen
  \bibfield  {author} {\bibinfo {author} {\bibfnamefont {A.}~\bibnamefont {Laeuger}}, \bibinfo {author} {\bibfnamefont {C.}~\bibnamefont {Weller}}, \bibinfo {author} {\bibfnamefont {D.}~\bibnamefont {Li}},\ and\ \bibinfo {author} {\bibfnamefont {Y.}~\bibnamefont {Chen}},\ }\href {https://doi.org/10.1103/gkj4-m1c1} {\bibfield  {journal} {\bibinfo  {journal} {Phys. Rev. D}\ }\textbf {\bibinfo {volume} {112}},\ \bibinfo {pages} {084042} (\bibinfo {year} {2025})}\BibitemShut {NoStop}%
\bibitem [{Dha()}]{Dhani:2025xno}%
  \BibitemOpen
  \href@noop {} {}\bibinfo {note} {A. Dhani, A. Camilletti, A. L. De Santis, A. Cozzumbo, D. Radice, D. Logoteta, A. Perego, J. Harms, and M. Branchesi, \href{https://arxiv.org/abs/2507.14071}{arXiv:2507.14071}.}\BibitemShut {Stop}%
\bibitem [{\citenamefont {Dhani}\ \emph {et~al.}()\citenamefont {Dhani}, \citenamefont {Camilletti} \emph {et~al.}}]{Dhani:2025axt}%
  \BibitemOpen
  \bibfield  {author} {\bibinfo {author} {\bibfnamefont {A.}~\bibnamefont {Dhani}}, \bibinfo {author} {\bibfnamefont {A.}~\bibnamefont {Camilletti}}, \emph {et~al.},\ }\href@noop {} {\ }\Eprint {https://arxiv.org/abs/2507.19431} {arXiv:2507.19431} \BibitemShut {NoStop}%
\bibitem [{\citenamefont {Bandyopadhyay}\ \emph {et~al.}(2024)\citenamefont {Bandyopadhyay}, \citenamefont {Radice}, \citenamefont {Prakash}, \citenamefont {Dhani}, \citenamefont {Logoteta}, \citenamefont {Perego},\ and\ \citenamefont {Kashyap}}]{Bandyopadhyay:2023ohl}%
  \BibitemOpen
  \bibfield  {author} {\bibinfo {author} {\bibfnamefont {H.}~\bibnamefont {Bandyopadhyay}}, \bibinfo {author} {\bibfnamefont {D.}~\bibnamefont {Radice}}, \bibinfo {author} {\bibfnamefont {A.}~\bibnamefont {Prakash}}, \bibinfo {author} {\bibfnamefont {A.}~\bibnamefont {Dhani}}, \bibinfo {author} {\bibfnamefont {D.}~\bibnamefont {Logoteta}}, \bibinfo {author} {\bibfnamefont {A.}~\bibnamefont {Perego}},\ and\ \bibinfo {author} {\bibfnamefont {R.}~\bibnamefont {Kashyap}},\ }\href {https://doi.org/10.1088/1361-6382/ad56ed} {\bibfield  {journal} {\bibinfo  {journal} {Classical Quantum Gravity}\ }\textbf {\bibinfo {volume} {41}},\ \bibinfo {pages} {145006} (\bibinfo {year} {2024})}\BibitemShut {NoStop}%
\bibitem [{\citenamefont {Nee}\ \emph {et~al.}(2023)\citenamefont {Nee}, \citenamefont {V{\"o}lkel},\ and\ \citenamefont {Pfeiffer}}]{Nee:2023osy}%
  \BibitemOpen
  \bibfield  {author} {\bibinfo {author} {\bibfnamefont {P.~J.}\ \bibnamefont {Nee}}, \bibinfo {author} {\bibfnamefont {S.~H.}\ \bibnamefont {V{\"o}lkel}},\ and\ \bibinfo {author} {\bibfnamefont {H.~P.}\ \bibnamefont {Pfeiffer}},\ }\href {https://doi.org/10.1103/PhysRevD.108.044032} {\bibfield  {journal} {\bibinfo  {journal} {Phys. Rev. D}\ }\textbf {\bibinfo {volume} {108}},\ \bibinfo {pages} {044032} (\bibinfo {year} {2023})}\BibitemShut {NoStop}%
\bibitem [{\citenamefont {Br{\"u}gmann}\ \emph {et~al.}(2008)\citenamefont {Br{\"u}gmann}, \citenamefont {Gonzalez}, \citenamefont {Hannam}, \citenamefont {Husa}, \citenamefont {Sperhake},\ and\ \citenamefont {Tichy}}]{Bruegmann:2006ulg}%
  \BibitemOpen
  \bibfield  {author} {\bibinfo {author} {\bibfnamefont {B.}~\bibnamefont {Br{\"u}gmann}}, \bibinfo {author} {\bibfnamefont {J.~A.}\ \bibnamefont {Gonzalez}}, \bibinfo {author} {\bibfnamefont {M.}~\bibnamefont {Hannam}}, \bibinfo {author} {\bibfnamefont {S.}~\bibnamefont {Husa}}, \bibinfo {author} {\bibfnamefont {U.}~\bibnamefont {Sperhake}},\ and\ \bibinfo {author} {\bibfnamefont {W.}~\bibnamefont {Tichy}},\ }\href {https://doi.org/10.1103/PhysRevD.77.024027} {\bibfield  {journal} {\bibinfo  {journal} {Phys. Rev. D}\ }\textbf {\bibinfo {volume} {77}},\ \bibinfo {pages} {024027} (\bibinfo {year} {2008})}\BibitemShut {NoStop}%
\bibitem [{\citenamefont {Thierfelder}\ \emph {et~al.}(2011)\citenamefont {Thierfelder}, \citenamefont {Bernuzzi},\ and\ \citenamefont {Br{\"u}gmann}}]{Thierfelder:2011yi}%
  \BibitemOpen
  \bibfield  {author} {\bibinfo {author} {\bibfnamefont {M.}~\bibnamefont {Thierfelder}}, \bibinfo {author} {\bibfnamefont {S.}~\bibnamefont {Bernuzzi}},\ and\ \bibinfo {author} {\bibfnamefont {B.}~\bibnamefont {Br{\"u}gmann}},\ }\href {https://doi.org/10.1103/PhysRevD.84.044012} {\bibfield  {journal} {\bibinfo  {journal} {Phys. Rev. D}\ }\textbf {\bibinfo {volume} {84}},\ \bibinfo {pages} {044012} (\bibinfo {year} {2011})}\BibitemShut {NoStop}%
\bibitem [{\citenamefont {Pretorius}(2005)}]{Pretorius:2005gq}%
  \BibitemOpen
  \bibfield  {author} {\bibinfo {author} {\bibfnamefont {F.}~\bibnamefont {Pretorius}},\ }\href {https://doi.org/10.1103/PhysRevLett.95.121101} {\bibfield  {journal} {\bibinfo  {journal} {Phys. Rev. Lett.}\ }\textbf {\bibinfo {volume} {95}},\ \bibinfo {pages} {121101} (\bibinfo {year} {2005})}\BibitemShut {NoStop}%
\bibitem [{\citenamefont {Baumgarte}\ and\ \citenamefont {Shapiro}(2010)}]{baumgarte2010numerical}%
  \BibitemOpen
  \bibfield  {author} {\bibinfo {author} {\bibfnamefont {T.~W.}\ \bibnamefont {Baumgarte}}\ and\ \bibinfo {author} {\bibfnamefont {S.~L.}\ \bibnamefont {Shapiro}},\ }\href@noop {} {\emph {\bibinfo {title} {Numerical Relativity: Solving Einstein's Equations on the Computer}}}\ (\bibinfo  {publisher} {Cambridge University Press, Cambridge, England},\ \bibinfo {year} {2010})\BibitemShut {NoStop}%
\bibitem [{\citenamefont {Rezzolla}\ and\ \citenamefont {Zanotti}(2013)}]{rezzolla2013relativistic}%
  \BibitemOpen
  \bibfield  {author} {\bibinfo {author} {\bibfnamefont {L.}~\bibnamefont {Rezzolla}}\ and\ \bibinfo {author} {\bibfnamefont {O.}~\bibnamefont {Zanotti}},\ }\href@noop {} {\emph {\bibinfo {title} {Relativistic Hydrodynamics}}}\ (\bibinfo  {publisher} {Oxford University Press, New York},\ \bibinfo {year} {2013})\BibitemShut {NoStop}%
\bibitem [{\citenamefont {{Shibata}}(2016)}]{ShibateNRbook}%
  \BibitemOpen
  \bibfield  {author} {\bibinfo {author} {\bibfnamefont {M.}~\bibnamefont {{Shibata}}},\ }\href {https://doi.org/10.1142/9692} {\emph {\bibinfo {title} {{Numerical Relativity}}}}\ (\bibinfo  {publisher} {World Scientific, Singapore},\ \bibinfo {year} {2016})\BibitemShut {NoStop}%
\bibitem [{\citenamefont {Palenzuela}(2020)}]{Palenzuela2020}%
  \BibitemOpen
  \bibfield  {author} {\bibinfo {author} {\bibfnamefont {C.}~\bibnamefont {Palenzuela}},\ }\href {https://doi.org/10.3389/fspas.2020.00058} {\bibfield  {journal} {\bibinfo  {journal} {Front. Astron. Space Sci.}\ }\textbf {\bibinfo {volume} {7}},\ \bibinfo {pages} {58} (\bibinfo {year} {2020})}\BibitemShut {NoStop}%
\bibitem [{\citenamefont {Galeazzi}\ \emph {et~al.}(2012)\citenamefont {Galeazzi}, \citenamefont {Yoshida},\ and\ \citenamefont {Eriguchi}}]{Galeazzi2011}%
  \BibitemOpen
  \bibfield  {author} {\bibinfo {author} {\bibfnamefont {F.}~\bibnamefont {Galeazzi}}, \bibinfo {author} {\bibfnamefont {S.}~\bibnamefont {Yoshida}},\ and\ \bibinfo {author} {\bibfnamefont {Y.}~\bibnamefont {Eriguchi}},\ }\href {https://doi.org/10.1051/0004-6361/201016316} {\bibfield  {journal} {\bibinfo  {journal} {Astron. Astrophys.}\ }\textbf {\bibinfo {volume} {541}},\ \bibinfo {pages} {A156} (\bibinfo {year} {2012})}\BibitemShut {NoStop}%
\bibitem [{\citenamefont {Ury\ifmmode~\bar{u}\else \={u}\fi{}}\ \emph {et~al.}(2016)\citenamefont {Ury\ifmmode~\bar{u}\else \={u}\fi{}}, \citenamefont {Tsokaros}, \citenamefont {Galeazzi}, \citenamefont {Hotta}, \citenamefont {Sugimura}, \citenamefont {Taniguchi},\ and\ \citenamefont {Yoshida}}]{Uryu2016}%
  \BibitemOpen
  \bibfield  {author} {\bibinfo {author} {\bibfnamefont {K.}~\bibnamefont {Ury\ifmmode~\bar{u}\else \={u}\fi{}}}, \bibinfo {author} {\bibfnamefont {A.}~\bibnamefont {Tsokaros}}, \bibinfo {author} {\bibfnamefont {F.}~\bibnamefont {Galeazzi}}, \bibinfo {author} {\bibfnamefont {H.}~\bibnamefont {Hotta}}, \bibinfo {author} {\bibfnamefont {M.}~\bibnamefont {Sugimura}}, \bibinfo {author} {\bibfnamefont {K.}~\bibnamefont {Taniguchi}},\ and\ \bibinfo {author} {\bibfnamefont {S.}~\bibnamefont {Yoshida}},\ }\href {https://doi.org/10.1103/PhysRevD.93.044056} {\bibfield  {journal} {\bibinfo  {journal} {Phys. Rev. D}\ }\textbf {\bibinfo {volume} {93}},\ \bibinfo {pages} {044056} (\bibinfo {year} {2016})}\BibitemShut {NoStop}%
\bibitem [{\citenamefont {Ury\ifmmode~\bar{u}\else \={u}\fi{}}\ \emph {et~al.}(2017)\citenamefont {Ury\ifmmode~\bar{u}\else \={u}\fi{}}, \citenamefont {Tsokaros}, \citenamefont {Baiotti}, \citenamefont {Galeazzi}, \citenamefont {Taniguchi},\ and\ \citenamefont {Yoshida}}]{Uryu2017}%
  \BibitemOpen
  \bibfield  {author} {\bibinfo {author} {\bibfnamefont {K.}~\bibnamefont {Ury\ifmmode~\bar{u}\else \={u}\fi{}}}, \bibinfo {author} {\bibfnamefont {A.}~\bibnamefont {Tsokaros}}, \bibinfo {author} {\bibfnamefont {L.}~\bibnamefont {Baiotti}}, \bibinfo {author} {\bibfnamefont {F.}~\bibnamefont {Galeazzi}}, \bibinfo {author} {\bibfnamefont {K.}~\bibnamefont {Taniguchi}},\ and\ \bibinfo {author} {\bibfnamefont {S.}~\bibnamefont {Yoshida}},\ }\href {https://doi.org/10.1103/PhysRevD.96.103011} {\bibfield  {journal} {\bibinfo  {journal} {Phys. Rev. D}\ }\textbf {\bibinfo {volume} {96}},\ \bibinfo {pages} {103011} (\bibinfo {year} {2017})}\BibitemShut {NoStop}%
\bibitem [{\citenamefont {Bauswein}\ and\ \citenamefont {Stergioulas}(2017)}]{Bauswein2017}%
  \BibitemOpen
  \bibfield  {author} {\bibinfo {author} {\bibfnamefont {A.}~\bibnamefont {Bauswein}}\ and\ \bibinfo {author} {\bibfnamefont {N.}~\bibnamefont {Stergioulas}},\ }\href {https://doi.org/10.1093/mnras/stx1983} {\bibfield  {journal} {\bibinfo  {journal} {Mon. Not. R. Astron. Soc.}\ }\textbf {\bibinfo {volume} {471}},\ \bibinfo {pages} {4956} (\bibinfo {year} {2017})}\BibitemShut {NoStop}%
\bibitem [{\citenamefont {{Komatsu}}\ \emph {et~al.}(1989)\citenamefont {{Komatsu}}, \citenamefont {{Eriguchi}},\ and\ \citenamefont {{Hachisu}}}]{Komatsu1989}%
  \BibitemOpen
  \bibfield  {author} {\bibinfo {author} {\bibfnamefont {H.}~\bibnamefont {{Komatsu}}}, \bibinfo {author} {\bibfnamefont {Y.}~\bibnamefont {{Eriguchi}}},\ and\ \bibinfo {author} {\bibfnamefont {I.}~\bibnamefont {{Hachisu}}},\ }\href {https://doi.org/10.1093/mnras/237.2.355} {\bibfield  {journal} {\bibinfo  {journal} {Mon. Not. R. Astron. Soc.}\ }\textbf {\bibinfo {volume} {237}},\ \bibinfo {pages} {355} (\bibinfo {year} {1989})}\BibitemShut {NoStop}%
\bibitem [{\citenamefont {{Stergioulas}}\ and\ \citenamefont {{Friedman}}(1995)}]{Stergioulas1995}%
  \BibitemOpen
  \bibfield  {author} {\bibinfo {author} {\bibfnamefont {N.}~\bibnamefont {{Stergioulas}}}\ and\ \bibinfo {author} {\bibfnamefont {J.~L.}\ \bibnamefont {{Friedman}}},\ }\href {https://doi.org/10.1086/175605} {\bibfield  {journal} {\bibinfo  {journal} {\apj}\ }\textbf {\bibinfo {volume} {444}},\ \bibinfo {pages} {306} (\bibinfo {year} {1995})}\BibitemShut {NoStop}%
\bibitem [{\citenamefont {{Stergioulas}}(2003)}]{Stergioulas2003}%
  \BibitemOpen
  \bibfield  {author} {\bibinfo {author} {\bibfnamefont {N.}~\bibnamefont {{Stergioulas}}},\ }\href {https://doi.org/10.12942/lrr-2003-3} {\bibfield  {journal} {\bibinfo  {journal} {Living Rev. in Relativity}\ }\textbf {\bibinfo {volume} {6}},\ \bibinfo {eid} {3} (\bibinfo {year} {2003})}\BibitemShut {NoStop}%
\bibitem [{\citenamefont {{Cook}}\ \emph {et~al.}(1992)\citenamefont {{Cook}}, \citenamefont {{Shapiro}},\ and\ \citenamefont {{Teukolsky}}}]{Cook1992}%
  \BibitemOpen
  \bibfield  {author} {\bibinfo {author} {\bibfnamefont {G.~B.}\ \bibnamefont {{Cook}}}, \bibinfo {author} {\bibfnamefont {S.~L.}\ \bibnamefont {{Shapiro}}},\ and\ \bibinfo {author} {\bibfnamefont {S.~A.}\ \bibnamefont {{Teukolsky}}},\ }\href {https://doi.org/10.1086/171849} {\bibfield  {journal} {\bibinfo  {journal} {\apj}\ }\textbf {\bibinfo {volume} {398}},\ \bibinfo {pages} {203} (\bibinfo {year} {1992})}\BibitemShut {NoStop}%
\bibitem [{\citenamefont {{Cook}}\ \emph {et~al.}(1994)\citenamefont {{Cook}}, \citenamefont {{Shapiro}},\ and\ \citenamefont {{Teukolsky}}}]{Cook1994}%
  \BibitemOpen
  \bibfield  {author} {\bibinfo {author} {\bibfnamefont {G.~B.}\ \bibnamefont {{Cook}}}, \bibinfo {author} {\bibfnamefont {S.~L.}\ \bibnamefont {{Shapiro}}},\ and\ \bibinfo {author} {\bibfnamefont {S.~A.}\ \bibnamefont {{Teukolsky}}},\ }\href {https://doi.org/10.1086/173934} {\bibfield  {journal} {\bibinfo  {journal} {\apj}\ }\textbf {\bibinfo {volume} {424}},\ \bibinfo {pages} {823} (\bibinfo {year} {1994})}\BibitemShut {NoStop}%
\bibitem [{\citenamefont {Steppohn}(2025{\natexlab{a}})}]{osteppohn2025repo}%
  \BibitemOpen
  \bibfield  {author} {\bibinfo {author} {\bibfnamefont {O.}~\bibnamefont {Steppohn}},\ }\href@noop {} {\bibinfo {title} {{dRNS}}},\ \bibinfo {howpublished} {\url{https://github.com/spacetimecurv/dRNS}} (\bibinfo {year} {2025}{\natexlab{a}}),\ \bibinfo {note} {gitHub repository}\BibitemShut {NoStop}%
\bibitem [{\citenamefont {Tichy}(2009)}]{Tichy:2009yr}%
  \BibitemOpen
  \bibfield  {author} {\bibinfo {author} {\bibfnamefont {W.}~\bibnamefont {Tichy}},\ }\href {https://doi.org/10.1088/0264-9381/26/17/175018} {\bibfield  {journal} {\bibinfo  {journal} {Classical Quantum Gravity}\ }\textbf {\bibinfo {volume} {26}},\ \bibinfo {pages} {175018} (\bibinfo {year} {2009})}\BibitemShut {NoStop}%
\bibitem [{\citenamefont {Tichy}(2012)}]{Tichy:2012rp}%
  \BibitemOpen
  \bibfield  {author} {\bibinfo {author} {\bibfnamefont {W.}~\bibnamefont {Tichy}},\ }\href {https://doi.org/10.1103/PhysRevD.86.064024} {\bibfield  {journal} {\bibinfo  {journal} {Phys. Rev. D}\ }\textbf {\bibinfo {volume} {86}},\ \bibinfo {pages} {064024} (\bibinfo {year} {2012})}\BibitemShut {NoStop}%
\bibitem [{\citenamefont {Dietrich}\ \emph {et~al.}(2015{\natexlab{a}})\citenamefont {Dietrich}, \citenamefont {Moldenhauer}, \citenamefont {Johnson-McDaniel}, \citenamefont {Bernuzzi}, \citenamefont {Markakis}, \citenamefont {Br{\"u}gmann},\ and\ \citenamefont {Tichy}}]{Dietrich:2015pxa}%
  \BibitemOpen
  \bibfield  {author} {\bibinfo {author} {\bibfnamefont {T.}~\bibnamefont {Dietrich}}, \bibinfo {author} {\bibfnamefont {N.}~\bibnamefont {Moldenhauer}}, \bibinfo {author} {\bibfnamefont {N.~K.}\ \bibnamefont {Johnson-McDaniel}}, \bibinfo {author} {\bibfnamefont {S.}~\bibnamefont {Bernuzzi}}, \bibinfo {author} {\bibfnamefont {C.~M.}\ \bibnamefont {Markakis}}, \bibinfo {author} {\bibfnamefont {B.}~\bibnamefont {Br{\"u}gmann}},\ and\ \bibinfo {author} {\bibfnamefont {W.}~\bibnamefont {Tichy}},\ }\href {https://doi.org/10.1103/PhysRevD.92.124007} {\bibfield  {journal} {\bibinfo  {journal} {Phys. Rev. D}\ }\textbf {\bibinfo {volume} {92}},\ \bibinfo {pages} {124007} (\bibinfo {year} {2015}{\natexlab{a}})}\BibitemShut {NoStop}%
\bibitem [{\citenamefont {Tichy}\ \emph {et~al.}(2019)\citenamefont {Tichy}, \citenamefont {Rashti}, \citenamefont {Dietrich}, \citenamefont {Dudi},\ and\ \citenamefont {Br{\"u}gmann}}]{Tichy:2019ouu}%
  \BibitemOpen
  \bibfield  {author} {\bibinfo {author} {\bibfnamefont {W.}~\bibnamefont {Tichy}}, \bibinfo {author} {\bibfnamefont {A.}~\bibnamefont {Rashti}}, \bibinfo {author} {\bibfnamefont {T.}~\bibnamefont {Dietrich}}, \bibinfo {author} {\bibfnamefont {R.}~\bibnamefont {Dudi}},\ and\ \bibinfo {author} {\bibfnamefont {B.}~\bibnamefont {Br{\"u}gmann}},\ }\href {https://doi.org/10.1103/PhysRevD.100.124046} {\bibfield  {journal} {\bibinfo  {journal} {Phys. Rev. D}\ }\textbf {\bibinfo {volume} {100}},\ \bibinfo {pages} {124046} (\bibinfo {year} {2019})}\BibitemShut {NoStop}%
\bibitem [{\citenamefont {York}(1999)}]{York:1998hy}%
  \BibitemOpen
  \bibfield  {author} {\bibinfo {author} {\bibfnamefont {J.~W.}\ \bibnamefont {York}, \bibfnamefont {Jr.}},\ }\href {https://doi.org/10.1103/PhysRevLett.82.1350} {\bibfield  {journal} {\bibinfo  {journal} {Phys. Rev. Lett.}\ }\textbf {\bibinfo {volume} {82}},\ \bibinfo {pages} {1350} (\bibinfo {year} {1999})}\BibitemShut {NoStop}%
\bibitem [{\citenamefont {Pfeiffer}\ and\ \citenamefont {York}(2003)}]{Pfeiffer:2002iy}%
  \BibitemOpen
  \bibfield  {author} {\bibinfo {author} {\bibfnamefont {H.~P.}\ \bibnamefont {Pfeiffer}}\ and\ \bibinfo {author} {\bibfnamefont {J.~W.}\ \bibnamefont {York}, \bibfnamefont {Jr.}},\ }\href {https://doi.org/10.1103/PhysRevD.67.044022} {\bibfield  {journal} {\bibinfo  {journal} {Phys. Rev. D}\ }\textbf {\bibinfo {volume} {67}},\ \bibinfo {pages} {044022} (\bibinfo {year} {2003})}\BibitemShut {NoStop}%
\bibitem [{\citenamefont {Read}\ \emph {et~al.}(2009)\citenamefont {Read}, \citenamefont {Lackey}, \citenamefont {Owen},\ and\ \citenamefont {Friedman}}]{Read:2008iy}%
  \BibitemOpen
  \bibfield  {author} {\bibinfo {author} {\bibfnamefont {J.~S.}\ \bibnamefont {Read}}, \bibinfo {author} {\bibfnamefont {B.~D.}\ \bibnamefont {Lackey}}, \bibinfo {author} {\bibfnamefont {B.~J.}\ \bibnamefont {Owen}},\ and\ \bibinfo {author} {\bibfnamefont {J.~L.}\ \bibnamefont {Friedman}},\ }\href {https://doi.org/10.1103/PhysRevD.79.124032} {\bibfield  {journal} {\bibinfo  {journal} {Phys. Rev. D}\ }\textbf {\bibinfo {volume} {79}},\ \bibinfo {pages} {124032} (\bibinfo {year} {2009})}\BibitemShut {NoStop}%
\bibitem [{\citenamefont {Douchin}\ and\ \citenamefont {Haensel}(2001)}]{Douchin:2001sv}%
  \BibitemOpen
  \bibfield  {author} {\bibinfo {author} {\bibfnamefont {F.}~\bibnamefont {Douchin}}\ and\ \bibinfo {author} {\bibfnamefont {P.}~\bibnamefont {Haensel}},\ }\href {https://doi.org/10.1051/0004-6361:20011402} {\bibfield  {journal} {\bibinfo  {journal} {Astron. Astrophys.}\ }\textbf {\bibinfo {volume} {380}},\ \bibinfo {pages} {151} (\bibinfo {year} {2001})}\BibitemShut {NoStop}%
\bibitem [{\citenamefont {Bauswein}\ \emph {et~al.}(2010)\citenamefont {Bauswein}, \citenamefont {Janka},\ and\ \citenamefont {Oechslin}}]{Bauswein:2010dn}%
  \BibitemOpen
  \bibfield  {author} {\bibinfo {author} {\bibfnamefont {A.}~\bibnamefont {Bauswein}}, \bibinfo {author} {\bibfnamefont {H.~T.}\ \bibnamefont {Janka}},\ and\ \bibinfo {author} {\bibfnamefont {R.}~\bibnamefont {Oechslin}},\ }\href {https://doi.org/10.1103/PhysRevD.82.084043} {\bibfield  {journal} {\bibinfo  {journal} {Phys. Rev. D}\ }\textbf {\bibinfo {volume} {82}},\ \bibinfo {pages} {084043} (\bibinfo {year} {2010})}\BibitemShut {NoStop}%
\bibitem [{\citenamefont {Dietrich}\ \emph {et~al.}(2015{\natexlab{b}})\citenamefont {Dietrich}, \citenamefont {Bernuzzi}, \citenamefont {Ujevic},\ and\ \citenamefont {Br{\"u}gmann}}]{Dietrich:2015iva}%
  \BibitemOpen
  \bibfield  {author} {\bibinfo {author} {\bibfnamefont {T.}~\bibnamefont {Dietrich}}, \bibinfo {author} {\bibfnamefont {S.}~\bibnamefont {Bernuzzi}}, \bibinfo {author} {\bibfnamefont {M.}~\bibnamefont {Ujevic}},\ and\ \bibinfo {author} {\bibfnamefont {B.}~\bibnamefont {Br{\"u}gmann}},\ }\href {https://doi.org/10.1103/PhysRevD.91.124041} {\bibfield  {journal} {\bibinfo  {journal} {Phys. Rev. D}\ }\textbf {\bibinfo {volume} {91}},\ \bibinfo {pages} {124041} (\bibinfo {year} {2015}{\natexlab{b}})}\BibitemShut {NoStop}%
\bibitem [{\citenamefont {Bernuzzi}\ and\ \citenamefont {Dietrich}(2016)}]{Bernuzzi:2016pie}%
  \BibitemOpen
  \bibfield  {author} {\bibinfo {author} {\bibfnamefont {S.}~\bibnamefont {Bernuzzi}}\ and\ \bibinfo {author} {\bibfnamefont {T.}~\bibnamefont {Dietrich}},\ }\href {https://doi.org/10.1103/PhysRevD.94.064062} {\bibfield  {journal} {\bibinfo  {journal} {Phys. Rev. D}\ }\textbf {\bibinfo {volume} {94}},\ \bibinfo {pages} {064062} (\bibinfo {year} {2016})}\BibitemShut {NoStop}%
\bibitem [{\citenamefont {Bernuzzi}\ and\ \citenamefont {Hilditch}(2010)}]{Bernuzzi:2009ex}%
  \BibitemOpen
  \bibfield  {author} {\bibinfo {author} {\bibfnamefont {S.}~\bibnamefont {Bernuzzi}}\ and\ \bibinfo {author} {\bibfnamefont {D.}~\bibnamefont {Hilditch}},\ }\href {https://doi.org/10.1103/PhysRevD.81.084003} {\bibfield  {journal} {\bibinfo  {journal} {Phys. Rev. D}\ }\textbf {\bibinfo {volume} {81}},\ \bibinfo {pages} {084003} (\bibinfo {year} {2010})}\BibitemShut {NoStop}%
\bibitem [{\citenamefont {Hilditch}\ \emph {et~al.}(2013)\citenamefont {Hilditch}, \citenamefont {Bernuzzi}, \citenamefont {Thierfelder}, \citenamefont {Cao}, \citenamefont {Tichy},\ and\ \citenamefont {Br{\"u}gmann}}]{Hilditch:2012fp}%
  \BibitemOpen
  \bibfield  {author} {\bibinfo {author} {\bibfnamefont {D.}~\bibnamefont {Hilditch}}, \bibinfo {author} {\bibfnamefont {S.}~\bibnamefont {Bernuzzi}}, \bibinfo {author} {\bibfnamefont {M.}~\bibnamefont {Thierfelder}}, \bibinfo {author} {\bibfnamefont {Z.}~\bibnamefont {Cao}}, \bibinfo {author} {\bibfnamefont {W.}~\bibnamefont {Tichy}},\ and\ \bibinfo {author} {\bibfnamefont {B.}~\bibnamefont {Br{\"u}gmann}},\ }\href {https://doi.org/10.1103/PhysRevD.88.084057} {\bibfield  {journal} {\bibinfo  {journal} {Phys. Rev. D}\ }\textbf {\bibinfo {volume} {88}},\ \bibinfo {pages} {084057} (\bibinfo {year} {2013})}\BibitemShut {NoStop}%
\bibitem [{\citenamefont {Bona}\ \emph {et~al.}(1995)\citenamefont {Bona}, \citenamefont {Masso}, \citenamefont {Seidel},\ and\ \citenamefont {Stela}}]{Bona:1994dr}%
  \BibitemOpen
  \bibfield  {author} {\bibinfo {author} {\bibfnamefont {C.}~\bibnamefont {Bona}}, \bibinfo {author} {\bibfnamefont {J.}~\bibnamefont {Masso}}, \bibinfo {author} {\bibfnamefont {E.}~\bibnamefont {Seidel}},\ and\ \bibinfo {author} {\bibfnamefont {J.}~\bibnamefont {Stela}},\ }\href {https://doi.org/10.1103/PhysRevLett.75.600} {\bibfield  {journal} {\bibinfo  {journal} {Phys. Rev. Lett.}\ }\textbf {\bibinfo {volume} {75}},\ \bibinfo {pages} {600} (\bibinfo {year} {1995})}\BibitemShut {NoStop}%
\bibitem [{\citenamefont {Alcubierre}\ \emph {et~al.}(2003)\citenamefont {Alcubierre}, \citenamefont {Br{\"u}gmann}, \citenamefont {Diener}, \citenamefont {Koppitz}, \citenamefont {Pollney}, \citenamefont {Seidel},\ and\ \citenamefont {Takahashi}}]{Alcubierre:2002kk}%
  \BibitemOpen
  \bibfield  {author} {\bibinfo {author} {\bibfnamefont {M.}~\bibnamefont {Alcubierre}}, \bibinfo {author} {\bibfnamefont {B.}~\bibnamefont {Br{\"u}gmann}}, \bibinfo {author} {\bibfnamefont {P.}~\bibnamefont {Diener}}, \bibinfo {author} {\bibfnamefont {M.}~\bibnamefont {Koppitz}}, \bibinfo {author} {\bibfnamefont {D.}~\bibnamefont {Pollney}}, \bibinfo {author} {\bibfnamefont {E.}~\bibnamefont {Seidel}},\ and\ \bibinfo {author} {\bibfnamefont {R.}~\bibnamefont {Takahashi}},\ }\href {https://doi.org/10.1103/PhysRevD.67.084023} {\bibfield  {journal} {\bibinfo  {journal} {Phys. Rev. D}\ }\textbf {\bibinfo {volume} {67}},\ \bibinfo {pages} {084023} (\bibinfo {year} {2003})}\BibitemShut {NoStop}%
\bibitem [{\citenamefont {van Meter}\ \emph {et~al.}(2006)\citenamefont {van Meter}, \citenamefont {Baker}, \citenamefont {Koppitz},\ and\ \citenamefont {Choi}}]{vanMeter:2006vi}%
  \BibitemOpen
  \bibfield  {author} {\bibinfo {author} {\bibfnamefont {J.}~\bibnamefont {van Meter}}, \bibinfo {author} {\bibfnamefont {J.~G.}\ \bibnamefont {Baker}}, \bibinfo {author} {\bibfnamefont {M.}~\bibnamefont {Koppitz}},\ and\ \bibinfo {author} {\bibfnamefont {D.-I.}\ \bibnamefont {Choi}},\ }\href {https://doi.org/10.1103/PhysRevD.73.124011} {\bibfield  {journal} {\bibinfo  {journal} {Phys. Rev. D}\ }\textbf {\bibinfo {volume} {73}},\ \bibinfo {pages} {124011} (\bibinfo {year} {2006})}\BibitemShut {NoStop}%
\bibitem [{\citenamefont {Marti}\ \emph {et~al.}(1991)\citenamefont {Marti}, \citenamefont {Ibanez},\ and\ \citenamefont {Miralles}}]{Marti:1991wi}%
  \BibitemOpen
  \bibfield  {author} {\bibinfo {author} {\bibfnamefont {J.~M.}\ \bibnamefont {Marti}}, \bibinfo {author} {\bibfnamefont {J.~M.}\ \bibnamefont {Ibanez}},\ and\ \bibinfo {author} {\bibfnamefont {J.~A.}\ \bibnamefont {Miralles}},\ }\href {https://doi.org/10.1103/PhysRevD.43.3794} {\bibfield  {journal} {\bibinfo  {journal} {Phys. Rev. D}\ }\textbf {\bibinfo {volume} {43}},\ \bibinfo {pages} {3794} (\bibinfo {year} {1991})}\BibitemShut {NoStop}%
\bibitem [{\citenamefont {Banyuls}\ \emph {et~al.}(1997)\citenamefont {Banyuls}, \citenamefont {Font}, \citenamefont {Ibáñez}, \citenamefont {Martí},\ and\ \citenamefont {Miralles}}]{Banyuls:1997zz}%
  \BibitemOpen
  \bibfield  {author} {\bibinfo {author} {\bibfnamefont {F.}~\bibnamefont {Banyuls}}, \bibinfo {author} {\bibfnamefont {J.~A.}\ \bibnamefont {Font}}, \bibinfo {author} {\bibfnamefont {J.~M.}\ \bibnamefont {Ibáñez}}, \bibinfo {author} {\bibfnamefont {J.~M.}\ \bibnamefont {Martí}},\ and\ \bibinfo {author} {\bibfnamefont {J.~A.}\ \bibnamefont {Miralles}},\ }\href {https://doi.org/10.1086/303604} {\bibfield  {journal} {\bibinfo  {journal} {Astrophys. J.}\ }\textbf {\bibinfo {volume} {476}},\ \bibinfo {pages} {221} (\bibinfo {year} {1997})}\BibitemShut {NoStop}%
\bibitem [{\citenamefont {Anton}\ \emph {et~al.}(2006)\citenamefont {Anton}, \citenamefont {Zanotti}, \citenamefont {Miralles}, \citenamefont {Marti}, \citenamefont {Ibanez}, \citenamefont {Font},\ and\ \citenamefont {Pons}}]{Anton:2005gi}%
  \BibitemOpen
  \bibfield  {author} {\bibinfo {author} {\bibfnamefont {L.}~\bibnamefont {Anton}}, \bibinfo {author} {\bibfnamefont {O.}~\bibnamefont {Zanotti}}, \bibinfo {author} {\bibfnamefont {J.~A.}\ \bibnamefont {Miralles}}, \bibinfo {author} {\bibfnamefont {J.~M.}\ \bibnamefont {Marti}}, \bibinfo {author} {\bibfnamefont {J.~M.}\ \bibnamefont {Ibanez}}, \bibinfo {author} {\bibfnamefont {J.~A.}\ \bibnamefont {Font}},\ and\ \bibinfo {author} {\bibfnamefont {J.~A.}\ \bibnamefont {Pons}},\ }\href {https://doi.org/10.1086/498238} {\bibfield  {journal} {\bibinfo  {journal} {Astrophys. J.}\ }\textbf {\bibinfo {volume} {637}},\ \bibinfo {pages} {296} (\bibinfo {year} {2006})}\BibitemShut {NoStop}%
\bibitem [{\citenamefont {Harten}\ \emph {et~al.}(1983)\citenamefont {Harten}, \citenamefont {Lax},\ and\ \citenamefont {Leer}}]{doi:10.1137/1025002}%
  \BibitemOpen
  \bibfield  {author} {\bibinfo {author} {\bibfnamefont {A.}~\bibnamefont {Harten}}, \bibinfo {author} {\bibfnamefont {P.~D.}\ \bibnamefont {Lax}},\ and\ \bibinfo {author} {\bibfnamefont {B.~v.}\ \bibnamefont {Leer}},\ }\href {https://doi.org/10.1137/1025002} {\bibfield  {journal} {\bibinfo  {journal} {SIAM Rev.}\ }\textbf {\bibinfo {volume} {25}},\ \bibinfo {pages} {35} (\bibinfo {year} {1983})}\BibitemShut {NoStop}%
\bibitem [{\citenamefont {Nessyahu}\ and\ \citenamefont {Tadmor}(1990)}]{NESSYAHU1990408}%
  \BibitemOpen
  \bibfield  {author} {\bibinfo {author} {\bibfnamefont {H.}~\bibnamefont {Nessyahu}}\ and\ \bibinfo {author} {\bibfnamefont {E.}~\bibnamefont {Tadmor}},\ }\href {https://doi.org/https://doi.org/10.1016/0021-9991(90)90260-8} {\bibfield  {journal} {\bibinfo  {journal} {J. Comput. Phys.}\ }\textbf {\bibinfo {volume} {87}},\ \bibinfo {pages} {408} (\bibinfo {year} {1990})}\BibitemShut {NoStop}%
\bibitem [{\citenamefont {{Kurganov}}\ and\ \citenamefont {{Tadmor}}(2000)}]{2000JCoPh.160..241K}%
  \BibitemOpen
  \bibfield  {author} {\bibinfo {author} {\bibfnamefont {A.}~\bibnamefont {{Kurganov}}}\ and\ \bibinfo {author} {\bibfnamefont {E.}~\bibnamefont {{Tadmor}}},\ }\href {https://doi.org/10.1006/jcph.2000.6459} {\bibfield  {journal} {\bibinfo  {journal} {J. Comput. Phys.}\ }\textbf {\bibinfo {volume} {160}},\ \bibinfo {pages} {241} (\bibinfo {year} {2000})}\BibitemShut {NoStop}%
\bibitem [{\citenamefont {{Borges}}\ \emph {et~al.}(2008)\citenamefont {{Borges}}, \citenamefont {{Carmona}}, \citenamefont {{Costa}},\ and\ \citenamefont {{Don}}}]{2008JCoPh.227.3191B}%
  \BibitemOpen
  \bibfield  {author} {\bibinfo {author} {\bibfnamefont {R.}~\bibnamefont {{Borges}}}, \bibinfo {author} {\bibfnamefont {M.}~\bibnamefont {{Carmona}}}, \bibinfo {author} {\bibfnamefont {B.}~\bibnamefont {{Costa}}},\ and\ \bibinfo {author} {\bibfnamefont {W.~S.}\ \bibnamefont {{Don}}},\ }\href {https://doi.org/10.1016/j.jcp.2007.11.038} {\bibfield  {journal} {\bibinfo  {journal} {J. Comput. Phys.}\ }\textbf {\bibinfo {volume} {227}},\ \bibinfo {pages} {3191} (\bibinfo {year} {2008})}\BibitemShut {NoStop}%
\bibitem [{\citenamefont {Haas}\ \emph {et~al.}(2016)\citenamefont {Haas} \emph {et~al.}}]{Haas:2016cop}%
  \BibitemOpen
  \bibfield  {author} {\bibinfo {author} {\bibfnamefont {R.}~\bibnamefont {Haas}} \emph {et~al.},\ }\href {https://doi.org/10.1103/PhysRevD.93.124062} {\bibfield  {journal} {\bibinfo  {journal} {Phys. Rev. D}\ }\textbf {\bibinfo {volume} {93}},\ \bibinfo {pages} {124062} (\bibinfo {year} {2016})}\BibitemShut {NoStop}%
\bibitem [{\citenamefont {Steppohn}(2025{\natexlab{b}})}]{steppohnthesis}%
  \BibitemOpen
  \bibfield  {author} {\bibinfo {author} {\bibfnamefont {O.}~\bibnamefont {Steppohn}},\ }\emph {\bibinfo {title} {Black hole spectroscopy of collapsing and merging neutron stars}},\ \href@noop {} {Master's thesis},\ \bibinfo  {school} {University of Potsdam}, \bibinfo {address} {Potsdam, Germany} (\bibinfo {year} {2025}{\natexlab{b}}),\ \bibinfo {note} {master's thesis}\BibitemShut {NoStop}%
\bibitem [{\citenamefont {Topolski}\ \emph {et~al.}(2025)\citenamefont {Topolski}, \citenamefont {Tootle},\ and\ \citenamefont {Rezzolla}}]{Topolski:2024jva}%
  \BibitemOpen
  \bibfield  {author} {\bibinfo {author} {\bibfnamefont {K.}~\bibnamefont {Topolski}}, \bibinfo {author} {\bibfnamefont {S.~D.}\ \bibnamefont {Tootle}},\ and\ \bibinfo {author} {\bibfnamefont {L.}~\bibnamefont {Rezzolla}},\ }\href {https://doi.org/10.1103/PhysRevD.111.064023} {\bibfield  {journal} {\bibinfo  {journal} {Phys. Rev. D}\ }\textbf {\bibinfo {volume} {111}},\ \bibinfo {pages} {064023} (\bibinfo {year} {2025})}\BibitemShut {NoStop}%
\bibitem [{\citenamefont {{Shibata}}\ \emph {et~al.}(2003)\citenamefont {{Shibata}}, \citenamefont {{Taniguchi}},\ and\ \citenamefont {{Ury{\={u}}}}}]{Shibata2003}%
  \BibitemOpen
  \bibfield  {author} {\bibinfo {author} {\bibfnamefont {M.}~\bibnamefont {{Shibata}}}, \bibinfo {author} {\bibfnamefont {K.}~\bibnamefont {{Taniguchi}}},\ and\ \bibinfo {author} {\bibfnamefont {K.}~\bibnamefont {{Ury{\={u}}}}},\ }\href {https://doi.org/10.1103/PhysRevD.68.084020} {\bibfield  {journal} {\bibinfo  {journal} {\prd}\ }\textbf {\bibinfo {volume} {68}},\ \bibinfo {eid} {084020} (\bibinfo {year} {2003})}\BibitemShut {NoStop}%
\bibitem [{\citenamefont {Lehner}\ \emph {et~al.}(2016)\citenamefont {Lehner}, \citenamefont {Liebling}, \citenamefont {Palenzuela}, \citenamefont {Caballero}, \citenamefont {O'Connor}, \citenamefont {Anderson},\ and\ \citenamefont {Neilsen}}]{Lehner2016}%
  \BibitemOpen
  \bibfield  {author} {\bibinfo {author} {\bibfnamefont {L.}~\bibnamefont {Lehner}}, \bibinfo {author} {\bibfnamefont {S.~L.}\ \bibnamefont {Liebling}}, \bibinfo {author} {\bibfnamefont {C.}~\bibnamefont {Palenzuela}}, \bibinfo {author} {\bibfnamefont {O.~L.}\ \bibnamefont {Caballero}}, \bibinfo {author} {\bibfnamefont {E.}~\bibnamefont {O'Connor}}, \bibinfo {author} {\bibfnamefont {M.}~\bibnamefont {Anderson}},\ and\ \bibinfo {author} {\bibfnamefont {D.}~\bibnamefont {Neilsen}},\ }\href {https://doi.org/10.1088/0264-9381/33/18/184002} {\bibfield  {journal} {\bibinfo  {journal} {Classical Quantum Gravity}\ }\textbf {\bibinfo {volume} {33}},\ \bibinfo {pages} {184002} (\bibinfo {year} {2016})}\BibitemShut {NoStop}%
\bibitem [{\citenamefont {Baumgarte}\ \emph {et~al.}(2000)\citenamefont {Baumgarte}, \citenamefont {Shapiro},\ and\ \citenamefont {Shibata}}]{Baumgarte1999}%
  \BibitemOpen
  \bibfield  {author} {\bibinfo {author} {\bibfnamefont {T.~W.}\ \bibnamefont {Baumgarte}}, \bibinfo {author} {\bibfnamefont {S.~L.}\ \bibnamefont {Shapiro}},\ and\ \bibinfo {author} {\bibfnamefont {M.}~\bibnamefont {Shibata}},\ }\href {https://doi.org/10.1086/312425} {\bibfield  {journal} {\bibinfo  {journal} {Astrophys. J. Lett.}\ }\textbf {\bibinfo {volume} {528}},\ \bibinfo {pages} {L29} (\bibinfo {year} {2000})}\BibitemShut {NoStop}%
\bibitem [{\citenamefont {Hotokezaka}\ \emph {et~al.}(2011)\citenamefont {Hotokezaka}, \citenamefont {Kyutoku}, \citenamefont {Okawa}, \citenamefont {Shibata},\ and\ \citenamefont {Kiuchi}}]{Hotokezaka2011}%
  \BibitemOpen
  \bibfield  {author} {\bibinfo {author} {\bibfnamefont {K.}~\bibnamefont {Hotokezaka}}, \bibinfo {author} {\bibfnamefont {K.}~\bibnamefont {Kyutoku}}, \bibinfo {author} {\bibfnamefont {H.}~\bibnamefont {Okawa}}, \bibinfo {author} {\bibfnamefont {M.}~\bibnamefont {Shibata}},\ and\ \bibinfo {author} {\bibfnamefont {K.}~\bibnamefont {Kiuchi}},\ }\href {https://doi.org/10.1103/PhysRevD.83.124008} {\bibfield  {journal} {\bibinfo  {journal} {Phys. Rev. D}\ }\textbf {\bibinfo {volume} {83}},\ \bibinfo {pages} {124008} (\bibinfo {year} {2011})}\BibitemShut {NoStop}%
\bibitem [{\citenamefont {Bauswein}\ and\ \citenamefont {Janka}(2012)}]{Bauswein2011}%
  \BibitemOpen
  \bibfield  {author} {\bibinfo {author} {\bibfnamefont {A.}~\bibnamefont {Bauswein}}\ and\ \bibinfo {author} {\bibfnamefont {H.~T.}\ \bibnamefont {Janka}},\ }\href {https://doi.org/10.1103/PhysRevLett.108.011101} {\bibfield  {journal} {\bibinfo  {journal} {Phys. Rev. Lett.}\ }\textbf {\bibinfo {volume} {108}},\ \bibinfo {pages} {011101} (\bibinfo {year} {2012})},\ \Eprint {https://arxiv.org/abs/1106.1616} {arXiv:1106.1616 [astro-ph.SR]} \BibitemShut {NoStop}%
\bibitem [{\citenamefont {Teukolsky}\ and\ \citenamefont {Press}(1974)}]{Teukolsky:1974yv}%
  \BibitemOpen
  \bibfield  {author} {\bibinfo {author} {\bibfnamefont {S.~A.}\ \bibnamefont {Teukolsky}}\ and\ \bibinfo {author} {\bibfnamefont {W.~H.}\ \bibnamefont {Press}},\ }\href {https://doi.org/10.1086/153180} {\bibfield  {journal} {\bibinfo  {journal} {Astrophys. J.}\ }\textbf {\bibinfo {volume} {193}},\ \bibinfo {pages} {443} (\bibinfo {year} {1974})}\BibitemShut {NoStop}%
\bibitem [{\citenamefont {Chandrasekhar}\ and\ \citenamefont {Detweiler}(1975)}]{Chandrasekhar:1975zza}%
  \BibitemOpen
  \bibfield  {author} {\bibinfo {author} {\bibfnamefont {S.}~\bibnamefont {Chandrasekhar}}\ and\ \bibinfo {author} {\bibfnamefont {S.~L.}\ \bibnamefont {Detweiler}},\ }\href {https://doi.org/10.1098/rspa.1975.0112} {\bibfield  {journal} {\bibinfo  {journal} {Proc. R. Soc. A}\ }\textbf {\bibinfo {volume} {344}},\ \bibinfo {pages} {441} (\bibinfo {year} {1975})}\BibitemShut {NoStop}%
\bibitem [{\citenamefont {{Leaver}}(1985)}]{Leaver1985}%
  \BibitemOpen
  \bibfield  {author} {\bibinfo {author} {\bibfnamefont {E.~W.}\ \bibnamefont {{Leaver}}},\ }\href {https://doi.org/10.1098/rspa.1985.0119} {\bibfield  {journal} {\bibinfo  {journal} {Proc. R. Soc. A}\ }\textbf {\bibinfo {volume} {402}},\ \bibinfo {pages} {285} (\bibinfo {year} {1985})}\BibitemShut {NoStop}%
\bibitem [{\citenamefont {Thornburg}(2004)}]{Thornburg:2003sf}%
  \BibitemOpen
  \bibfield  {author} {\bibinfo {author} {\bibfnamefont {J.}~\bibnamefont {Thornburg}},\ }\href {https://doi.org/10.1088/0264-9381/21/2/026} {\bibfield  {journal} {\bibinfo  {journal} {Classical Quantum Gravity}\ }\textbf {\bibinfo {volume} {21}},\ \bibinfo {pages} {743} (\bibinfo {year} {2004})}\BibitemShut {NoStop}%
\bibitem [{\citenamefont {Price}(1972)}]{Price:1971fb}%
  \BibitemOpen
  \bibfield  {author} {\bibinfo {author} {\bibfnamefont {R.~H.}\ \bibnamefont {Price}},\ }\href {https://doi.org/10.1103/PhysRevD.5.2419} {\bibfield  {journal} {\bibinfo  {journal} {Phys. Rev. D}\ }\textbf {\bibinfo {volume} {5}},\ \bibinfo {pages} {2419} (\bibinfo {year} {1972})}\BibitemShut {NoStop}%
\bibitem [{\citenamefont {Leaver}(1986)}]{Leaver:1986gd}%
  \BibitemOpen
  \bibfield  {author} {\bibinfo {author} {\bibfnamefont {E.~W.}\ \bibnamefont {Leaver}},\ }\href {https://doi.org/10.1103/PhysRevD.34.384} {\bibfield  {journal} {\bibinfo  {journal} {Phys. Rev. D}\ }\textbf {\bibinfo {volume} {34}},\ \bibinfo {pages} {384} (\bibinfo {year} {1986})}\BibitemShut {NoStop}%
\bibitem [{\citenamefont {Gundlach}\ \emph {et~al.}(1994)\citenamefont {Gundlach}, \citenamefont {Price},\ and\ \citenamefont {Pullin}}]{Gundlach:1993tp}%
  \BibitemOpen
  \bibfield  {author} {\bibinfo {author} {\bibfnamefont {C.}~\bibnamefont {Gundlach}}, \bibinfo {author} {\bibfnamefont {R.~H.}\ \bibnamefont {Price}},\ and\ \bibinfo {author} {\bibfnamefont {J.}~\bibnamefont {Pullin}},\ }\href {https://doi.org/10.1103/PhysRevD.49.883} {\bibfield  {journal} {\bibinfo  {journal} {Phys. Rev. D}\ }\textbf {\bibinfo {volume} {49}},\ \bibinfo {pages} {883} (\bibinfo {year} {1994})}\BibitemShut {NoStop}%
\bibitem [{\citenamefont {Barack}(1999)}]{Barack:1998bw}%
  \BibitemOpen
  \bibfield  {author} {\bibinfo {author} {\bibfnamefont {L.}~\bibnamefont {Barack}},\ }\href {https://doi.org/10.1103/PhysRevD.59.044017} {\bibfield  {journal} {\bibinfo  {journal} {Phys. Rev. D}\ }\textbf {\bibinfo {volume} {59}},\ \bibinfo {pages} {044017} (\bibinfo {year} {1999})}\BibitemShut {NoStop}%
\bibitem [{\citenamefont {Virtanen}\ \emph {et~al.}(2020)\citenamefont {Virtanen} \emph {et~al.}}]{2020SciPy-NMeth}%
  \BibitemOpen
  \bibfield  {author} {\bibinfo {author} {\bibfnamefont {P.}~\bibnamefont {Virtanen}} \emph {et~al.},\ }\href {https://doi.org/10.1038/s41592-019-0686-2} {\bibfield  {journal} {\bibinfo  {journal} {Nat. Methods}\ }\textbf {\bibinfo {volume} {17}},\ \bibinfo {pages} {261} (\bibinfo {year} {2020})}\BibitemShut {NoStop}%
\bibitem [{\citenamefont {Reitze}\ \emph {et~al.}(2019)\citenamefont {Reitze} \emph {et~al.}}]{Reitze:2019iox}%
  \BibitemOpen
  \bibfield  {author} {\bibinfo {author} {\bibfnamefont {D.}~\bibnamefont {Reitze}} \emph {et~al.},\ }\href@noop {} {\bibfield  {journal} {\bibinfo  {journal} {Bull. Am. Astron. Soc.}\ }\textbf {\bibinfo {volume} {51}},\ \bibinfo {pages} {035} (\bibinfo {year} {2019})}\BibitemShut {NoStop}%
\bibitem [{\citenamefont {Abac}\ \emph {et~al.}()\citenamefont {Abac} \emph {et~al.}}]{Abac:2025saz}%
  \BibitemOpen
  \bibfield  {author} {\bibinfo {author} {\bibfnamefont {A.}~\bibnamefont {Abac}} \emph {et~al.},\ }\href@noop {} {\ }\Eprint {https://arxiv.org/abs/arXiv:2503.12263} {arXiv:2503.12263} \BibitemShut {NoStop}%
\bibitem [{\citenamefont {Punturo}\ \emph {et~al.}(2010)\citenamefont {Punturo} \emph {et~al.}}]{Punturo:2010zz}%
  \BibitemOpen
  \bibfield  {author} {\bibinfo {author} {\bibfnamefont {M.}~\bibnamefont {Punturo}} \emph {et~al.},\ }\href {https://doi.org/10.1088/0264-9381/27/19/194002} {\bibfield  {journal} {\bibinfo  {journal} {Classical Quantum Gravity}\ }\textbf {\bibinfo {volume} {27}},\ \bibinfo {pages} {194002} (\bibinfo {year} {2010})}\BibitemShut {NoStop}%
\bibitem [{\citenamefont {Maggiore}\ \emph {et~al.}(2020)\citenamefont {Maggiore} \emph {et~al.}}]{ET:2019dnz}%
  \BibitemOpen
  \bibfield  {author} {\bibinfo {author} {\bibfnamefont {M.}~\bibnamefont {Maggiore}} \emph {et~al.} (\bibinfo {collaboration} {ET Collaboration}),\ }\href {https://doi.org/10.1088/1475-7516/2020/03/050} {\bibfield  {journal} {\bibinfo  {journal} {J. Cosmol. Astropart. Phys.}\ }\textbf {\bibinfo {volume} {03}},\ \bibinfo {pages} {050}}\BibitemShut {NoStop}%
\bibitem [{\citenamefont {Evans}\ \emph {et~al.}()\citenamefont {Evans} \emph {et~al.}}]{Evans:2021gyd}%
  \BibitemOpen
  \bibfield  {author} {\bibinfo {author} {\bibfnamefont {M.}~\bibnamefont {Evans}} \emph {et~al.},\ }\href@noop {} {\ }\Eprint {https://arxiv.org/abs/arXiv:2109.09882} {arXiv:2109.09882} \BibitemShut {NoStop}%
\bibitem [{\citenamefont {Cotesta}\ \emph {et~al.}(2022)\citenamefont {Cotesta}, \citenamefont {Carullo}, \citenamefont {Berti},\ and\ \citenamefont {Cardoso}}]{Cotesta:2022pci}%
  \BibitemOpen
  \bibfield  {author} {\bibinfo {author} {\bibfnamefont {R.}~\bibnamefont {Cotesta}}, \bibinfo {author} {\bibfnamefont {G.}~\bibnamefont {Carullo}}, \bibinfo {author} {\bibfnamefont {E.}~\bibnamefont {Berti}},\ and\ \bibinfo {author} {\bibfnamefont {V.}~\bibnamefont {Cardoso}},\ }\href {https://doi.org/10.1103/PhysRevLett.129.111102} {\bibfield  {journal} {\bibinfo  {journal} {Phys. Rev. Lett.}\ }\textbf {\bibinfo {volume} {129}},\ \bibinfo {pages} {111102} (\bibinfo {year} {2022})}\BibitemShut {NoStop}%
\bibitem [{\citenamefont {Isi}\ and\ \citenamefont {Farr}(2023)}]{Isi:2023nif}%
  \BibitemOpen
  \bibfield  {author} {\bibinfo {author} {\bibfnamefont {M.}~\bibnamefont {Isi}}\ and\ \bibinfo {author} {\bibfnamefont {W.~M.}\ \bibnamefont {Farr}},\ }\href {https://doi.org/10.1103/PhysRevLett.131.169001} {\bibfield  {journal} {\bibinfo  {journal} {Phys. Rev. Lett.}\ }\textbf {\bibinfo {volume} {131}},\ \bibinfo {pages} {169001} (\bibinfo {year} {2023})}\BibitemShut {NoStop}%
\bibitem [{\citenamefont {Carullo}\ \emph {et~al.}(2023)\citenamefont {Carullo}, \citenamefont {Cotesta}, \citenamefont {Berti},\ and\ \citenamefont {Cardoso}}]{Carullo:2023gtf}%
  \BibitemOpen
  \bibfield  {author} {\bibinfo {author} {\bibfnamefont {G.}~\bibnamefont {Carullo}}, \bibinfo {author} {\bibfnamefont {R.}~\bibnamefont {Cotesta}}, \bibinfo {author} {\bibfnamefont {E.}~\bibnamefont {Berti}},\ and\ \bibinfo {author} {\bibfnamefont {V.}~\bibnamefont {Cardoso}},\ }\href {https://doi.org/10.1103/PhysRevLett.131.169002} {\bibfield  {journal} {\bibinfo  {journal} {Phys. Rev. Lett.}\ }\textbf {\bibinfo {volume} {131}},\ \bibinfo {pages} {169002} (\bibinfo {year} {2023})}\BibitemShut {NoStop}%
\bibitem [{\citenamefont {Finch}\ and\ \citenamefont {Moore}(2022)}]{Finch:2022ynt}%
  \BibitemOpen
  \bibfield  {author} {\bibinfo {author} {\bibfnamefont {E.}~\bibnamefont {Finch}}\ and\ \bibinfo {author} {\bibfnamefont {C.~J.}\ \bibnamefont {Moore}},\ }\href {https://doi.org/10.1103/PhysRevD.106.043005} {\bibfield  {journal} {\bibinfo  {journal} {Phys. Rev. D}\ }\textbf {\bibinfo {volume} {106}},\ \bibinfo {pages} {043005} (\bibinfo {year} {2022})}\BibitemShut {NoStop}%
\bibitem [{\citenamefont {Dhani}\ \emph {et~al.}(2024)\citenamefont {Dhani}, \citenamefont {Radice}, \citenamefont {Sch{\"u}tte-Engel}, \citenamefont {Gardner}, \citenamefont {Sathyaprakash}, \citenamefont {Logoteta}, \citenamefont {Perego},\ and\ \citenamefont {Kashyap}}]{Dhani:2023ijt}%
  \BibitemOpen
  \bibfield  {author} {\bibinfo {author} {\bibfnamefont {A.}~\bibnamefont {Dhani}}, \bibinfo {author} {\bibfnamefont {D.}~\bibnamefont {Radice}}, \bibinfo {author} {\bibfnamefont {J.}~\bibnamefont {Sch{\"u}tte-Engel}}, \bibinfo {author} {\bibfnamefont {S.}~\bibnamefont {Gardner}}, \bibinfo {author} {\bibfnamefont {B.}~\bibnamefont {Sathyaprakash}}, \bibinfo {author} {\bibfnamefont {D.}~\bibnamefont {Logoteta}}, \bibinfo {author} {\bibfnamefont {A.}~\bibnamefont {Perego}},\ and\ \bibinfo {author} {\bibfnamefont {R.}~\bibnamefont {Kashyap}},\ }\href {https://doi.org/10.1103/PhysRevD.109.044071} {\bibfield  {journal} {\bibinfo  {journal} {Phys. Rev. D}\ }\textbf {\bibinfo {volume} {109}},\ \bibinfo {pages} {044071} (\bibinfo {year} {2024})}\BibitemShut {NoStop}%
\bibitem [{\citenamefont {Finn}\ and\ \citenamefont {Chernoff}(1993)}]{Finn:1992xs}%
  \BibitemOpen
  \bibfield  {author} {\bibinfo {author} {\bibfnamefont {L.~S.}\ \bibnamefont {Finn}}\ and\ \bibinfo {author} {\bibfnamefont {D.~F.}\ \bibnamefont {Chernoff}},\ }\href {https://doi.org/10.1103/PhysRevD.47.2198} {\bibfield  {journal} {\bibinfo  {journal} {Phys. Rev. D}\ }\textbf {\bibinfo {volume} {47}},\ \bibinfo {pages} {2198} (\bibinfo {year} {1993})}\BibitemShut {NoStop}%
\bibitem [{\citenamefont {Krolak}\ \emph {et~al.}(1993)\citenamefont {Krolak}, \citenamefont {Lobo},\ and\ \citenamefont {Meers}}]{Krolak:1993zy}%
  \BibitemOpen
  \bibfield  {author} {\bibinfo {author} {\bibfnamefont {A.}~\bibnamefont {Krolak}}, \bibinfo {author} {\bibfnamefont {J.~A.}\ \bibnamefont {Lobo}},\ and\ \bibinfo {author} {\bibfnamefont {B.~J.}\ \bibnamefont {Meers}},\ }\href {https://doi.org/10.1103/PhysRevD.48.3451} {\bibfield  {journal} {\bibinfo  {journal} {Phys. Rev. D}\ }\textbf {\bibinfo {volume} {48}},\ \bibinfo {pages} {3451} (\bibinfo {year} {1993})}\BibitemShut {NoStop}%
\bibitem [{\citenamefont {Flanagan}\ and\ \citenamefont {Hughes}(1998)}]{Flanagan:1997kp}%
  \BibitemOpen
  \bibfield  {author} {\bibinfo {author} {\bibfnamefont {E.~E.}\ \bibnamefont {Flanagan}}\ and\ \bibinfo {author} {\bibfnamefont {S.~A.}\ \bibnamefont {Hughes}},\ }\href {https://doi.org/10.1103/PhysRevD.57.4566} {\bibfield  {journal} {\bibinfo  {journal} {Phys. Rev. D}\ }\textbf {\bibinfo {volume} {57}},\ \bibinfo {pages} {4566} (\bibinfo {year} {1998})}\BibitemShut {NoStop}%
\bibitem [{\citenamefont {Cutler}\ and\ \citenamefont {Vallisneri}(2007)}]{Cutler:2007mi}%
  \BibitemOpen
  \bibfield  {author} {\bibinfo {author} {\bibfnamefont {C.}~\bibnamefont {Cutler}}\ and\ \bibinfo {author} {\bibfnamefont {M.}~\bibnamefont {Vallisneri}},\ }\href {https://doi.org/10.1103/PhysRevD.76.104018} {\bibfield  {journal} {\bibinfo  {journal} {Phys. Rev. D}\ }\textbf {\bibinfo {volume} {76}},\ \bibinfo {pages} {104018} (\bibinfo {year} {2007})}\BibitemShut {NoStop}%
\bibitem [{\citenamefont {Capuano}\ \emph {et~al.}(2025)\citenamefont {Capuano}, \citenamefont {Vaglio}, \citenamefont {Chandramouli}, \citenamefont {Pitte}, \citenamefont {Kuntz},\ and\ \citenamefont {Barausse}}]{Capuano:2025kkl}%
  \BibitemOpen
  \bibfield  {author} {\bibinfo {author} {\bibfnamefont {L.}~\bibnamefont {Capuano}}, \bibinfo {author} {\bibfnamefont {M.}~\bibnamefont {Vaglio}}, \bibinfo {author} {\bibfnamefont {R.~S.}\ \bibnamefont {Chandramouli}}, \bibinfo {author} {\bibfnamefont {C.~L.}\ \bibnamefont {Pitte}}, \bibinfo {author} {\bibfnamefont {A.}~\bibnamefont {Kuntz}},\ and\ \bibinfo {author} {\bibfnamefont {E.}~\bibnamefont {Barausse}},\ }\href {https://doi.org/10.1103/86yd-x1sl} {\bibfield  {journal} {\bibinfo  {journal} {Phys. Rev. D}\ }\textbf {\bibinfo {volume} {112}},\ \bibinfo {pages} {104031} (\bibinfo {year} {2025})}\BibitemShut {NoStop}%
\bibitem [{\citenamefont {V{\"o}lkel}\ and\ \citenamefont {Dhani}(2025)}]{Volkel:2025jdx}%
  \BibitemOpen
  \bibfield  {author} {\bibinfo {author} {\bibfnamefont {S.~H.}\ \bibnamefont {V{\"o}lkel}}\ and\ \bibinfo {author} {\bibfnamefont {A.}~\bibnamefont {Dhani}},\ }\href {https://doi.org/10.1103/g6sz-dw28} {\bibfield  {journal} {\bibinfo  {journal} {Phys. Rev. D}\ }\textbf {\bibinfo {volume} {112}},\ \bibinfo {pages} {084076} (\bibinfo {year} {2025})}\BibitemShut {NoStop}%
\bibitem [{\citenamefont {Steppohn}(2025{\natexlab{c}})}]{zenodorepo}%
  \BibitemOpen
  \bibfield  {author} {\bibinfo {author} {\bibfnamefont {O.}~\bibnamefont {Steppohn}},\ }\href@noop {} {\bibinfo {title} {Black hole spectroscopy of collapsing and merging neutron stars}},\ \bibinfo {howpublished} {Zenodo} (\bibinfo {year} {2025}{\natexlab{c}}),\ \bibinfo {note} {doi:10.5281/zenodo.16792099, \url{https://doi.org/10.5281/zenodo.16792099}}\BibitemShut {NoStop}%
\end{thebibliography}%

\newpage 

\appendix

\section{ISOLATED NEUTRON STAR EQUILIBRIUM MODELS}\label{app:RNSseq}

In the following, we list the parameters of all 130 dRNS models, which were created with the procedure in Sec.~\ref{dRNSinit}. The parameters are dimensionless, based on the rescalings of~\cite{Cook1992,Cook1994}. Each model in the table obtains an identifier according to the following procedure: 1.~The first half of the identifier is a letter corresponding to the respective rotation parameter of the model, e.g., the letter $A$ for $\hat{A}=1.0$, the letter $B$ for $\hat{A}=0.9$, all the way to $J$ for $\hat{A}=0.1$. 2. Next to the letter, we give the model a number based on its rotation parameter, e.g., the number 1 for $r_p/r_e=0.35$, the number 2 for $r_p/r_e=0.4$, up to the number 13 for $r_p/r_e=0.95$. We group all models with the same rotation parameter under a \textit{Series} with the letter used in the model name, i.e., the models with $\hat{A}=1.0$ under \textit{Series A}, the models with $\hat{A}=0.9$ under \textit{Series B}, and so on. Next to the model identifier, we show the respective class based on the classification scheme presented in Sec.~\ref{classAnalysis}. Some models are denoted with an asterisk, which indicates that for this particular model the central density had to be increased to induce the collapse of the NS. The adjusted value is given in the $\rho_c$ column behind the interpolated value from the corresponding model sequence. A double asterisk is given behind two model identifiers because these models showed problematic output for the baryonic mass around the remnant. It seemed as if the baryonic mass increased to its precollapse value and then went down again. To prevent the contamination of these results to our overall analysis, we reran these two models at a different central density to prevent the issue from occurring again.
\onecolumngrid
\begin{longtable}{p{0.13\textwidth}p{0.14\textwidth}p{0.08125\textwidth}p{0.08125\textwidth}p{0.08125\textwidth}p{0.08125\textwidth}p{0.08125\textwidth}p{0.08125\textwidth}p{0.08125\textwidth}p{0.08125\textwidth}}
\multicolumn{10}{c}{\parbox{\linewidth}{TABLE \thetable. This table groups all model parameters by their respective series. Shown from left to right are the model identifier and its class, the central density $\rho_c$, the central energy density $\varepsilon_c$, the mass of the NS $M/M_\odot$, its equatorial radius scaled by the mass $R_e/M$, the binding energy $T/|W|$, the central angular velocity $\Omega_c$, the dimensionless spin $J/M^2$, the rescaled rotation parameter $\hat{A}$ and the axis ratio.}}
\label{tab:inittab}\\[2.0em]
\toprule
\addlinespace
Model - Class&$\rho_c$&$\varepsilon_c$&$M/M_{\odot}$&$R_e/M$&$T/|W|$&$\Omega_c$&$J/M^2$&$\hat{A}$&$r_p/r_e$\\
\addlinespace
\midrule
\endfirsthead

\toprule
\addlinespace
Model - Class&$\rho_c$&$e_c$&$M/M_{\odot}$&$R_e/M$&$T/|W|$&$\Omega_c$&$J/M^2$&$\hat{A}$&$r_p/r_e$ \\
\addlinespace
\midrule
\endhead

\midrule
\addlinespace
\multicolumn{2}{r}{\textit{Continued on next page}} \\
\addlinespace
\midrule
\endfoot

\bottomrule
\endlastfoot
\addlinespace
\textit{Series A}&\\
A1$^*$ - II&0.001 ($^*$0.0014)&0.0011&2.728&12.323&0.228&0.044&0.889&1.0&0.35\\
A2$^*$ - {III}&0.0013 ($^*0.0016$)&0.0015&2.533&11.59&0.207&0.045&0.847&1.0&0.4\\
A3$^*$ - {III}&0.0016 ($^*0.0018$)&0.0018&2.367&10.886&0.185&0.045&0.8&1.0&0.45\\
A4$^*$ - {II}&0.0018 ($^*0.002$)&0.0021&2.23&10.351&0.163&0.045&0.752&1.0&0.5\\
A5$^*$ - {II}&0.002 ($^*0.0021$)&0.0024&2.117&9.847&0.142&0.045&0.701&1.0&0.55\\
A6 - {II}&0.0022&0.0027&2.024&9.463&0.122&0.043&0.65&1.0&0.6\\
A7 - {I}&0.0024&0.0029&1.946&9.107&0.103&0.042&0.597&1.0&0.65\\
A8 - {I}&0.0025&0.0031&1.88&8.81&0.085&0.039&0.543&1.0&0.7\\
A9 - {I}&0.0026&0.0033&1.824&8.563&0.069&0.036&0.487&1.0&0.75\\
A10 - {I}&0.0028&0.0036&1.775&8.31&0.053&0.033&0.427&1.0&0.8\\
A11 - {III}&0.0029&0.0037&1.734&8.122&0.039&0.029&0.363&1.0&0.85\\
A12 - {II}&0.003&0.0039&1.697&7.947&0.025&0.024&0.291&1.0&0.9\\
A13 - {II}&0.0031&0.004&1.666&7.785&0.012&0.017&0.202&1.0&0.95\\
\addlinespace
\midrule
\addlinespace
\textit{Series B}&\\
B1$^*$ - {III}&0.001 ($^*0.0014$)&0.0011&2.69&12.096&0.224&0.044&0.877&0.9&0.35\\
B2 - {III}&0.0013&0.0014&2.508&11.394&0.204&0.045&0.836&0.9&0.4\\
B3$^*$ - {II}&0.0015 ($^*0.0017$)&0.0017&2.35&10.805&0.183&0.046&0.792&0.9&0.45\\
B4$^*$ - {I}&0.0018 ($^*0.002$)&0.0021&2.218&10.239&0.161&0.046&0.743&0.9&0.5\\
B5 - {II}&0.002&0.0024&2.109&9.81&0.141&0.045&0.695&0.9&0.55\\
B6 - {I}&0.0022&0.0026&2.018&9.406&0.121&0.044&0.643&0.9&0.6\\
B7 - {I}&0.0023&0.0029&1.941&9.068&0.102&0.042&0.591&0.9&0.65\\
B8 - {I}&0.0025&0.0031&1.876&8.785&0.085&0.039&0.538&0.9&0.7\\
B9 - {I}&0.0026&0.0033&1.821&8.524&0.068&0.037&0.482&0.9&0.75\\
B10 - {I}&0.0028&0.0035&1.773&8.306&0.053&0.033&0.423&0.9&0.8\\
B11 - {II}&0.0029&0.0037&1.732&8.103&0.038&0.029&0.36&0.9&0.85\\
B12 - {II}&0.003&0.0039&1.697&7.936&0.025&0.024&0.289&0.9&0.9\\
B13 - {II}&0.0031&0.004&1.665&7.78&0.012&0.017&0.201&0.9&0.95\\
\addlinespace
\midrule
\addlinespace
\textit{Series C}&\\
C1$^*$ - {II}&0.0009 ($^*0.0013$)&0.001&2.604&11.96&0.215&0.044&0.859&0.8&0.35\\
C2$^*$ - {I}&0.0012 ($^*0.0014$)&0.0013&2.446&11.116&0.196&0.046&0.813&0.8&0.4\\
C3$^*$ - {I}&0.0014 ($^*0.0016$)&0.0017&2.307&10.574&0.176&0.046&0.771&0.8&0.45\\
C4$^*$ - {I}&0.0017 ($^*0.0019$)&0.0019&2.187&10.118&0.156&0.046&0.726&0.8&0.5\\
C5 - {II}&0.0019&0.0022&2.086&9.683&0.137&0.045&0.678&0.8&0.55\\
C6 - {I}&0.0021&0.0025&2.001&9.319&0.118&0.044&0.628&0.8&0.6\\
C7 - {I}&0.0023&0.0028&1.928&9.013&0.1&0.042&0.578&0.8&0.65\\
C8 - {I}&0.0024&0.003&1.867&8.727&0.082&0.04&0.525&0.8&0.7\\
C9 - {I}&0.0026&0.0033&1.814&8.489&0.066&0.037&0.471&0.8&0.75\\
C10 - {I}&0.0027&0.0035&1.769&8.267&0.051&0.033&0.414&0.8&0.8\\
C11 - {III}&0.0029&0.0037&1.729&8.083&0.037&0.029&0.352&0.8&0.85\\
C12 - {III}&0.003&0.0039&1.694&7.91&0.024&0.024&0.282&0.8&0.9\\
C13 - {III}&0.0031&0.004&1.664&7.768&0.012&0.017&0.196&0.8&0.95\\
\addlinespace
\midrule
\addlinespace
\textit{Series D}&\\
D1$^*$ - {II}&0.0008 ($^*0.0012$)&0.0009&2.507&11.565&0.204&0.045&0.827&0.7&0.35\\
D2$^*$ - {I}&0.0011 ($^*0.0013$)&0.0012&2.371&10.951&0.186&0.046&0.788&0.7&0.4\\
D3$^*$ - {I}&0.0013 ($^*0.0015$)&0.0015&2.251&10.415&0.168&0.047&0.746&0.7&0.45\\
D4 - {III}&0.0016&0.0018&2.146&9.932&0.149&0.047&0.701&0.7&0.5\\
D5 - {II}&0.0018&0.0021&2.055&9.588&0.131&0.045&0.656&0.7&0.55\\
D6 - {I}&0.002&0.0024&1.978&9.217&0.113&0.044&0.608&0.7&0.6\\
D7 - {I}&0.0022&0.0027&1.911&8.911&0.096&0.042&0.559&0.7&0.65\\
D8 - {I}&0.0024&0.003&1.854&8.657&0.079&0.04&0.509&0.7&0.7\\
D9 - {I}&0.0025&0.0032&1.804&8.444&0.064&0.037&0.456&0.7&0.75\\
D10 - {I}&0.0027&0.0034&1.762&8.244&0.049&0.033&0.401&0.7&0.8\\
D11 - {II}&0.0028&0.0036&1.724&8.056&0.036&0.029&0.341&0.7&0.85\\
D12 - {II}&0.003&0.0038&1.691&7.901&0.023&0.024&0.273&0.7&0.9\\
D13 - {III}&0.0031&0.004&1.663&7.754&0.011&0.017&0.19&0.7&0.95\\
\addlinespace
\midrule
\addlinespace
\textit{Series E}&\\
E1$^*$ - {III}&0.0008 ($^*0.001$)&0.0009&2.397&11.051&0.189&0.047&0.786&0.6&0.35\\
E2$^*$ - {I}&0.001 ($^*0.0012$)&0.0012&2.284&10.594&0.174&0.048&0.751&0.6&0.4\\
E3$^*$ - {I}&0.0013 ($^*0.0015$)&0.0015&2.184&10.105&0.157&0.048&0.71&0.6&0.45\\
E4 - {II} &0.0015&0.0017&2.094&9.733&0.14&0.047&0.669&0.6&0.5\\
E5 - {III}&0.0017&0.002&2.016&9.384&0.123&0.046&0.626&0.6&0.55\\
E6 - {I}&0.0019&0.0023&1.947&9.098&0.106&0.045&0.581&0.6&0.6\\
E7 - {I}&0.0021&0.0026&1.888&8.826&0.09&0.043&0.534&0.6&0.65\\
E8 - {I}&0.0023&0.0029&1.836&8.572&0.075&0.04&0.486&0.6&0.7\\
E9 - {I}&0.0025&0.0031&1.791&8.388&0.06&0.037&0.436&0.6&0.75\\
E10 - {I}&0.0027&0.0034&1.752&8.188&0.047&0.034&0.383&0.6&0.8\\
E11 - {III}&0.0028&0.0036&1.718&8.023&0.034&0.029&0.326&0.6&0.85\\
E12 - {II}&0.0029&0.0038&1.688&7.887&0.022&0.024&0.261&0.6&0.9\\
E13 - {III}&0.0031&0.004&1.661&7.757&0.011&0.017&0.181&0.6&0.95\\
\addlinespace
\midrule
\addlinespace
\textit{Series F}&\\
F1$^*$ - {I}&0.0008 ($^*0.001$)&0.0009&2.274&10.548&0.171&0.049&0.736&0.5&0.35\\
F2$^*$ - {I}&0.001 ($^*0.0012$)&0.0011&2.185&10.142&0.157&0.05&0.702&0.5&0.4\\
F3$^*$ - {I}&0.0012 ($^*0.0014$)&0.0014&2.105&9.788&0.142&0.049&0.665&0.5&0.45\\
F4$^{**}$ - {I}&0.0015 ($^{**}0.0016$)&0.0017&2.032&9.465&0.127&0.048&0.627&0.5&0.5\\
F5 - {I}&0.0017&0.002&1.966&9.164&0.112&0.047&0.586&0.5&0.55\\
F6 - {I}&0.0019&0.0022&1.909&8.924&0.097&0.045&0.544&0.5&0.6\\
F7 - {I}&0.0021&0.0025&1.858&8.693&0.082&0.043&0.501&0.5&0.65\\
F8 - {I}&0.0023&0.0028&1.813&8.503&0.068&0.04&0.456&0.5&0.7\\
F9 - {I}&0.0025&0.0031&1.774&8.294&0.055&0.037&0.409&0.5&0.75\\
F10 - {I}&0.0026&0.0033&1.739&8.123&0.043&0.034&0.359&0.5&0.8\\
F11 - {III}&0.0028&0.0035&1.709&8.005&0.031&0.029&0.306&0.5&0.85\\
F12 - {III}&0.0029&0.0038&1.682&7.869&0.02&0.024&0.245&0.5&0.9\\
F13 - {III}&0.0031&0.004&1.658&7.739&0.01&0.017&0.17&0.5&0.95\\
\addlinespace
\midrule
\addlinespace
\textit{Series G}&\\
G1$^*$ - {I}&0.0008 ($^*0.001$)&0.0009&2.142&9.934&0.148&0.053&0.669&0.4&0.35\\
G2 - {I}&0.001&0.0011&2.076&9.692&0.137&0.052&0.64&0.4&0.4\\
G3 - {I}&0.0012&0.0013&2.015&9.463&0.124&0.051&0.608&0.4&0.45\\
G4 - {I}&0.0014&0.0016&1.958&9.185&0.111&0.05&0.572&0.4&0.5\\
G5 - {III}&0.0016&0.0019&1.907&8.977&0.098&0.048&0.535&0.4&0.55\\
G6 - {III}&0.0018&0.0022&1.861&8.735&0.085&0.046&0.496&0.4&0.6\\
G7 - {I}&0.002&0.0025&1.821&8.544&0.072&0.044&0.456&0.4&0.65\\
G8 - {I}&0.0022&0.0027&1.784&8.362&0.06&0.041&0.415&0.4&0.7\\
G9 - {I}&0.0024&0.003&1.752&8.215&0.048&0.038&0.372&0.4&0.75\\
G10 - {III}&0.0026&0.0033&1.723&8.073&0.037&0.034&0.327&0.4&0.8\\
G11 - {III}&0.0028&0.0035&1.698&7.936&0.027&0.03&0.278&0.4&0.85\\
G12 - {III}&0.0029&0.0038&1.675&7.826&0.018&0.024&0.223&0.4&0.9\\
G13 - {III}&0.0031&0.004&1.655&7.72&0.008&0.017&0.155&0.4&0.95\\
\addlinespace
\midrule
\addlinespace
\textit{Series H}&\\
H1 - {I}&0.0008&0.0009&2.005&9.343&0.121&0.056&0.584&0.3&0.35\\
H2$^{**}$ - {I}&0.001 ($^{**}0.0011$)&0.0011&1.959&9.242&0.111&0.054&0.56&0.3&0.4\\
H3 - {III}&0.0012&0.0013&1.916&9.002&0.1&0.054&0.529&0.3&0.45\\
H4 - {I}&0.0014&0.0016&1.877&8.843&0.09&0.052&0.498&0.3&0.5\\
H5 - {II}&0.0016&0.0018&1.84&8.687&0.079&0.05&0.466&0.3&0.55\\
H6 - {II}&0.0018&0.0021&1.807&8.533&0.069&0.047&0.433&0.3&0.6\\
H7 - {I}&0.002&0.0024&1.777&8.352&0.058&0.045&0.397&0.3&0.65\\
H8 - {II}&0.0022&0.0027&1.75&8.211&0.049&0.042&0.361&0.3&0.7\\
H9 - {III}&0.0024&0.003&1.725&8.1&0.039&0.038&0.324&0.3&0.75\\
H10 - {III}&0.0026&0.0032&1.704&7.991&0.03&0.034&0.284&0.3&0.8\\
H11 - {III}&0.0027&0.0035&1.684&7.884&0.022&0.03&0.241&0.3&0.85\\
H12 - {III}&0.0029&0.0037&1.667&7.799&0.014&0.024&0.193&0.3&0.9\\
H13 - {III}&0.003&0.004&1.651&7.716&0.007&0.017&0.134&0.3&0.95\\
\addlinespace
\midrule
\addlinespace
\textit{Series I}&\\
I1 - {II}&0.0008&0.0009&1.867&8.782&0.087&0.06&0.473&0.2&0.35\\
I2 - {III}&0.001&0.0011&1.839&8.731&0.08&0.058&0.452&0.2&0.4\\
I3 - {II}&0.0012&0.0013&1.814&8.56&0.072&0.057&0.426&0.2&0.45\\
I4 - {III}&0.0014&0.0016&1.79&8.459&0.064&0.054&0.401&0.2&0.5\\
I5 - {III}&0.0016&0.0018&1.767&8.358&0.056&0.052&0.374&0.2&0.55\\
I6 - {III}&0.0018&0.0021&1.747&8.257&0.049&0.049&0.347&0.2&0.6\\
I7 - {II}&0.002&0.0024&1.728&8.156&0.042&0.046&0.319&0.2&0.65\\
I8 - {III}&0.0022&0.0027&1.71&8.056&0.035&0.043&0.289&0.2&0.7\\
I9 - {II}&0.0024&0.003&1.695&7.957&0.028&0.039&0.258&0.2&0.75\\
I10 - {II}&0.0026&0.0032&1.681&7.883&0.022&0.035&0.227&0.2&0.8\\
I11 - {III}&0.0027&0.0035&1.668&7.808&0.016&0.03&0.192&0.2&0.85\\
I12 - {III}&0.0029&0.0037&1.657&7.751&0.01&0.024&0.154&0.2&0.9\\
I13 - {III}&0.003&0.004&1.647&7.693&0.005&0.017&0.107&0.2&0.95\\
\addlinespace
\midrule
\addlinespace
\textit{Series J}&\\
J1 - {II}&0.0008&0.0009&1.741&8.207&0.047&0.065&0.321&0.1&0.35\\
J2 - {III}&0.001&0.0011&1.729&8.157&0.042&0.062&0.305&0.1&0.4\\
J3 - {III}&0.0012&0.0014&1.718&8.112&0.038&0.06&0.288&0.1&0.45\\
J4 - {III}&0.0014&0.0017&1.707&8.03&0.034&0.058&0.27&0.1&0.5\\
J5 - {III}&0.0016&0.0019&1.697&7.988&0.03&0.054&0.251&0.1&0.55\\
J6 - {III}&0.0018&0.0022&1.688&7.941&0.026&0.051&0.232&0.1&0.6\\
J7 - {II}&0.002&0.0025&1.679&7.89&0.022&0.048&0.213&0.1&0.65\\
J8 - {III}&0.0022&0.0027&1.671&7.86&0.018&0.044&0.193&0.1&0.7\\
J9 - {III}&0.0024&0.003&1.664&7.801&0.015&0.04&0.172&0.1&0.75\\
J10 - {III}&0.0026&0.0033&1.658&7.761&0.011&0.035&0.151&0.1&0.8\\
J11 - {III}&0.0027&0.0035&1.652&7.736&0.008&0.03&0.128&0.1&0.85\\
J12 - {III}&0.0029&0.0038&1.646&7.687&0.005&0.025&0.102&0.1&0.9\\
J13 - {III}&0.0031&0.004&1.642&7.654&0.003&0.017&0.071&0.1&0.95
\end{longtable}
\twocolumngrid
\newpage
\clearpage
\section{MODE STABILITY AND STARTING TIME DEPENDENCE}\label{app:Modestab}
\begin{figure}
    \centering
    \includegraphics[width=0.5\textwidth]{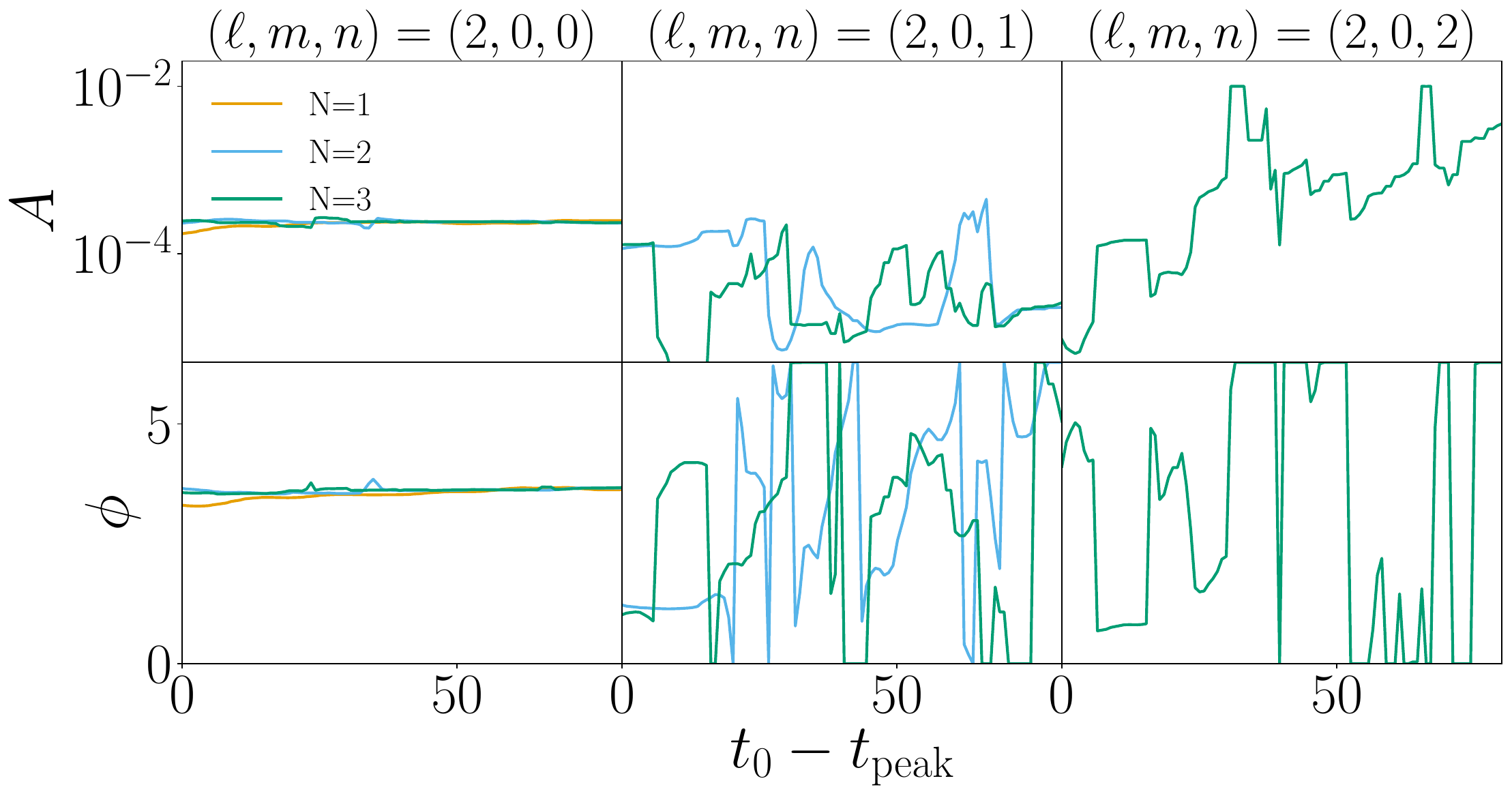}
    \caption{Each column shows the amplitude (upper row) and phase (lower row) against the starting time for the fundamental mode (left column), first overtone (middle column), and second overtone (right column) of the isolated NS model C8.}
    \label{fig:ModestabC8}
\end{figure}
\begin{figure}
    \centering
    \includegraphics[width=0.5\textwidth]{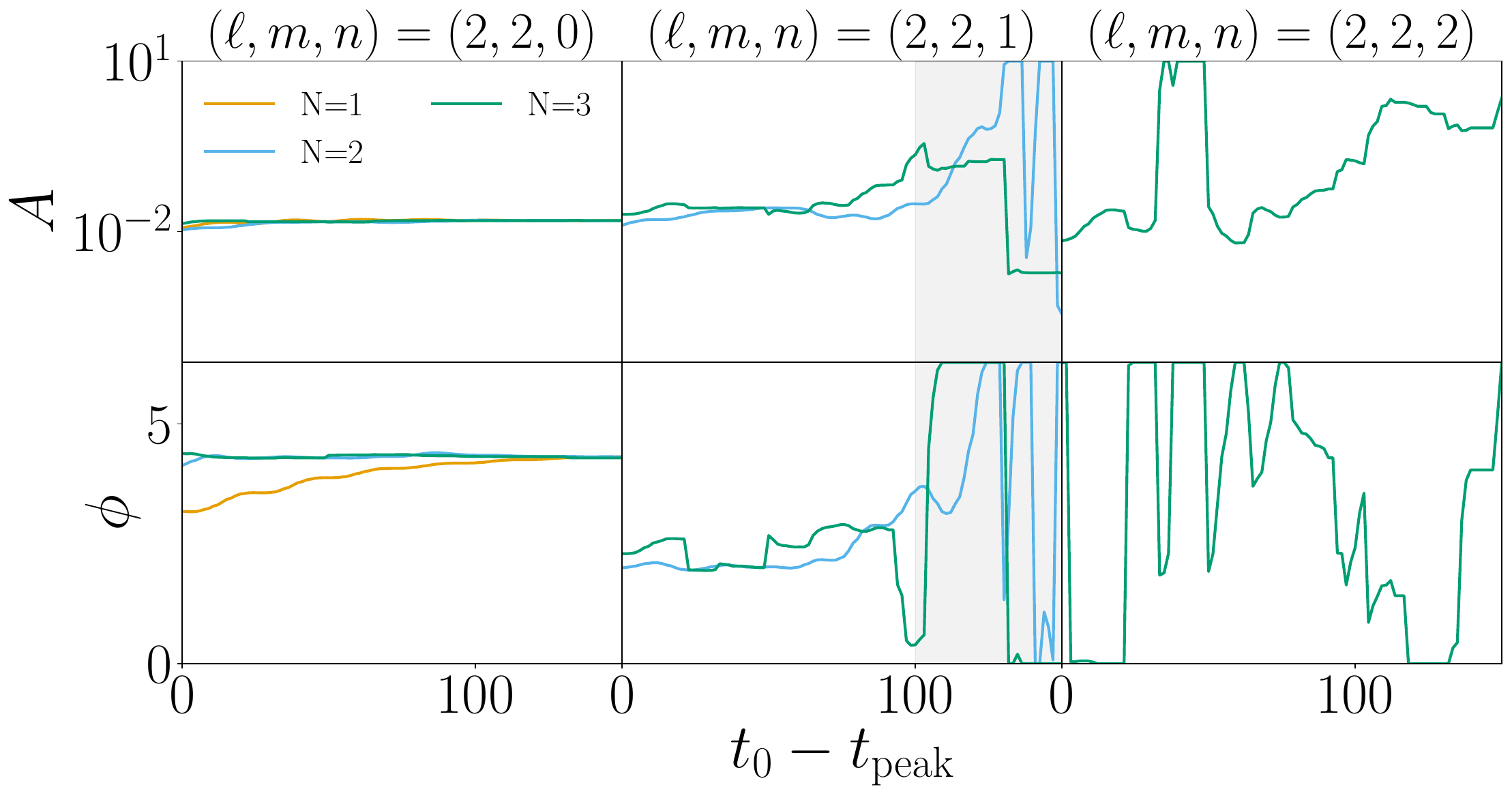}
    \caption{Similar to~\ref{fig:ModestabC8} for the $M_\mathrm{tot}=3.8M_\odot$ BNS system. In the middle column, for the first overtone, we show a gray shaded region to indicate the rough onset of instability of the mode.}
    \label{fig:ModestabM38}
\end{figure}

\begin{figure*}[t]
    \centering
    \includegraphics[width=\linewidth]{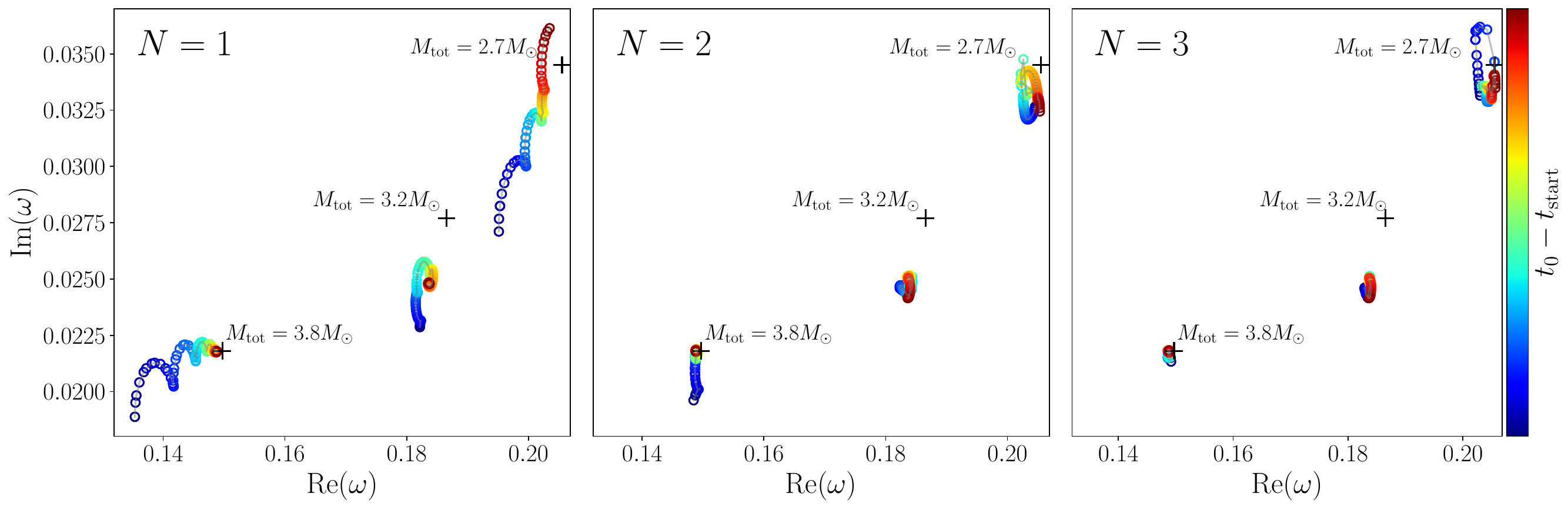}\\[1ex]
    \includegraphics[width=0.977\linewidth,right]{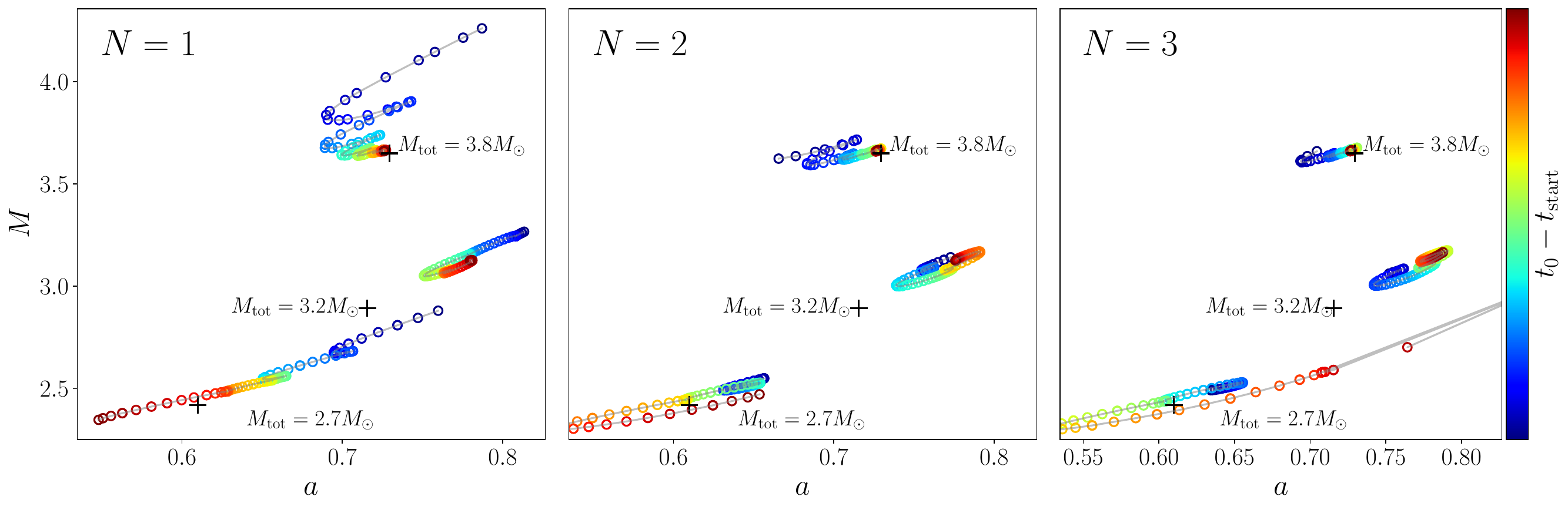}
    \caption{Each panel shows the fitting results for the one-, two-, and three-mode model in the $\mathrm{Re}(\omega)-\mathrm{Im}(\omega)$ plane (top panels) 
    and $M-a$ plane (bottom panels)
    at different starting times (as indicated by the color bar) for the three simulated BNS systems. Black crosses show the Kerr predictions for the 220 mode
    frequencies of each system (top panels), as well as the mass and spin of each system as determined by the horizon finder (bottom panels).}
    \label{fig:cycloid}
\end{figure*}

The mode stability depends generally on the studied system. Here, we start with a stability analysis of the \texttt{Class I} model C8, shown in Fig.~\ref{fig:ModestabC8}. Since we consider three different ringdown models with $N=1,2,3$ modes, we can examine the stability of the fundamental mode and the first two overtones if used in the respective model. We observe a stable fundamental mode amplitude and phase across all ringdown models, where stability improves upon the inclusion of additional modes. The same cannot be said about the first and second overtone which are quite unstable for the multimode models. In particular, from Fig.~\ref{fig:ModestabC8} it is apparent, that partial stability is only present at the start of the ringdown signal. Afterwards, the amplitude and phase of the first and second overtone blow up and hit the respective boundaries set by the prior. Increasing the bounds does not help with stability, but only delays the point when the bounds are hit. The real and imaginary part of the mode frequency on the other hand vary largely and can easily be off by more than 30\%. Other fitting algorithms might be able to extract overtones that are more stable, although these could be artificial. As there is no applicable ringdown model that treats the nonlinear effects better, overtones remain hard to extract. This suggests something already pointed out in the literature for the analysis of vacuum simulations~\cite{Giesler2019,Cheung2023,Baibhav2023,Giesler2024Overtones,Gao2025}, namely, that adding more modes can drastically reduce the mismatches between parameters, even if these modes might be physically unrealistic or even absent. A similar behavior is observed across all \texttt{Class I} ringdowns. \\

Figure~\ref{fig:ModestabM38} shows the mode stability for the $M_\mathrm{tot}=3.8M_\odot$ BNS. Again, the fundamental mode amplitude and phase are stable, although the phase is quite off at the beginning of the $N=1$ ringdown model, which improves for the multimode models. However, at least for the first overtone, we can observe a semistable amplitude and phase for the first half of the starting time window. Afterward, the amplitude and phase grow in value and at some point hit their respective bounds. So at least for the first half of the ringdown, we can claim that a first overtone might be physically present, which is not the case for the other two BNS cases. But when looking at $\delta\omega_R$ and $\delta\omega_I$, we observe deviations of $\sim10\%$ and $\gtrsim30\%$, respectively, which weakens the argument. On the other hand, the second overtone seems to be unstable over the entire range with wrong values for the frequency and damping time. These results are in agreement with the above discussion for our \texttt{Class I} model. Also, the phase of the fundamental mode and the first overtone should be phase anti-aligned by a factor of $\pi$, which is something we could at least observe for the $3.8M_\odot$ binary; see Fig.~\ref{fig:ModestabM38}. In Ref.~\cite{Volkel:2025jdx}, it was suggested, that the bias of inferred parameters should oscillate with the fundamental mode frequency. We found that this is indeed the case for some unmodeled effects when one considers the amplitude subtracted by the mean of its late-time value.\\

In addition to our discussion on mode stability, we investigated the starting time evolution for the theory-agnostic and -specific fits of the three selected BNS ringdown signals. The top panels of Fig.~\ref{fig:cycloid} show the theory-agnostic results in the $\mathrm{Re}(\omega)-\mathrm{Im}(\omega)$ plane for the three ringdown models used. When considering only the fundamental mode, we can see a cycloidal pattern across the starting time range for the heavy and light BNS. While the heavy $M_\mathrm{tot}=3.8M_\odot$ binary converges to the Kerr predictions at late times, the other systems are further from the vacuum frequencies, even at late starting times. Convergence for the heavy BNS happens faster when considering two or three modes. While this is also the case for the other two systems, they still do not converge to the predictions, but there is a visible offset. A similar picture emerges, when looking at the starting time evolution of the theory-specific fits in the $M-a$ plane, shown in the bottom panels of Fig.~\ref{fig:cycloid}. The cycloidal pattern is also visible for some systems when using a multimode ringdown model. As before, including additional modes improves the convergence, although the $M_\mathrm{tot}=2.7M_\odot$ and $M_\mathrm{tot}=3.2M_\odot$ systems do not convergence to the ``true'' mass and spin as determined by the simulation. 
\end{document}